\documentclass[prd,amsmath,aps,floats,amssymb, floatfix, superscriptaddress,nofootinbib,twocolumn,preprintnumbers]{revtex4-1}  

\setcitestyle{authoryear,round}

\usepackage[T1]{fontenc}
\usepackage{aecompl}

\usepackage{newtxtext,newtxmath}
\usepackage{mdframed,bm}
\usepackage[T1]{fontenc}
\usepackage{ae,aecompl}
\usepackage{graphicx}	
\usepackage{amsmath}	
\usepackage{amssymb}	
\usepackage{booktabs}
\usepackage{color}
\usepackage{enumitem}
\usepackage[colorlinks=true]{hyperref}

\hypersetup{
    urlcolor=black,
    citecolor=blue,
    linkcolor=blue,
  }

\usepackage{physics}
\usepackage{cleveref}
\usepackage{comment}
\usepackage[dvipsnames]{xcolor}
\usepackage{xspace}
\usepackage{xcolor}
\usepackage{float}
\restylefloat{table}
\newcommand{\lcdm}{$\Lambda$CDM\xspace}
\newcommand{\wcdm}{$w$CDM\xspace}
\usepackage{physics}

\newcommand{\aap}{A\&A}
\newcommand{\apjs}{ApJS}
\newcommand{\aj}{A.J.}
\newcommand{\mnras}{MNRAS}

\newcommand{\jcap}{JCAP}
\newcommand{\apjl}{ApJ}
\newcommand{\pasj}{PASJ}
\newcommand{\pasp}{PASP}
\newcommand{\araa}{ARAA}

\newcommand{\aaps}{AAPS}

\newcommand{\omegam}{\ensuremath{\Omega_\mathrm{m}}}
\newcommand{\buzzard}{\textsc{Buzzard}}

\newcommand{\omegab}{\ensuremath{\Omega_\mathrm{b}}}

\newcommand{\as}{\ensuremath{A_\mathrm{s}}}
\newcommand{\ns}{\ensuremath{n_\mathrm{s}}}
\newcommand{\inprep}{\emph{in preparation}}
\newcommand{\blockfont}[1]{{\textsc{#1}}\xspace}

\newcommand{\metacal}{\blockfont{Metacalibration}}

\DeclareRobustCommand{\PRF}[3]{#2}

\allowdisplaybreaks
\begin{document}

\title{Dark Energy Survey Year 3 Results: \\ Cosmology from Cosmic Shear and Robustness to Modeling Uncertainty}

\author{L.~F.~Secco}\email{secco@uchicago.edu}
\affiliation{Kavli Institute for Cosmological Physics, University of Chicago, Chicago, IL 60637, USA}
\affiliation{Department of Physics and Astronomy, University of Pennsylvania, Philadelphia, PA 19104, USA}
\author{S.~Samuroff}\email{s.samuroff@northeastern.edu}
\affiliation{McWilliams Center for Cosmology, Department of Physics, Carnegie Mellon University, Pittsburgh, PA 15213, USA}
\affiliation{Department of Physics, Northeastern University, Boston, MA, 02115, USA}
\author{E.~Krause}
\affiliation{Department of Astronomy/Steward Observatory, University of Arizona, 933 North Cherry Avenue, Tucson, AZ 85721-0065, USA}
\author{B.~Jain}
\affiliation{Department of Physics and Astronomy, University of Pennsylvania, Philadelphia, PA 19104, USA}
\author{J.~Blazek}
\affiliation{Department of Physics, Northeastern University, Boston, MA, 02115, USA}
\affiliation{Laboratory of Astrophysics, \'Ecole Polytechnique F\'ed\'erale de Lausanne (EPFL), 1290 Versoix, Switzerland}
\author{M.~Raveri}
\affiliation{Department of Physics and Astronomy, University of Pennsylvania, Philadelphia, PA 19104, USA}
\author{A.~Campos}
\affiliation{McWilliams Center for Cosmology, Department of Physics, Carnegie Mellon University, Pittsburgh, PA 15213, USA}
\author{A.~Amon}
\affiliation{Kavli Institute for Particle Astrophysics \&  Cosmology, P. O. Box 2450, Stanford University, Stanford, CA 94305, USA}
\author{A.~Chen}
\affiliation{Department of Physics, University of Michigan, Ann Arbor, MI 48109, USA}
\author{C.~Doux}
\affiliation{Department of Physics and Astronomy, University of Pennsylvania, Philadelphia, PA 19104, USA}
% ...

\collaboration{DES Collaboration}

\date{\today}

\label{firstpage}
\begin{abstract}
This work and its companion paper, \citet{y3-cosmicshear1}, present cosmic shear measurements and cosmological constraints from over 100 million source galaxies in the Dark Energy Survey (DES) Year 3 data.
We constrain the lensing amplitude parameter $S_8\equiv\sigma_8\sqrt{\omegam/0.3}$ at the 3\% level in \lcdm: 
$S_8=0.759^{+0.025}_{-0.023}$ (68\% CL). Our constraint is at the 2\% level when using angular scale cuts that are optimized for the \lcdm analysis: $S_8=0.772^{+0.018}_{-0.017}$ (68\% CL). With cosmic shear alone, we find no statistically significant constraint on the dark energy equation-of-state parameter at our present statistical power.  
We carry out our analysis blind, and compare our measurement with constraints from two other contemporary weak lensing experiments: the Kilo-Degree Survey (KiDS) and Hyper-Suprime Camera Subaru Strategic Program (HSC). We additionally quantify the agreement between our data and external constraints from the Cosmic Microwave Background (CMB).
Our DES Y3 result under the assumption of \lcdm is found to be in statistical agreement with Planck 2018, although favors a lower $S_8$ than the CMB-inferred value by $2.3\sigma$ (a $p$-value of 0.02).
This paper explores the robustness of these cosmic shear results to modeling of intrinsic alignments, the matter power spectrum and baryonic physics. We additionally explore the statistical preference of our data for intrinsic alignment models of different complexity. The fiducial cosmic shear model is tested using synthetic data, and we report no biases greater than 0.3$\sigma$ in the plane of $S_8\times\omegam$ caused by uncertainties in the theoretical models.

\vspace{25pt}

\end{abstract}

\preprint{DES-2019-0480}
\preprint{FERMILAB-PUB-21-253-AE}
\maketitle



\section{Introduction}

Discoveries and advances in modern cosmology have resulted in a remarkably simple standard cosmological model, known as \lcdm. The model is specified by a spatially flat universe, governed by the general theory of relativity, which contains baryonic matter,  dark matter, and a dark energy component that causes the expansion of the Universe to accelerate. Although remarkably simple, it appears to be sufficient to describe a great many observations, including the stability of cold disk galaxies, flat galaxy rotation curves, observations of strong gravitational lensing in clusters, the acceleration of the expansion of the Universe as inferred by type Ia supernovae (SNe Ia), and the pattern of temperature fluctuations in the Cosmic Microwave Background (CMB). Yet despite all this, \lcdm is fundamentally mysterious in the sense that the physical nature of its two main components, dark matter and dark energy, is still completely unknown.

The success of \lcdm has, however, been shaken in recent years by new experimental results. We have seen tentative hints that the model might fail to simultaneously describe the late- (low redshift) and early-time (high redshift) Universe. To take one prominent example, constraints on the local expansion parameter $H_0$ obtained from the local distance ladder and SNe Ia appear to be in tension with those inferred by the CMB \citep{Aghanim:2018eyx} at a statistically significant level \citep{Riess:2020fzl}, with varying levels of significance being reported by different probes (\citealt{freedman19, allam20}). In a separate but analogous tension, the value of the  $S_8\equiv \sigma_8 (\omegam/0.3)^{1/2}$ parameter --- the amplitude of mass fluctuations $\sigma_8$ scaled by the square root of matter density $\Omega_\mathrm{m}$ --- differs when inferred  via cosmological lensing  \citep{TroxelY1, hikage2019, asgari20} from the value obtained using Planck (assuming \lcdm; \citealt{Aghanim:2018eyx}) at the level of $2-3\sigma$. Other probes of the late Universe, in particular spectroscopic galaxy clustering \citep{troester20}, redshift-space distortions \citep{allam20} and the abundance of galaxy clusters \citep{mantz15,y1clusters}, also all tend to prefer relatively low values of $S_8$. Although the evidence is by no means definitive, we are perhaps beginning to see hints of new physics, and so stress-testing \lcdm with new measurements is extremely important.

Cosmic shear, or cosmological weak lensing (the two-point correlation function of gravitational shear), is one of the most informative of the the low redshift probes.  
It has two main advantages, as a means to infer the properties of the large scale Universe \citep{Bartelmann:1999yn,Huterer:2001yu,Hu_Jain_2004,Frieman08}. First, the signal is insensitive to galaxy bias, which is a significant source of uncertainty in cosmological analyses based on galaxy clustering and galaxy-galaxy lensing. Second, weak lensing is sensitive both to the \textit{geometry} of the Universe through the lensing kernel (which is a function of $H_0$ and ratios of angular diameter distances), and also to the \textit{growth of structure} and its evolution in redshift. Since geometry and structure growth are tightly related to the evolution of dark energy and its equation-of-state parameter $w$, this sensitivity carries over to the cosmic shear signal.

Cosmic shear was first measured over twenty years ago, roughly simultaneously by a number of groups \citep{wittman2000,bacon00,vanwaerbeke00,kaiser00}. Although too noisy to constrain cosmological parameters, these observations represented the first steps towards fulfilling the potential pointed out by theoretical 
studies years earlier \citep{miraldaescude1991, jain1997, Hu1999}. The intervening two decades have seen steady improvements in signal-to-noise and cosmological constraining power, as new ground- and space-based lensing data sets have become available \citep{Hoekstra:2002cj,Refregier:2002ux,Jarvis:2002vs,Hamana:2002yd,Brown:2002wt,Rhodes:2003wj,Heymans:2004zp,Jarvis:2005ck,Hetterscheidt:2006up,Massey:2007gh,Leauthaud:2007fb,Benjamin:2007ys,Fu:2007qq,Schrabback:2009ba,Huff:2011gq,Huff:2011aa,Lin:2011bc,Jee:2012hr,Erben:2012zw,Miller:2012am,Kilbinger:2012qz,Kitching:2014dtq,Jee:2015jta,Asgari:2016xuw,Yoon:2018eua,hikage2019,hamana20,asgari20}. 
As the volume and quality of lensing data have improved, so too have the methods used to study it, with the development of an array of sophisticated statistical and theoretical tools. There has, for example, been a coherent effort to test and improve shape measurement algorithms using increasingly complex image simulations \citep{heymans06, massey07, bridle06, kitching12, mandelbaum2014}. Methods for estimating the distribution of source galaxies along the line of sight have also gradually evolved to become highly sophisticated, incorporating various sources of information (\citealt*{gatti18}; \citealt*{prat19}; \citealt*{y3-sompzbuzzard}; \citealt{sanchez19,alarcon19,wright20}). 

Alongside the Dark Energy Survey (DES)\footnote{ \url{https://www.darkenergysurvey.org/}}, the major lensing surveys of the current generation are the Kilo-Degree Survey (KiDS; \citealt{deJong})\footnote{  \url{http://kids.strw.leidenuniv.nl/DR4/index.php}}  and the Hyper-Suprime Camera Subaru Strategic Program (HSC; \citealt{aihara18})\footnote{\url{https://www.naoj.org/Projects/HSC}}. We show the approximate, nominal footprints of these surveys in Fig.~\ref{fig: footprint}. Each of these three collaborations have, in recent years, released cosmic shear analyses analogous to the one presented in this paper. Lensing analyses based on HSC data were carried out over a footprint of 136.9 deg$^2$ split into six fields (red patches in Fig.~\ref{fig: footprint}; \citealt{mandelbaum17}); they presented consistent cosmology results using two types of statistics: real space correlation functions \citep{hamana20} and harmonic space power spectra \citep{hikage2019}. More recently, the KiDS collaboration released results based on approximately 1000 deg$^2$ of data (blue patches in Fig. \ref{fig: footprint}; \citealt{giblin20}), and presented an analysis of band-power spectra, correlation functions and the complete orthogonal sets of E-/B-mode integrals (COSEBIs \citealt{asgari20}). They further combined their cosmic shear results with external spectroscopic data from BOSS \citep{alam2015} to obtain a 3$\times$2pt constraint \citep{heymans2020}, which is internally consistent with their cosmic shear results, but differs from Planck in the full parameter space by $\sim 2\sigma$. 

The trend observed in earlier cosmic shear studies is that the amplitude of the cosmic shear signal (tied to the amplitude of matter fluctuations through the $S_8$ parameter) is
lower than that extrapolated from the CMB. In order to demonstrate whether this discrepancy is physical and significant, we must have a high degree of confidence in our modeling of the data and its possible systematic errors. Among the most significant of these sources of systematic error are intrinsic alignments (IAs), or astrophysically sourced correlations of galaxy shapes, which mimic cosmic shear. Given how difficult it is to disentangle IAs from lensing, the most common approach is to forward-model their effect, assuming a model for the IA power spectrum with a number of free parameters. Depending on the galaxy sample, however, IA model insufficiency can easily translate into a bias in cosmological parameters \citep{krause2016,blazek19}. In addition to IAs, effects such as nonlinear growth and the impact of baryons on the large-scale distribution of dark matter can alter the matter power spectrum in a significant way, and so bias the inferred lensing amplitude if neglected \citep{schneider2019quantifying,deRose_aemulus,martinelli20,Huang2020,2021ApJ...908...13Y}. Although it is clear that these effects are scale-dependent, finding the angular scales where our modeling is sufficient is by no means straightforward. This paper describes the choices made in modeling and scale cuts, and validates that the potential biases on cosmological parameters are smaller than the statistical uncertainties.

Our companion paper \citep{y3-cosmicshear1} presents a detailed investigation of observational errors that can similarly bias cosmological inference. Undiagnosed biases in the shear measurement process, for example, can lead one to incorrectly infer the lensing amplitude. Likewise, errors in the estimation of galaxy redshift distributions $n(z)$ can subtly alter the interpretation of the lensing measurement, both in terms of cosmology and of IAs.  \citet{y3-cosmicshear1} demonstrate that these measurement systematic errors are well controlled in the Y3 cosmic shear analysis. \textit{We note that the main cosmological constraints presented in both papers are identical}.

The cosmic shear analysis presented in this paper, and in \citet{y3-cosmicshear1}, is part of a series of Year 3 cosmological results from large-scale structure produced by the Dark Energy Survey Collaboration.
This work relies on many companion papers that validate the data, catalogs and theoretical methods; those papers, as well as this one, feed into the main ``$3\times2$pt'' constraints, which combine cosmic shear with galaxy-galaxy lensing and galaxy clustering in \lcdm~and \wcdm \citep{y3-3x2ptkp}, as well as extended cosmological parameter spaces \citep{y3-extensions}. These include: 

\begin{itemize}
    \item  The construction and validation of the \blockfont{Gold} catalog of objects in DES Y3 is described in \citet*{y3-gold}.
    
    \smallskip
    
    \item The Point-Spread Function (PSF) modeling algorithm and its validation tests are described in \citet{y3-piff}.

    \smallskip
    
    \item A suite of image simulations, used to test the shape measurement pipeline and ultimately determine the shear calibration uncertainties is described in \citet{y3-imagesims}.

    \smallskip
    
    \item The \metacal~shape catalog, and the tests that validate its science-readiness, are described in
    \citet*{y3-shapecatalog}. This paper also discusses the (first layer) catalog-level blinding implemented in Y3.
    
    \smallskip
    
    \item The characterization of the source redshift distribution, and the related systematic and statistical uncertainties, are detailed in five papers. Namely, \citet*{y3-sompz} and \citet*{y3-sompzbuzzard} present the baseline methodology for estimating wide-field redshift distributions using Self-Organizing Maps; \citet*{y3-sourcewz} outline an alternative method using cross correlations with spectroscopic galaxies; 
    \citet*{y3-shearratio} presents a complementary likelihood using small scale galaxy-galaxy lensing, improving constraints on redshifts and IA;  
    finally \citet*{y3-hyperrank} validates our fiducial error parameterization using a more complete alternative based on distribution realizations.
    In addition to this, \citet*{y3-deepfields} and \citet{y3-balrog} respectively describe the DES deep fields and the Balrog image simulations, both of which are crucial in testing and implementing the Y3 redshift methodology. 
    
    \smallskip

    \item The data covariance matrix is described in \citet{y3-covariances}. This paper also presents various validation tests based on DES Y3 simulations, and demonstrates its suitability for likelihood analyses.
    
    \smallskip
    
    \item The numerical metrics used to assess tension between our DES results and external data sets are described in \citet{y3-tensions}. That work considers a number of alternatives, and sets out the methodology used in this paper and \citet{y3-3x2ptkp}. 
      
    \smallskip
    
    \item The simultaneous blinding of the multiple DES Y3 probes at the two-point correlation function level is described in \citet{JessieBlinding};  
    
    \smallskip 
    \item \citet{y3-simvalidation} presents a set of cosmological simulations which are used as an end-to-end validation of our analysis framework on mock $N$-body data.
    
    \smallskip
    \item Finally, tests of the theoretical and numerical methods, as well as modeling assumptions for all $3\times2$pt analyses are described in \citet{y3-generalmethods}.

\end{itemize}

This paper is organized as follows: Sec. \ref{sec:data} describes the DES Y3 data, and the catalog construction and calibration. Sec. \ref{sec:measurements} describes the two-point measurements upon which our results are based, as well as the covariance estimation and blinding scheme. In Sec. \ref{sec:model}, we describe the theoretical modeling of the cosmic shear two-point data vector. We demonstrate our model is robust to various forms of systematic error, using simulated data, in Sec. \ref{section: unblinding}. Our baseline results and an exploration of the IA model complexity present in our data are then presented in Sec. \ref{section: baseline cosmology}. In Sec. \ref{section: robustness to modeling} we present a series of reanalyses, using slightly different modeling choices, in order to verify the robustness of our findings. The consistency of DES Y3 cosmic shear data with external probes such as other weak lensing surveys and the CMB is examined in Sec. \ref{sec:external_data}. Finally, Sec. \ref{sec:conclusions} summarizes our findings and discusses their significance in the context of the field.

\section{DES Y3 Data \& Sample Selection}\label{sec:data}
This section briefly describes the DES Y3 data, and defines the galaxy samples used in this paper. We also discuss a number of related topics, including calibrating selection biases.

\subsection{Data Collection \& the \blockfont{Gold} Selection}

DES has now completed its six-year campaign, covering a footprint of around 5000 deg$^2$ to a depth of $r \sim 24.4$. The DES data were collected using the 570 megapixel Dark Energy Camera (DECam; \citet{flaugher15}), at the Blanco telescope at the Cerro Tololo Inter-American Observatory (CTIO), Chile, using five photometric filters $grizY$, which cover a region of the optical and near infrared spectrum between $0.40$ and $1.06$ $\mu \mathrm{m}$. 
DES SV, Y1 and Y3 cover sequentially larger fractions of the full Y6 footprint, with Y3, the data set used in this analysis, encompassing 4143 deg$^2$ after masking, with the ``Wide Survey'' footprint covered with 4 overlapping images in each band (compared with the final survey depth of $\sim$ 8). 
The images undergo a series of reduction and pre-processing steps, including background subtraction \citep{bernstein18,morganson18,eckert20}, and masking out cosmic rays, satellite trails and bright stars. Object detection is performed on the $riz$ coadd images using Source Extractor \citep{bertin96}.
For the detected galaxies, derived photometric measurements are generated using Multi-Object Fitting (MOF; \citealt{y1gold}) to mitigate blending. 
The final Y3 selection with baseline masking is referred to as the \blockfont{Gold} catalog, and is described in detail in \citet*{y3-gold}.

\begin{figure*}
	\includegraphics[width=\textwidth]{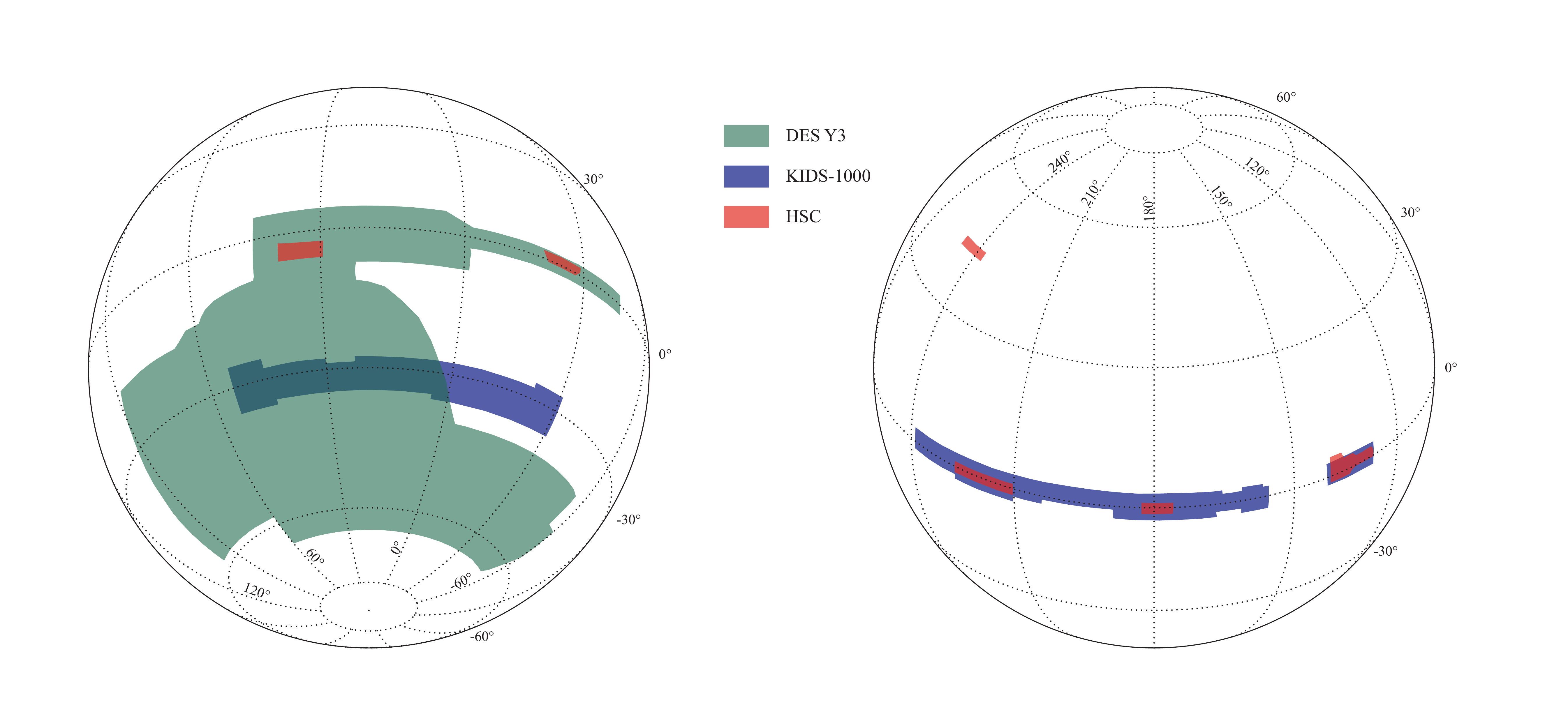}
    \caption{The approximate footprints of Stage-III dark energy experiments: Dark Energy Survey Year 3 (DES Y3; green), Kilo-Degree Survey (KiDS-1000; blue) and first-year Hyper Suprime-Cam Subaru Strategic Program (HSC; red). The left and right panels show orthographic projections of the northern and southern sky respectively. The parallels and meridians show declination and right ascension. The different survey areas not only affect the final analysis choices, but also reflect the individual science strategies and the complementarity of Stage-III surveys.}\label{fig: footprint}
\end{figure*}

\subsection{Shape Catalog \& Image Simulations}

The DES Y3 shape catalog is created using the \metacal algorithm \citep{Sheldon_Huff_2017,Huff_Mandelbaum_2017}. 
The basic shape measurement entails fitting a single elliptical Gaussian to each detected galaxy. The fit is repeated on artificially sheared copies of the given galaxy, in order to construct a shear response matrix  $R_{\gamma}$ via a numerical derivative; a selection response $R_S$ is also computed in a similar way. These multiplicative responses are the essence of \metacal.
After quality cuts, the Y3 \metacal~catalog contains over 100 million galaxies, with a mean redshift of $z=0.63$ and a weighted number density\footnote{The effective number density here is as defined by \citet{heymans12}. The equivalent value using \citet{chang13}'s definition is $5.32 \; \mathrm{arcmin}^{-2}$ (see \citealt*{y3-shapecatalog} for details.)} $n_{\rm eff} = 5.59 \; \mathrm{arcmin}^{-2}$; for discussion of the cuts and why they are necessary, see \citet*{y3-shapecatalog}.

Although \metacal~greatly reduces the biases inherent to shear estimation, the process is not perfect. We must still rely on image simulations for validation and for deriving priors on the residual biases (predominantly due to blending, and its impact on the redshift distribution). These simulations, and the conclusions we draw from them for Y3, are discussed in \citet{y3-imagesims}.
In addition to tests using simulations, the catalogs are subject to a number of null tests, applied directly to the data. Using both pseudo-$C_\ell$s and COSEBIs \citep{Schneider2010A&A...520A.116S}, we find no evidence for non-zero B-modes in Y3.

\subsection{Photometric Redshift Calibration}\label{sec: photoz calibration}

We estimate and calibrate the redshift distributions of our source sample with a combination of three different methods. Our base methodology is known as  \textit{Self-Organizing Map $p(z)$} (SOMPZ; \citealt*{y3-sompz}). The most important aspect of this methodology is that knowledge from precise redshifts (from spectroscopic samples) and higher-quality photometric data (from DES deep fields \citealt*{y3-deepfields}) informs the bulk of the DES observations (the wide fields), essentially acting as a Bayesian prior. The connection between the deep and wide field data is determined empirically using an image simulation framework known as Balrog \citep{y3-balrog}.

Additionally, clustering redshifts (WZ; \citealt*{y3-sourcewz}) employ cross-correlations of galaxy densities to improve redshift constraints, and shear ratios \citep*{y3-shearratio} help to constrain redshifts (and also intrinsic alignment parameters), utilizing galaxy-shear correlation functions at small scales. While SOMPZ and WZ are applied upstream to generate and select $n(z)$ estimates, the shear-ratio information, on the other hand, is incorporated at the point of evaluation of cosmological likelihoods (see Sec. \ref{section:SR}). Details of each of these methods in the context of Y3 can be found in \citet*{y3-sompz}, and robustness tests of redshift distributions in the context of cosmic shear are presented in  \citet{y3-cosmicshear1}.

\section{Cosmic Shear Measurement}\label{sec:measurements}

 In this section we present the measured real-space cosmic shear two-point correlations ($\xi_\pm(\theta)$, see Eq. \ref{eq: xipm Hankel transform}), which form the basis of our results and are shown in Fig. \ref{fig: measured xipm}. Defining the signal-to-noise of our measurement as
 \begin{equation}
     \textrm{S/N}\equiv\frac{\xi^{\rm data}_\pm(\theta)^T\mathbf{C}^{-1}\xi^{\rm model}_\pm(\theta)}{\sqrt{\xi^{\rm model}_\pm(\theta)^T\mathbf{C}^{-1}\xi^{\rm model}_\pm(\theta)}},
 \end{equation}
where $\mathbf{C}$ is the data covariance matrix, the S/N of the cosmic shear detection in DES Y3 after scale cuts is 27. 
For the fiducial \lcdm model, our chi-square at the maximum posterior is $\chi^2=237.7$, with 222 effective degrees-of-freedom (d.o.f.), which gives us a $p-$value of 0.22 (see Sec. \ref{sec: LCDM results}). We define these quantities in more detail in Sec. \ref{sec:model}.

\subsection{Tomography}\label{sec:measurements:tomography}

We define a set of four broad redshift bins for our source sample in the nominal range $0<z<3$, with actual number densities being fairly small above $z\gtrsim1.5$. These are constructed by iteratively adjusting the redshift bin edges, such that they each yield approximately the same number of source galaxies. 
The Y3 SOMPZ methodology (see Section \ref{sec: photoz calibration}) makes use of Balrog, which artificially inserts COSMOS galaxies into DES images. The artificial galaxies are assigned to cells in both the wide- and deep-field Self-Organizing Maps (SOMs), which allows one to map between the two, and so assign DES wide-field galaxies to bins (see \citealt*{y3-sompz}, Sec. 4.3).

The redshift distributions computed in this way, which feed into our modeling in the next section, are shown in Fig. \ref{fig: nz}. The galaxy number densities are 1.476, 1.479, 1.484 and 1.461 per square arcminute respectively in these four redshift bins.

\subsection{Two-Point Estimator \& Measurement }\label{section: 2pt estimator}

The spin-2 shear field can be expressed in terms of a real and an imaginary component, $\gamma = \gamma_1 + i\gamma_2$. There are two possible shear two-point functions that preserve parity invariance, and a ``natural'' convention for them is:  $\xi_{+}\equiv\left\langle \gamma\gamma^{*}\right\rangle$ and $\xi_{-}\equiv\left\langle \gamma\gamma\right\rangle$  \citep{Schneider_Lombardi_2003}, where the angle brackets denote averaging over galaxy pairs. In terms of tangential ($t$) and cross ($\times$) components defined along the line that connects each pair of galaxies $a,b$, we have:
\begin{equation}
\xi_{\pm}=\left\langle \gamma_{t,a}\gamma_{t,b}\right\rangle_{ab} \pm\left\langle \gamma_{\times,a}\gamma_{\times,b}\right\rangle_{ab}.\label{eq:xipm} 
\end{equation} 
\noindent
In practice we do not have direct access to the shear field, but rather estimate it via per-galaxy ellipticities (although see \citealt{bernstein16} for an alternative approach). Correlating galaxies in a pair of redshift bins $(i,j)$ we define,
\begin{equation}
\xi_{\pm}^{ij}(\theta)=
\frac{\underset{ab}{\sum} w_a w_b \left(\hat{e}_{t,a}^{i}\,\hat{e}_{t,b}^{j}
\pm
\hat{e}_{\times,a}^{i}\,\hat{e}_{\times,b}^{j}\right)}
{\underset{ab}{\sum}w_{a}w_{b} R_a R_b,
}\label{eq:xipm_estimator}
\end{equation} 
\noindent
with inverse variance weighting $w$\footnote{Although referred to as such, the catalog weights only approximate inverse variance weighting. See \citet*{y3-shapecatalog}, Sec 4.3 for details.} (unlike in Y1, where such weighting was not included) and response factors $R$ that account for shear and selection biases (see \citealt*{y3-shapecatalog} for details), and where the sums run over pairs of galaxies $a,b$, for which the angular separation falls within the range $|\boldsymbol{\theta}-\Delta\boldsymbol{\theta}|$ and $|\boldsymbol{\theta}+\Delta\boldsymbol{\theta}|$.
Both $\xi_+$ and $\xi_-$ are measured using twenty log-spaced $\theta$ bins between $2.5$ and $250$ arcminutes, with $i,j\in (1,2,3,4)$. As discussed later, not all of the twenty angular bins are utilized in our likelihood analysis. We also assume the response matrix is diagonal and that the selection part is scale independent.
The ellipticities that enter Eq. \eqref{eq:xipm_estimator} are corrected for residual mean shear, such that $\hat{e}^i_k\equiv e^i_k - \langle e_k \rangle_i$ for components $k\in (1,2)$ and redshift bin $i$, again following the Y1 methodology \citep{TroxelY1}.
We show the resulting two-point functions, which are measured using using TreeCorr\footnote{\url{https://github.com/rmjarvis/TreeCorr}} \citep{TreeCorr}, in Fig. \ref{fig: measured xipm}, alongside best fitting theory predictions.

\subsection{Data Covariance Matrix}\label{sec: data covariance}

We model the statistical uncertainties in our combined measurements of $\xi_\pm$ as a multivariate Gaussian distribution. The disconnected 4-point function part of the covariance matrix of that data vector (the Gaussian covariance part) is described in \citealt{y3-covariances} and includes analytic treatment of bin averaging and sky curvature. We also verify in that paper that expected fluctuations in $\Delta\chi^2$ between the  measurement and our maximum posterior model do not significantly impact on our estimates of cosmological parameters. Our modeling of the connected 4-point function part of the covariance matrix and the contribution from super-sample covariance uses the public  CosmoCov\footnote{\url{https://github.com/CosmoLike/CosmoCov}} \citep{2020MNRAS.497.2699F} code, which is based on the CosmoLike framework \citep{cosmolike}. 

We use the RMS per-component shape dispersion $\sigma_e$ and effective number densities $n_\textrm{eff}$ specified in Table 1 of \citealt{y3-cosmicshear1} to calculate the shape-noise contribution to the covariance, and additionally account for survey geometry effects. 
We follow previous cosmic shear analyses in using a covariance matrix that assumes a baseline cosmology (see \citealt{hikage2019} for a different approach). That is, we assume a fiducial set of input parameters for the initial covariance matrix and run cosmological chains using this first guess. The covariance is then recomputed at the best fit from this first iteration, and the final chains are run. We find this update to have negligible effects on the cosmic shear constraints presented in this paper.

\subsection{Blinding}\label{section: blinding}

We implement a three-stage blinding strategy, performing transformations to the catalog, data vector, and parameters in order to obscure the cosmological results of the analysis. By disconnecting the people carrying out the analysis from the impact their various choices are having on the eventual cosmological results, the aim is to avoid unconscious biases, either towards or away from previous results in the literature. Although the approaches differ somewhat, all of the major cosmic shear collaborations have adopted a similar philosophy regarding the necessity of blinding \citep{svcosmicshear,TroxelY1,hikage2019,hildebrandt20,asgari20}.

The first level of blinding follows a similar method to that used in Y1 \citep{y1shearcat}, and is discussed in \citet*{y3-shapecatalog} (their Sec. 2.3). In short, the process involves a transformation of the shear catalog, where galaxy shapes are scaled by a random multiplicative factor. 
The second level is a transformation of the data vector using the method described in \citet{JessieBlinding}. We compute model predictions at two sets of input parameters: an arbitrary reference cosmology $\boldsymbol{\Theta}_{\textrm{ref}}$ and a shifted cosmology  $\boldsymbol{\Theta}_{\textrm{ref}}+\Delta\boldsymbol{\Theta}$, where $\Delta\boldsymbol{\Theta}$ is drawn randomly in \wcdm parameter space. The difference between these model predictions is then applied to the measured $\xi_\pm$ data vector prior to its analysis. The final stage of blinding is at the parameter level, and entails obscuring the axes of contour plots (effectively equivalent to shifting contours randomly in parameter space, preserving constraining power but making external consistency testing impossible). A detailed checklist of the tests that must be fulfilled before each stage of blinding can be removed can be found in \citet{y3-3x2ptkp}. 

From a modeling perspective, passing the tests we describe in Sec. \ref{section: unblinding} and the further tests on synthetic data described in  \citet{y3-simvalidation} and \citet{y3-generalmethods} fulfills our unblinding requirements. A set of internal consistency tests for cosmic shear must also be passed and are described in \citet{y3-cosmicshear1}.

\begin{figure*}
    \includegraphics[width=\textwidth]{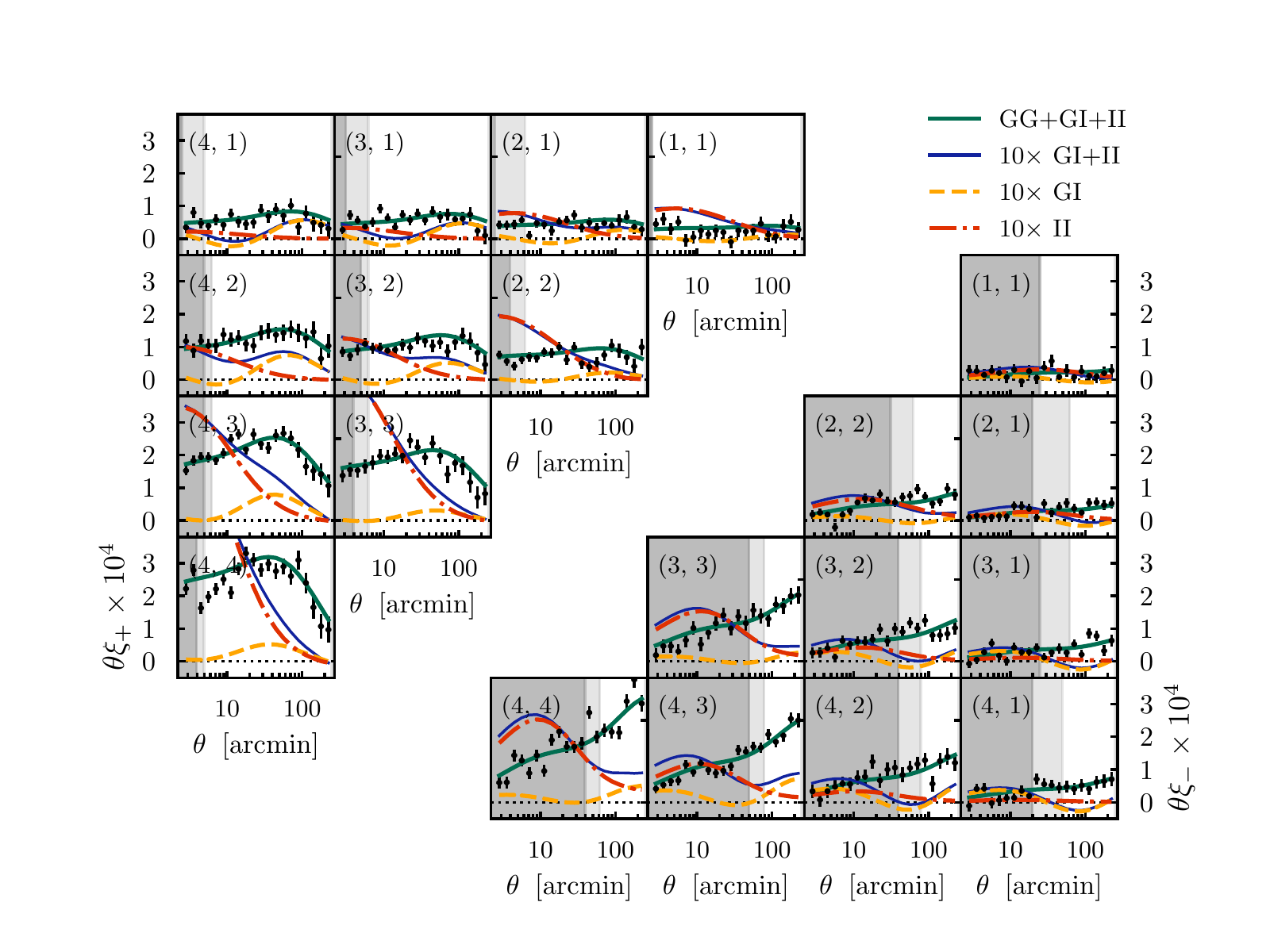}
    \caption{Cosmic shear two-point correlation measurements from DES Y3. We show here $\xi_+$ and $\xi_-$ (black data points, upper left and lower right halves respectively), with the different panels showing different combinations of redshift bins; in all cases the error bars come from our fiducial analytic covariance matrix. The lighter grey bands represent scales removed from our fiducial analysis, while the darker are the equivalent for the \lcdm~Optimized analysis. Also shown are the best-fit theory curve in \lcdm (solid green) and the intrinsic alignment contributions to the signal: GI (dashed yellow), II (dot-dashed red) and GI+II (solid blue). For clarity, we multiply the IA contributions by a factor of 10, and in most bins the total IA signal is $\sim 1 \%$ of GG+GI+II. The detection significance of the cosmic shear signal after fiducial scale cuts is 27. The $\chi^2$ per effective d.o.f of the \lcdm model is  $237.7/222 = 1.07$ (a $p$-value of 0.22).}
    \label{fig: measured xipm}
\end{figure*}

\section{Fiducial Model and Analysis Choices}\label{sec:model} 

A predictive physical model for cosmic shear has a number of requirements; first, any systematic deviations or effects omitted from the model must be comfortably subdominant to uncertainties on the data; and second, the implementation must be numerically stable at all points within the prior volume. 
We also aim for redundancy, and implement the full pipeline in two independent codes: \blockfont{CosmoSIS}\footnote{\url{https://bitbucket.org/joezuntz/cosmosis}} \citep{zuntz15} and CosmoLike \citep{cosmolike}, which are verified to be in agreement to within a negligible $\Delta \chi^2$ \citep*{y3-generalmethods}. 

In this section, we outline our baseline model for $\xi^{ij}_\pm(\theta)$ and discuss how it meets the above criteria. We subsequently show the cosmological constraints from these analysis choices in Sec. \ref{section: baseline cosmology}. To test the robustness of this baseline model, we later relax its main approximations and assumptions and show variations of analysis choices in Sec. \ref{section: robustness to modeling}. 

\subsection{Sampling and Parameter Inference}\label{section: 2pt_like}
For all parameter inference presented in this paper, we assume the likelihood of the data given the model $M$ with parameters $\boldsymbol{\mathrm{p}}$ to be a multivariate Gaussian:
\begin{equation}
\ln\mathcal{L}(\hat{\textbf{D}}|\boldsymbol{\mathrm{p}},M)=-\frac{1}{2}\chi^{2}+\textrm{const.,}
\end{equation}
\begin{equation}\label{eq: chi2}
\chi^{2}=\left(\hat{\textbf{D}}-\mathbf{T}_{M}(\boldsymbol{\mathrm{p}})\right)^{T}\mathbf{C}^{-1}\left(\hat{\textbf{D}}-\mathbf{T}_{M}(\boldsymbol{\mathrm{p}})\right)\end{equation}
\noindent
where \textbf{C} is the data covariance matrix and $\mathbf{T}_{M}(\boldsymbol{\mathrm{p}})$ is the theory prediction vector for a data vector $\hat{\textbf{D}}$, a concatenated version of all  elements of the tomographic cosmic shear data (with length $N_{D} = N_{\theta} N_{\rm z} (N_{\rm z} +1 )$, where $N_{\theta}$ is the number of angular bins included in each correlation function after  scale cuts ($N_{\theta}$ varies depending on the redshift bins, and equals 20 before cuts) and $ N_{\rm z}=4$ is the number of broad redshift bins). Since we are incorporating shear ratios at the inference level (see Sec. \ref{section:SR}), the final likelihood used in our analyses is the sum of two parts, $\mathrm{ln} \mathcal{L} = \mathrm{ln}\mathcal{L}_{\rm 2pt} + \mathrm{ln}\mathcal{L}_{\rm SR}$, which are assumed to be independent \citep*{y3-shearratio}.  

As we aim to perform a Bayesian analysis, the \textit{a posteriori} knowledge of the parameters given the observed data, denoted by $\mathcal{P}(\boldsymbol{\mathrm{p}}|\hat{\textbf{D}},M)$, depends not only on the likelihood but also on prior $\Pi(\boldsymbol{\mathrm{p}}|M)$. These pieces are related via Bayes' theorem:
\begin{equation}\label{eq: Bayes theorem}
\mathcal{P}(\boldsymbol{\mathrm{p}}|\hat{\textbf{D}},M)=\frac{\mathcal{L}(\hat{\textbf{D}}|\boldsymbol{\mathrm{p}},M)\Pi\left(\boldsymbol{\mathrm{p}}|M\right)}{P(\hat{\textbf{D}}|M)},
\end{equation}
where $P(\hat{\textbf{D}}|M)$ is the so-called evidence of the data. Sampling of the posterior is carried out using \blockfont{Polychord} (\citealt{handley15}; with 500 live points and tolerance 0.01). These settings have been tested to demonstrate the accuracy of the posteriors and Bayesian evidence estimates. 
Although sampling gives a rather noisy estimate of the best-fit point in the full parameter space, the \textit{Maximum a Posteriori} (MAP) quoted in Sec. \ref{sec: LCDM results}, we verify that standard \blockfont{Polychord} outputs, in practice, offer a reasonable estimate of that point when compared to a MAP optimizer\footnote{We run the \blockfont{MaxLike} sampler in Posterior mode for this test (\url{https://bitbucket.org/joezuntz/cosmosis/wiki/samplers/maxlike})}. Throughout this paper, we report parameter constraints using the MAP value and 1D marginalized summary statistics in the form:
\begin{equation*}
\textrm{Parameter = 1D mean}_{-\textrm{lower 34\% bound}}^{+\textrm{upper 34\% bound}}\textrm{ (MAP value)}.
\end{equation*}

The ratio of evidences is a well-defined quantity for model-testing within a single data set to indicate a preference for one modeling choice ($M_1$, with parameters $\boldsymbol{\mathrm{p}}_1$) over another ($M_2$, $\boldsymbol{\mathrm{p}}_2$): 

\begin{equation}\label{eq: evidence ratio}
R_{M_{1}/M_{2}}=\frac{P(\hat{\textbf{D}}|M_{1})}{P(\hat{\textbf{D}}|M_{2})}=\frac{\int d\boldsymbol{\mathrm{p}}\,\mathcal{L}(\hat{\textbf{D}}|\boldsymbol{\mathrm{p}}_{1},M_{1})P(\boldsymbol{\mathrm{p}}_{1}|M_{1})}{\int d\boldsymbol{\mathrm{p}}\,\mathcal{L}(\hat{\textbf{D}}|\boldsymbol{\mathrm{p}}_{2},M_{2})P(\boldsymbol{\mathrm{p}}_{2}|M_{2})}.
\end{equation}
\noindent
The evidence ratio has the advantage of naturally penalizing models of excessive parameter space volume, but needs to be interpreted using e.g. the Jeffreys scale \citep{jeffreys35}, which somewhat arbitrarily differentiates between ``strong'' and ``weak'' model preferences. Our main use of evidence ratios is to help assessing model preference in the context of IA complexity in Sec. \ref{sec: IA model selection}.

\subsection{Modeling Cosmic Shear}

The two-point cosmic shear correlations $\xi^{ij}_\pm(\theta)$ are related to the nonlinear matter power spectrum (and thus to the growth and evolution of structure). The key quantity that dictates how much a galaxy on a particular line of sight is distorted, is known as the \emph{convergence} $\kappa$. That is, the weighted mass overdensity $\delta$, integrated along the line-of-sight to the distance of the source $\chi_\mathrm{s}$:
\begin{equation}
    \kappa\left(\theta\right)= 
    \int_{0}^{\chi_{\mathrm{s}}}d\chi\,W(\chi)\delta(\theta,\chi).\label{eq: convergence}
\end{equation}

\noindent
The weight for a particular lens plane, quantified below in Eq. (\ref{eq: lensing kernel W}), is sensitive to the relative distances of the source and the lens; it is via this geometrical term that cosmic shear probes the expansion history of the Universe. Fitting all the auto- and cross-redshift bin correlations simultaneously significantly improves the cosmological constraining power  \citep{Hu1999}, both because it helps to untangle the signal at different epochs, and because it (partially) breaks the degeneracy with intrinsic alignments (see also Sec. \ref{section: IA modeling}).

Under the Limber approximation \citep{Limber53, Limber_LoVerde2008}, the 2D convergence power spectrum in tomographic bins $i$ and $j$, $C^{ij}_\kappa(\ell)$ is related to the full 3D matter power spectrum as: 
\begin{equation}\label{eq: Limber}
C_{\kappa}^{ij}(\ell)=\int_{0}^{\chi(z_\textrm{max})} d\chi\frac{W^{i}(\chi)W^{j}(\chi)}{\chi^{2}}P_\delta \left(\frac{\ell+1/2}{\chi},z(\chi)\right),
\end{equation}
\noindent
where $P_\delta$  is the nonlinear matter power spectrum and the lensing efficiency kernels are given by
\begin{equation}\label{eq: lensing kernel W}
W^{i}(\chi)=\frac{3H_{0}^{2}\Omega_{\mathrm{m}}}{2c^{2}}\frac{\chi}{a(\chi)}\int_{\chi}^{\chi_{\mathrm{H}}}d\chi'\,n^{i}\left(z(\chi')\right)\frac{dz}{d\chi'}\frac{\chi'-\chi}{\chi'}.
\end{equation}
The source galaxy redshift distribution $n^i(z)$ here is normalized to unity. Clearly, the amplitude of $C_\kappa$ responds directly to $\sigma_8^2$, and to \omegam~via the power spectrum, and to $\omegam h^2$ via the lensing kernel, which gives rise to a characteristic banana-shaped degeneracy in $\sigma_8\times\omegam$. The combination of parameters most strongly constrained by cosmic shear is a derived parameter, commonly referred to as $S_8\equiv\sigma_8 (\omegam/0.3)^{0.5}$.  

By decomposing $\kappa$ into E- and B-mode components  \citep{CrittendenEB, Schneider_vanW_Mell_2002} and in a full-sky formalism, one can express the angular two point shear correlations as:
\begin{align} \label{eq: xipm Hankel transform}
    \xi_{\pm}^{ij}(\theta)= & \sum_{\ell}\frac{2\ell+1}{2\pi\ell^{2}\left(\ell+1\right)^{2}}\left[G_{\ell,2}^{+}\left(\cos\theta\right)\pm G_{\ell,2}^{-}\left(\cos\theta\right)\right]\nonumber\\& \times\left[C_{EE}^{ij}(\ell)\pm C_{BB}^{ij}(\ell)\right],
\end{align}
\noindent
where the functions $G^\pm_\ell(x)$ are computed from Legendre polynomials $P_\ell(x)$ and averaged over angular bins (see \citet{y3-generalmethods} Eqs. 19 and 20). It is also worth bearing in mind that, in practice, the angular spectra in Eq. \eqref{eq: xipm Hankel transform} are not in fact pure cosmological convergence spectra $C_{\kappa}$, but rather shear spectra $C_\gamma$, which include contributions from intrinsic alignments (see Sec. \ref{section: IA modeling}), and additional higher order terms are explored later.

\begin{figure}
	\includegraphics[width=\columnwidth]{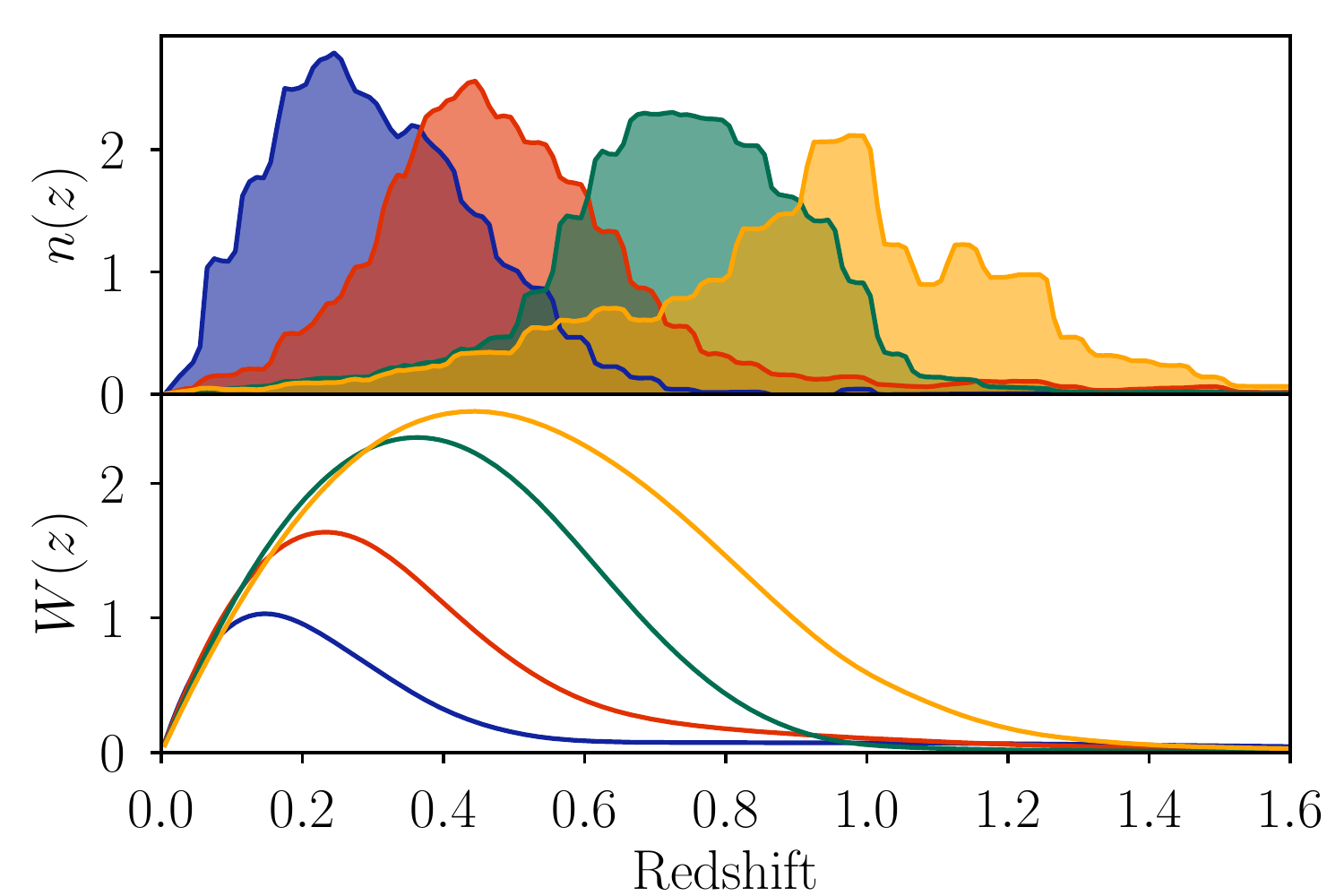}
    \caption{The estimated redshift distributions and lensing kernels for the fiducial source galaxy sample used in this work. Most of the sensitivity of the DES Y3 cosmic shear signal to large scale structure is in the range between $z=0.1$ and $z=0.5$, where individual kernels peak. Each distribution is independently normalized over the redshift range $z=0-3$. The total effective number density \citep{heymans12} of sources is $n_\textrm{eff}=5.59$ galaxies per square arcminute and is divided almost equally into the 4 redshift bins.}
    \label{fig: nz}
\end{figure}

\begin{figure}
	\includegraphics[width=\columnwidth]{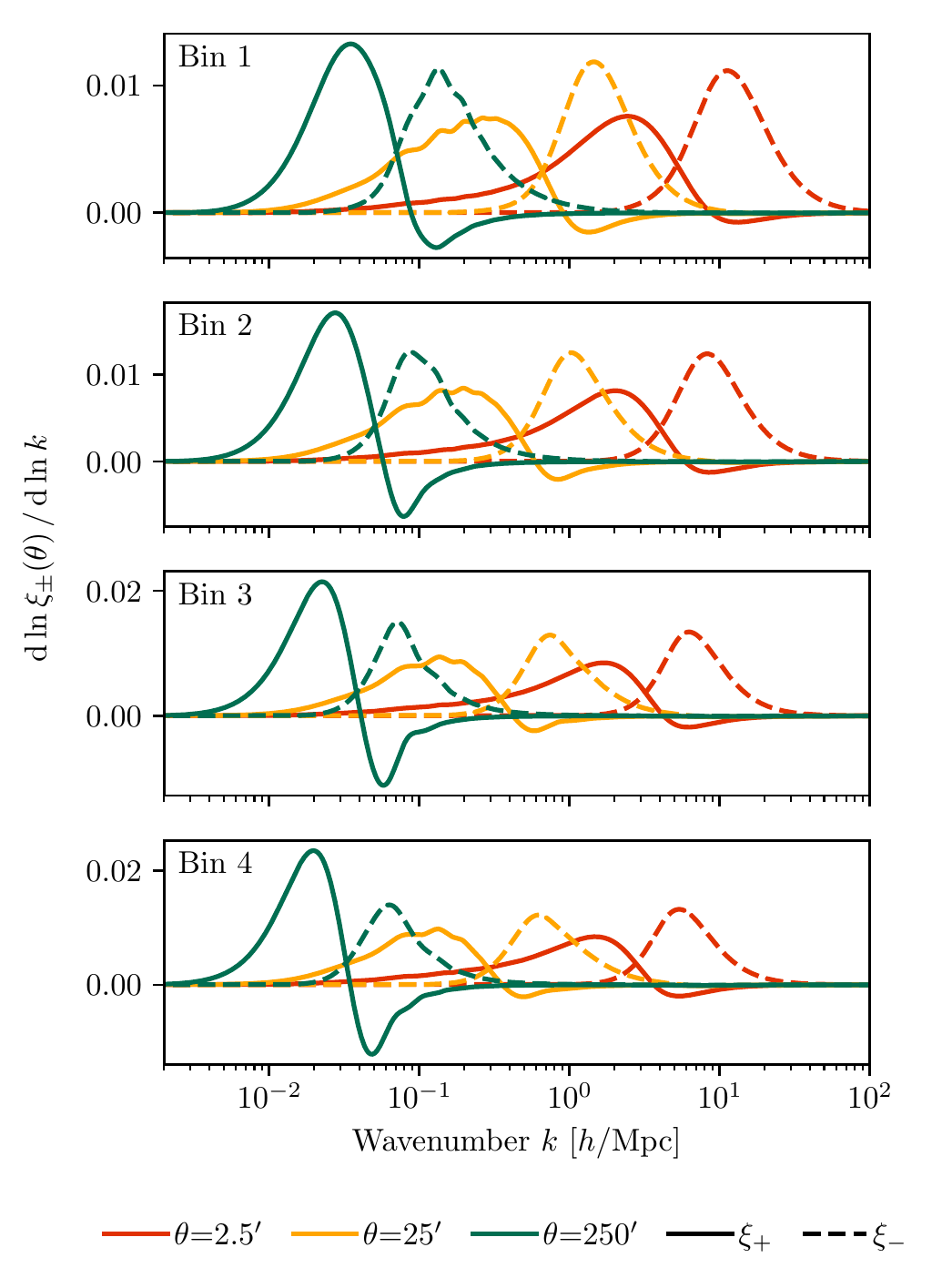}
    \caption{Window functions of $\xi_{+}$ (solid curves) and $\xi_{-}$ (dashed curves) over $k$-wavenumbers of the matter power spectrum at representative angular separations. Notice that our smallest angular scales (after cuts) in $\xi_{+}$ and $\xi_{-}$ are around 2.5 arcminutes and 30 arcminutes respectively, which means that only a relatively small contribution to the full signal comes from wavenumbers above $k\sim1\,h/$Mpc.}
    \label{fig: xipm k-window}
\end{figure} 

Since the lensing kernel in Eq. (\ref{eq: lensing kernel W}) acts as a redshift filter, which modulates sensitivity to $P_\delta$ in Eq. (\ref{eq: Limber}), it is informative to show its redshift dependence; we do so in the lower panel of Fig. \ref{fig: nz}, for the fiducial Y3 redshift distributions. As shown there, the DES Y3 cosmic shear signal is sensitive to a relatively broad range, with kernels peaking between approximately $z=0.1$ and $z=0.5$. 

Similarly, the polynomials in Eq. (\ref{eq: xipm Hankel transform}) mix together a range of physical distances into any given angular scale. We can elucidate this by writing $\xi_\pm^{ij}$ as an integral over $\ln k$ (e.g. \citealt{Tegmark_Zaldarriaga_2002}) to obtain : 
\begin{equation}
    \xi_\pm^{ij}\qty(\theta)=\int_{-\infty}^{+\infty} \dd{\ln k} \mathcal{P}(k)
\end{equation}
\noindent
where 
\begin{align}
    \mathcal{P}(k)=& \frac{k}{2\pi} \sum_{\ell}  \left[G_{\ell,2}^{+}\left(\cos\theta\right)\pm G_{\ell,2}^{-}\left(\cos\theta\right)\right]\nonumber\\&\times W^i\qty(\chi) W^j\qty(\chi) P_\delta\qty(k,z(\chi)),
\end{align}
with $\chi=(\ell+1/2)/k$. We show $\dv*{\ln \xi_\pm(\theta)}{\ln k} = \mathcal{P}(k)/\xi_\pm(\theta)$ for representative $\theta$ scales in Fig.~\ref{fig: xipm k-window}. This shows the sensitivity of our cosmic shear signal at a given angular scale to modes of the matter power spectrum. Our scale cuts, defined in Sec. \ref{section: scale cuts}, eventually remove most of the sensitivity to $k>1\,h$Mpc$^{-1}$.

The remainder of this Sec. motivates the ingredients introduced in Eq. (\ref{eq: Limber}) to (\ref{eq: xipm Hankel transform}) such as the non-linear power spectrum prescription and the set of scales for which the signal is not significantly contaminated by unmodeled physics.

\subsection{Nonlinear Power Spectrum}\label{sec: nonlinear Pk}

On the largest of physical scales, growth is linear and well described by a purely linear matter power spectrum $P_\delta^{\rm lin}(k)$. To evaluate $P_\delta^{\rm lin}$ we use the Boltzmann code CAMB\footnote{\url{ http://camb.info}} \citep{camb}, as implemented in \blockfont{CosmoSIS}. On smaller scales, however, this is not true, and one also needs a model for non-linear growth. Our fiducial model for the non-linear matter power spectrum $P_{\delta}(k)$ is the\textsc{HaloFit} functional prescription \citep{Smith2003, Takahashi2012}. We have made scale cuts to remove the parts of the data vector affected by baryonic effects, as described in Sec. \ref{section: scale cuts}; this largely removes the sensitivity to wavenumbers $k>1\,h $Mpc$^{-1}$. For $k<1\,h$Mpc$^{-1}$, \citet{Takahashi2012} reports an uncertainty on the \textsc{HaloFit} model of $5\%$. In \citet{y3-generalmethods} we demonstrate, by substituting \textsc{HaloFit} for \textsc{HMCode}\footnote{\url{https://github.com/alexander-mead/HMcode}} \citep{mead2015}, the \textsc{Euclid Emulator} \citep{euclidemu}, or the \textsc{Mira-Titan Emulator} \citep{Lawrence2017} that for cosmic shear alone we are insensitive to this choice. In the context of the Y3 $3\times 2$pt analysis, the distinction between these three models is more nuanced, and we refer the reader to \citet{y3-generalmethods} for a full justification of the use of \textsc{HaloFit}.

\subsection{Intrinsic Alignments}\label{section: IA modeling}

Galaxies are not idealized tracers of the underlying matter field, but rather astrophysical bodies, which are subject to local interactions. To account for this added complexity, the observed shape of a galaxy can be decomposed into two parts, the shear induced by gravitational lensing (G) and the intrinsic shape (I) induced by the local environment: $\gamma=\gamma^{\textrm{G}}+\gamma^{\textrm{I}}$. In this section, we consider only the correlated intrinsic component, and not the intrinsic ``shape noise'', which contributes to the covariance but not the signal. 

The term \emph{intrinsic alignments} covers two contributions from environmental interactions: (a) intrinsic shape - intrinsic shape correlations between galaxies that are physically close to each other, and (b) shear-intrinsic correlations between galaxies on neighbouring lines of sight. Known as II and GI contributions respectively, and contributing on similar angular scales to the cosmological lensing signal, these terms constitute a significant systematic in weak lensing analyses. 
Including IA contributions, the observed E-mode angular power spectrum is written
\begin{equation}
    C_{\gamma, \mathrm{EE}}^{ij}(\ell)=C_{\mathrm{GG}}^{ij}(\ell)+C_{\mathrm{GI}}^{ij}(\ell)+C_{\rm IG}^{ij}(\ell)+C_{\mathrm{II},\mathrm{EE}}^{ij}(\ell).
\end{equation}
Nonlinear models of IA, as discussed below, can also produce a non-zero B-mode power spectrum:
\begin{equation}
    C_{\gamma, \mathrm{BB}}^{ij}(\ell)=C_{\mathrm{II},\mathrm{BB}}^{ij}(\ell).
\end{equation}

\noindent
Assuming the Limber approximation as before, the two IA $C(\ell)$s are given by:
\begin{equation}\label{eq: Limber_GI}
C_{\rm GI}^{ij}(\ell)=\int_{0}^{\chi_\mathrm{H}} d\chi\frac{W^{i}(\chi)n^{j}(\chi)}{\chi^{2}}P_{\rm GI}\left(\frac{\ell + 1/2}{\chi},z(\chi)\right),
\end{equation}
\noindent
and
\begin{equation}\label{eq: Limber_II}
C_{\rm II}^{ij}(\ell)=\int_{0}^{\chi_\mathrm{H}} d\chi\frac{n^{i}(\chi)n^{j}(\chi)}{\chi^{2}}P_{\rm II}\left(\frac{\ell + 1/2}{\chi},z(\chi)\right),
\end{equation}
\noindent
These expressions are generic, and are valid regardless of
which model is used to predict $P_{\rm GI}$ and $P_{\rm II}$ (see the following subsections).

\subsubsection{IA and the tidal field}

It is typically assumed that the correlated component of galaxy shapes is determined by the large-scale cosmological tidal field. The simplest relationship, which should dominate on large scales and for central galaxies, involves the ``tidal alignment'' of galaxy shapes, producing a linear dependence \citep{catelan01, hirata04}. In this case, one can relate the intrinsic shape component to the gravitational potential at the assumed time of galaxy formation $\phi_*$:

\begin{equation}
\label{eq:IAtidal}
(\gamma_1^\textrm{I}, \gamma_2^\textrm{I})
= A_1(z)
\left( 
\frac{\partial^2 }{\partial x^2} - \frac{\partial^2 }{\partial y^2},  
2 \frac{\partial^2}{\partial x \partial y}
\right)
\phi_*,
\end{equation}
\noindent

where the proportionality factor $A_1(z)$ captures the response of intrinsic shape to the tidal field. More complex alignment processes, including ``tidal torquing,'' relevant for determining the angular momentum of spiral galaxies, are captured in a nonlinear perturbative framework, which we refer to as ``TATT'' (Tidal Alignment and Tidal Torquing; \citealt{blazek19}). In this more general model, we use nonlinear cosmological perturbation theory to express the intrinsic galaxy shape field, measured at the location of source galaxies \citep{blazek15}, as an expansion in the matter density field $\delta$ and tidal field $s_{ij}$:
\begin{align}\label{eq: all IA contributions to shear}
\bar{\gamma}^{\rm IA}_{ij} = A_1 s_{ij} +A_{1\delta} \delta s_{ij} + A_2 \sum_{k}s_{ik}s_{kj} + \cdots,
\end{align}
where $s_{ij}$ is the gravitational tidal field, which at any given position $\mathbf{x}$ is a $3\times3$ tensor (see \citealt{catelan00} for a formal definition).

Although the terms in the model can be associated with physical mechanisms, they can also be viewed as effective contributions to intrinsic shape correlations from small-scale physics. See also \citet{tugendhat17,schmitz18,vlah20} for further discussion of the perturbative approach and \citet{fortuna20} for a halo model treatment of IA.

\subsubsection{Model implementation: NLA and TATT}\label{sec:ias:modeling}

Within the TATT framework, three parameters capture the relevant responses to the large-scale tidal fields (see \citealt{blazek19} for more details): $A_1$, $A_2$, and $A_{1\delta}$, corresponding respectively to a linear response to the tidal field (tidal alignment), a quadratic response (tidal torquing), and a response to the product of the density and tidal fields. To date, the most frequently used intrinsic alignment model in the literature is known as the Nonlinear Alignment Model (NLA; \citealt{hirata07, bridle07}), an empirically-based modification of the Linear Alignment (LA) model of \citet{catelan01} and \citet{hirata04}, in which the fully nonlinear tidal field is used to calculate the tidal alignment term. Within the ``TATT'' framework, the NLA model corresponds to only $A_1$ being non-zero in Eq.~\ref{eq: all IA contributions to shear}. The GI and II power spectra then have the same shape as the nonlinear matter power spectrum, but are modulated by $A_1(z)$:

\begin{equation} \label{eq:theory:nla}
    P_{\rm GI}(k,z) = A_1(z) P_{\delta}(k,z), \;\;\;\;\; P_{\rm II}(k,z) = A_1^2(z) P_{\delta}(k,z).
\end{equation}

\noindent
Note that the nonlinear power spectrum and the IA amplitudes are functions of redshift. In the following, the $z$ dependence of the IA amplitudes and various $k$ dependent terms is left implicit.
More generally, in the TATT model, the GI and II power spectra are constructed with the relevant correlations of tidal and density fields:
\begin{align}
\label{eq:dEtot}
P_{\mathrm{GI}}(k) =& A_1 P_{\delta}(k) + A_{1\delta} P_{0|0E}(k)
+ A_2 P_{0|E2}(k)~, \\
\label{eq:EEtot}
P_{\mathrm{II},\mathrm{EE}}(k) =&
A_1^2 P_{\delta}(k) +
2 A_1 A_{1\delta} P_{0|0E}(k)
+ A_{1\delta}^2 P_{0E|0E}(k)\notag\\
&+ A_2^2 P_{E2|E2}(k)
+ 2 A_1 A_2 P_{0|E2}(k)\\
&+ 2 A_{1\delta} A_2 P_{0E|E2}(k)~,\notag\\
\label{eq:BBtot}
P_{\mathrm{II},\mathrm{BB}}(k) =& A_{1\delta}^2 P_{0B|0B}(k)
+ A_2^2 P_{B2|B2}(k)
+ 2 A_{1\delta} A_2 P_{0B|B2}(k) ~.
\end{align} 
In this work, these $k$-dependent terms are evaluated using \textsc{FAST-PT} \citep{mcewen16, fang17}, as implemented in \textsc{CosmoSIS}. The model, including the full expressions for these power spectra is set out in some depth in \citet{blazek19} (see their Eqs.~37--39 and appendix A), and we refer the reader to that paper for technical details. The $k$-dependent contributions are modulated by the redshift-dependent amplitudes $A_1$, $A_2$, and $A_{1\delta}$. We define the first two with the following convention:

\begin{equation}\label{eq:theory:tatt_c1}
    A_1(z) = -a_1 \bar{C}_{1} \frac{\rho_{\rm crit}\Omega_{\rm m}}{D(z)} \left(\frac{1+z}{1+z_{0}}\right)^{\eta_1},
\end{equation}
\begin{equation}\label{eq:theory:tatt_c2}
    A_2(z) = 5 a_2 \bar{C}_{1} \frac{\rho_{\rm crit}\Omega_{\rm m}}{D^2(z)} \left(\frac{1+z}{1+z_{0}}\right)^{\eta_2},
\end{equation}
\noindent
where $D(z)$ is the linear growth factor, $\rho_{\rm crit}$ is the critical density and $\bar{C}_1$ is a normalisation constant, by convention fixed at
$\bar{C}_1=5\times10^{-14}M_\odot h^{-2} \mathrm{Mpc}^2$,
obtained from SuperCOSMOS (see \citealt{brown02}). The leading factor of 5 in Eq.~(\ref{eq:theory:tatt_c2}) is included to account for the difference in the windowed variance produced by the TA and TT power spectra. With this factor included, the TA and TT contributions to $P_{\rm II}$ at $z=0$, averaged over this window, should be roughly equal if $a_1=a_2$, aiding in the interpretation of the best fitting values. Note that this is a matter of convention only, and does not affect our final cosmological results.
The denominator $z_{0}$ is a pivot redshift, which we fix to the value 0.62\footnote{The value was chosen in DES Y1 to be approximately equal to the mean source redshift. We choose to maintain that value to allow for an easier comparison of the IA amplitudes with those results.}. The dimensionless amplitudes $(a_1, a_2)$ and power law indices $(\eta_1,\eta_2)$ are free parameters in this model.

As mentioned above, the model also includes the $A_{1\delta}$ contribution, corresponding to the product of the density and tidal fields. This term was originally motivated by the modulation of the IA signal due to the galaxy density weighting (i.e.\ the fact that the shape field is preferentially sampled in overdense regions \citealt{blazek15}). In this case, within the TATT model, we have 
\begin{equation}
A_{1\delta} = b_{\mathrm{TA}} A_1,
\end{equation}
\noindent
where $b_{\rm TA}$ is the linear bias of source galaxies contributing to the tidal alignment signal. In our baseline analysis, rather than fixing $b_{\mathrm{TA}}$ to this bias value, we sample over it with a wide prior, allowing the $A_{1\delta}$ contribution to capture a broader range of nonlinear alignment contributions. We note that this is a departure from previous studies to have used this model \citep{blazek19, TroxelY1, samuroff18},
all of which held $b_{\mathrm{TA}}=1$ fixed.
The motivation for this change is set out in Sec. \ref{section: unblinding}. 
As can be seen from Eq. \eqref{eq:dEtot} - \eqref{eq:BBtot}, in the limit $a_2,b_{\mathrm{TA}}\rightarrow0$, the TATT model reduces to the NLA model. It is thus useful to think of NLA as a sub-space of the more complete TATT model, rather than a distinct, alternative model. 
Given the sensitivity of IAs to the details of the galaxy selection, and in the absence of informative priors, we choose to marginalize over all five IA parameters $(a_1,a_2,\eta_1,\eta_2, b_{\mathrm{TA}})$, governing the amplitude and redshift dependence of the IA terms, with wide flat priors (see Sec. \ref{section: priors}).
While a redshift evolution in the form of a power law, captured by the index $\eta_i$, is a common assumption, the $A_i(z)$ coefficients could, in theory, have a more complicated redshift dependence. We seek to test the impact of this assumption by rerunning our analysis with a more flexible parameterization, whereby the IA amplitude $A_{1,i}$ in each redshift bin is allowed to vary independently. The results of this exercise can be found in Sec. \ref{sec: IA robustness on data}.

It is finally worth remarking that the TATT model predicts a non-zero B-mode power spectrum $P_{\mathrm{II},BB}$. 
This extra signal component is incorporated into our modeling, and propagated into $\xi_\pm$ via Eq. (\ref{eq: xipm Hankel transform}). $P_{\mathrm{II},BB}$ is  expected to be small since the testing carried out in \citealt*{y3-shapecatalog} points to no statistically significant detection of B-modes in the DES Y3 shape catalog. 

A visual representation of the II and GI signals, as predicted at the best-fitting point in parameter space from our fiducial analysis, can be found in Fig. \ref{fig: measured xipm}.

\begin{figure*}
	\includegraphics[width=1.0\textwidth]{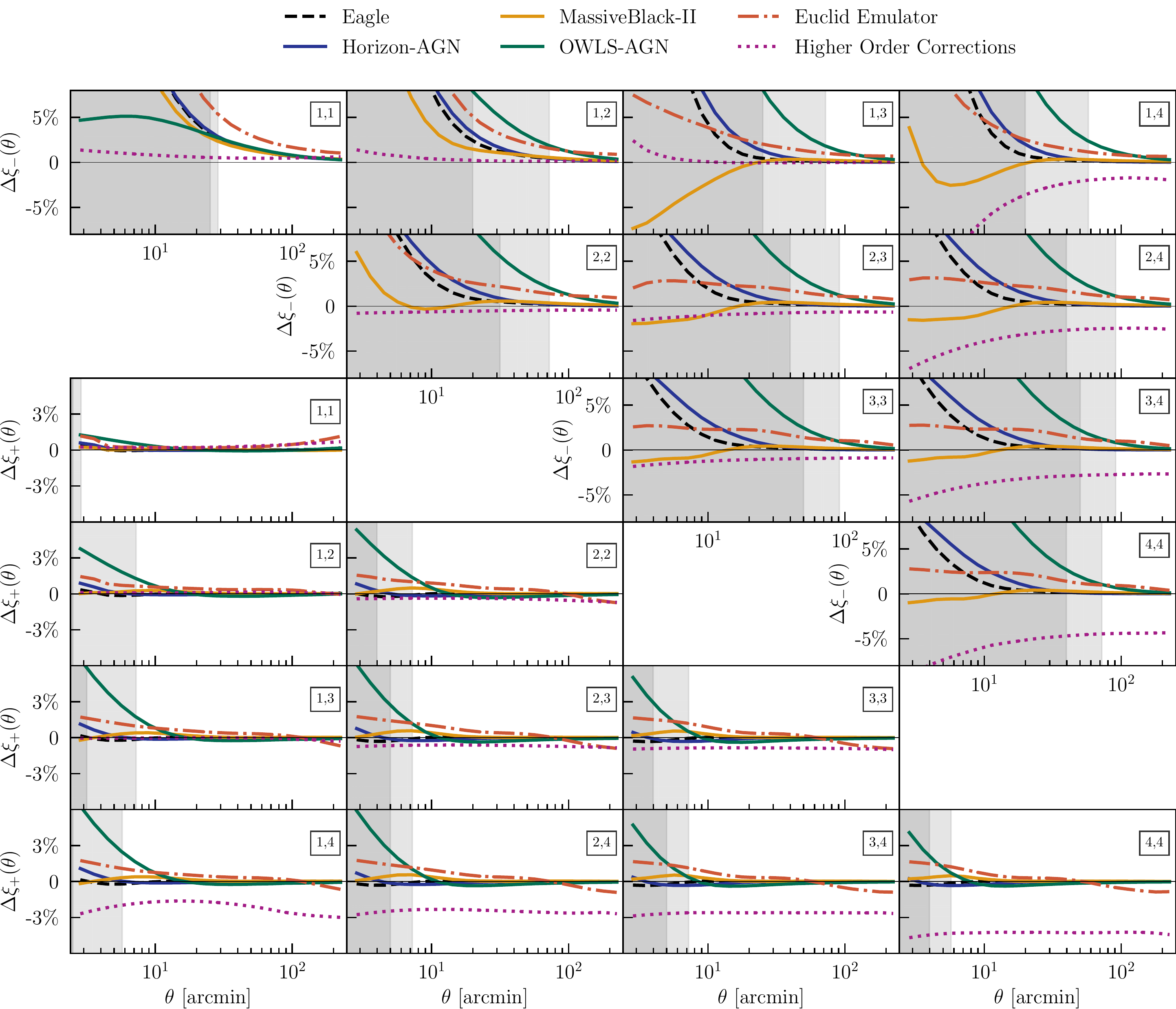}
    \caption{The theoretical contribution of different modeling systematics to $\xi_{\pm}$; for each systematic we show the fractional impact, relative to the fiducial case $\Delta \xi_\pm \equiv (\xi_\pm-\xi_\pm^{\textrm{syst.}})/\xi_\pm$. Fiducial scale cuts are shown as light shaded bands and are derived jointly for cosmic shear, $2\times2$pt and $3\times2$pt. Darker bands correspond to (less stringent) scale cuts that are optimized for cosmic shear and $3\times2$pt in \lcdm~only. Error bars on $\xi_{+(-)}$ are, at their smallest, around 15\% (20\%) of the signal, so the maximum contamination shown here $\mathcal{O}(1\sim5\%)$ is still significantly below the sensitivity of our data. We also find that none of these forms of contamination project translate into a bias in cosmological parameters at Y3 precision, despite appearing coherent in some redshift bins. \blockfont{Eagle} (black dashed), \blockfont{Horizon-AGN} (solid blue), \blockfont{MassiveBlack-II} (solid yellow) and OWLS-AGN (solid green) represent scenarios for baryonic physics, and obtained from the power spectra of hydrodynamic simulations, while Euclid Emulator (dot-dashed red) modifies the non-linear gravity-only power spectrum. Higher Order Corrections (dashed purple) is the theoretical impact of reduced shear and source magnification.}
    \label{fig: systematics}
\end{figure*}

\subsection{Modeling Nuisance Parameters: Shear Bias \& photo-$z$ Error}\label{sec:nuisance_parameters}

In addition to the parameters associated with the cosmological and IA models, there are a number of additional parameters included in our fiducial analysis, which are intended to absorb known sources of uncertainty.
First of all, we follow the majority of previous cosmic shear analyses, and parameterize errors in the redshift distributions as uniform shifts in their mean, which transform the assumed distribution in a given bin $i$ as:
\begin{equation}
    n^i(z) \rightarrow n^i(z-\Delta z^i).
\end{equation}
\noindent
This leaves us with four nuisance parameters $(\Delta z^{1-4})$, which are marginalized with informative priors (see Sec. \ref{section: priors}). This simple marginalization scheme was validated with a more complete method, known as \blockfont{HyperRank} \citet*{y3-hyperrank},  
which uses a large ensemble of possible redshift distributions (typically of order 1000) generated using the SOMPZ pipeline \citep*{y3-sompz}. The basic aim of this method is to capture the range of plausible variations in the full shape of the distributions (not only their mean), and also the correlations between bins. 
Although in our particular setup the simpler parameterization was found to be sufficient, the tests in \citet*{y3-hyperrank} and \citet{y3-cosmicshear1} show the crucial validation of that choice.

Another important source of uncertainty is the shape calibration process, and particularly its response to blending \citet{y3-imagesims}. The common way of describing such effects \citep{heymans06,huterer06} is a simple rescaling of the two-point function model prediction
\begin{equation}\label{eq:m_bias}
    \xi^{ij}_\pm \rightarrow (1+m^i)(i+m^j) \times \xi^{ij}_\pm,
\end{equation}
\noindent
where again the indices $i,j$ indicate redshift bins, and $m$ is assumed to be redshift and scale independent within each bin. This is our fiducial parameterization. 
As explained in Sec. \ref{sec: photoz calibration}, this approach is an approximation to the more complete methodology set out in \citet{y3-imagesims} (see in particular their Sec. 6.5), which incorporates the redshift-dependent impact of blending on the source redshift distributions. Nevertheless, an extensive series of tests indicate that Eq. (\ref{eq:m_bias}) is sufficient for the statistical power of DES Y3 (see \citealt{y3-imagesims,y3-cosmicshear1}).

\subsection{Incorporating Shear Ratios}\label{section:SR}

In the DES Y3 analysis, we also incorporate small-scale shear ratios (SR) at the likelihood evaluation level. \textit{This likelihood is included in all our present constraints unless explicitly noted otherwise}. This concept and its DES application is outlined in more detail by \citet*{y3-shearratio}. Essentially, SR is an additional lensing-based data vector with 9 data points made up of measured galaxy-galaxy lensing ratios. It is nearly independent of cosmology and galaxy bias, but responds to (and thus constrains) redshift distributions and intrinsic alignments. 
Note that this particular analysis feature is different from Y1 and most current cosmic shear results, a detail that should be kept in mind in any comparison of our findings with previous results.

We use SR on small scales that are not used in the $3\times2$pt likelihood ($<6$ Mpc$/h$), where uncertainties are dominated by galaxy shape noise, such that the likelihood can be treated as independent of that from the $\xi_\pm$ (and indeed the full $3\times2$pt) data. On these scales the SR measurement is only very weakly dependent on the matter power spectrum, and so immune to uncertainties in modeling $P_\delta$ on small scales \citep*{y3-shearratio}. In this work, we only employ SR as obtained from the fiducial \blockfont{MagLim} lens sample \citet{y3-2x2maglimforecast}, and marginalize over its required nuisance parameters. These are a free (linear) galaxy bias coefficient per lens bin (which are unconstrained by cosmic shear and have no impact on the final posteriors), as well as lens redshift parameters (three shift $\Delta z_l$ and three width parameters $\delta_z$; see \citealt{y3-2x2ptaltlensresults}), on which we have relatively tight priors. The main impact of including SR is an improvement in constraints on IAs and redshift nuisance parameters. Through internal degeneracies, this translates into significantly improved constraints on $S_8$, of around 30\% in our fiducial analysis setting. Further discussion can be found in \citet*{y3-shearratio} (Sec. 6 and Fig. 10) and \citet{y3-cosmicshear1} (Sec. X and Fig. 11). In particular, those papers also assess the impact of utilizing \textsc{redMaGiC}, an alternative DES Y3 lens sample \citep{RozoRM, y3-lenswz, y3-2x2ptbiasmodelling}, in order to obtain the shear ratios.

\subsection{Angular Scale Cuts}\label{section: scale cuts}

\subsubsection{Baryons}
The impact of baryons on the matter power spectrum on cosmological scales is a source of considerable uncertainty \citep{van2014impact,semboloni2013effect,harnois2015baryons}. While feedback processes from active galactic nuclei (AGN) and supernovae heat up the halo environment and tend to suppress matter clustering, metal enrichment can offer cooling channels that in fact increase power on small scales. These effects also depend on redshift and galaxy evolution \citep{semboloni2011quantifying}.

As a fiducial strategy for mitigating the uncertainties coming from baryonic physics, we employ a gravity-only matter power spectrum (\blockfont{CAMB} + \blockfont{HaloFit}) as described in Sec. \ref{sec: nonlinear Pk}, and remove angular scales from the data vector which are significantly affected by feedback processes.
We have additionally verified with synthetic data that marginalizing over baryonic halo model parameters with conservative priors does not lead to an actual gain in the $S_8\times\omegam$ sub-space. Specifically, we use \textsc{HMcode} \citep{mead2015} and free both the halo concentration amplitude $A$ with a flat prior [1.0, 7.5] and the bloating parameter $\eta_0$ with a flat prior [0.4, 1.0]. These priors were chosen such that all baryonic scenarios tested in \citet{mead2015} would be encapsulated by the allowed halo model modification. We find, with these fairly conservative priors, that attempting to model progressively smaller scales in synthetic data vectors results in tighter constraints on the baryon nuisance parameters $A$ and $\eta_0$, while leaving the constraining power on $S_8\times\omegam$ approximately unchanged.

\subsubsection{Determining scale cuts}

We define scale cuts based on the $\Delta \chi^2$ between noiseless synthetic cosmic shear data vectors, generated with and without a baryonic ``contamination'', according to Eq. \eqref{eq: chi2}. For the baryonic data vector, we use the OWLS \citep{Schaye2010,vanDaalen11} matter power spectrum. The AGN-feedback implementation in this suite of simulations represents one of the most extreme scenarios in the literature, and thus characterizes the most conservative case for the baryonic contamination we expect our analysis to be safe against. 

To determine scale cuts for the cosmic shear two-point functions $\xi_{\pm}^{ij}$, 
we evenly distribute the $\Delta \chi^2$ threshold among tomographic bins, and limit each of the shear two-point angular functions:
\begin{equation}
    \left ( \xi_{\pm, \mathrm{baryon}}^{ij}-\xi_{\pm, \mathrm{base}}^{ij} \right ) ^{\mathrm{t}} \mathbf{C}_{\mathrm{sub}}^{-1}\left ( \xi_{\pm, \mathrm{baryon}}^{ij}-\xi_{\pm, \mathrm{base}}^{ij}\right )<\frac{\Delta \chi^2_{\mathrm{threshold}}}{N}
    \label{eq: dchisqthreshold}
\end{equation}
\noindent
where $N=20$ is the number of cross-correlations between four tomographic redshift bins in the two shear correlations $\xi_\pm$. The $\mathbf{C}_{\mathrm{sub}}^{-1}$ sub-matrix of the inverse covariance matrix corresponds to the specific tomographic bin. For each element of $\xi^{ij}_{\pm}$, we find the minimum angular separation that satisfies Eq. \eqref{eq: dchisqthreshold} and exclude data points at smaller separations. 

For a given threshold, we run two chains using the fiducial pipeline: one on a baseline (systematics-free) synthetic data vector and one on a contaminated synthetic data vector which includes the modification of the OWLS-AGN matter power spectrum. Our most important criterion for selecting a threshold is that the peaks of the baseline and contaminated posteriors in the 2-D parameter subspace of $S_8\times\omegam$ must be separated by less than 0.3$\sigma$. We make this analysis additionally conservative by fixing redshift and shear uncertainties and the neutrino mass. It is also worth noting that scale cuts for the cosmic shear correlations $\xi_{\pm}^{ij}$ are determined in tandem with the $2\times2$pt and $3\times2$pt analyses. That means a candidate cut passing the $<0.3\sigma$ threshold for cosmic shear is then combined with cuts from the analogous $2\times2$pt analysis, and the finally selected scale cuts for cosmic shear and $2\times2$pt are those which also satisfy $3\times2$pt $<0.3\sigma$. In practice, this procedure ensures that the full DES Y3 analysis is optimized, as opposed to each probe alone being optimally constraining (although see the discussion in Sec. \ref{section:optimized_cuts}).

After an iterative procedure, we finally select the scale cuts that lead to a threshold of $\Delta \chi^2<0.5$ for cosmic shear systematics. These cuts are shown in grey bands in Figures \ref{fig: measured xipm} and \ref{fig: systematics}, and roughly correspond to bounds that are, at the smallest, $\theta_{\rm min, +}=2.4$ and $\theta_{\rm min, -}=30$ arcmin in $\xi_+$ and $\xi_-$ respectively, but differ significantly between redshift bin pairs. The large-scale limit of the data vector is 250 arcmin, chosen to match the DES Y1 analysis \citep{TroxelY1}. After cuts, we are left with 166 and 61 angular bins in $\xi_+$ and $\xi_-$, or a total of 227 data points.

Fig. \ref{fig: systematics} demonstrates that, by making our data immune to the feedback impact of OWLS-AGN, we are also safely excluding parts of the data vector that are affected by systematic uncertainty due to the matter power spectrum modeling (Euclid Emulator; \citealt{euclidemu}), baryonic feedback (\blockfont{MassiveBlack-II}, \blockfont{Horizon-AGN} and \blockfont{Eagle}; respectively \citealt{khandai15, Dubois15, Schaye15}) and higher order shear corrections including reduced shear and source magnification (Sec. \ref{sec: higher order shear} and \citealt{y3-generalmethods}). See Sec. \ref{section: unblinding} for more details. It is important to note that error bars on $\xi_{+}$ and $\xi_{-}$ are, at their smallest, around 15\% and 20\% of the signal, so the deviations seen in Fig. \ref{fig: systematics} are well below the sensitivity of our data. We refer the reader to \citet{y3-generalmethods} for an exploration of different baryonic power spectra, and in particular other extreme feedback cases such as \blockfont{Illustris} \citep{Illustris}.

\subsubsection{\lcdm Optimized scale cuts}\label{section:optimized_cuts}

As discussed above, our fiducial choice of scale cuts, by construction, optimizes the joint $3\times2$pt analysis of \citet{y3-3x2ptkp}, which includes not just cosmic shear, but also galaxy clustering and galaxy-lensing. Those scale cut tests were also required to pass the same tolerance threshold in both \lcdm~and \wcdm, making the exercise even more stringent. The end result is that the fiducial scale cuts used for Y3 cosmic shear (both in this work and in \citealt{y3-cosmicshear1}) are more conservative than strictly necessary for the baseline cosmic shear-only \lcdm~analysis. 
For this reason we repeat a subset of our analyses with an alternative set of relaxed scale cuts, designed to maximize the constraining power in cosmic shear and $3\times2$pt \lcdm alone (referred to as the ``\lcdm Optimized" analysis, disregarding $2\times2$pt-alone and $w$CDM). These optimized cuts increase the total number of data points from 227 to 273 (184 $\xi_+$ and 89 $\xi_-$). We discuss the improvement in constraining power that comes from this choice in Sec. \ref{section: baseline cosmology}. While the gain in angular scales is not uniform across redshift bins, the minimum scale is reduced by a factor of $20-70\%$ for different bin pairs. We show the \lcdm optimized scale cuts as darker grey bands on Fig. \ref{fig: systematics}
(compared to the lighter grey bands which correspond to the fiducial scale cuts and eliminate more data points on small scales).

\subsection{Choice of Priors}\label{section: priors}

The priors on cosmological and nuisance parameters are summarized in Table \ref{table: priors}.
Our choice of cosmological priors is relatively conservative, in order to ensure that the posterior distributions can span any point in parameter space we believe to be reasonable. Although the argument can be made for imposing more informative priors on our less well constrained parameters (primarily \omegab~and $h$) based on external data (e.g. Planck), we choose not to do so here. The main reasoning here is that we want to maintain the statistical independence of the DES Y3 cosmic shear data, to avoid double-counting information when combining our results with external data sets. 

In contrast, we \emph{do} impose informative priors on the measurement systematics (i.e.\ shear bias and photo-$z$ error), which are derived from image simulations (\citealt{y3-imagesims}; \citealt*{y3-shapecatalog}) and tests using Buzzard (\citealt*{y3-sompz}; \citealt*{y3-sompzbuzzard}). More details on how these priors were chosen, and the dominant uncertainties can be found in those papers. It is worth bearing in mind that, while cosmic shear has only limited potential to self-calibrate photo-$z$ errors \citep{samuroff17}, the combination with galaxy clustering and galaxy-galaxy lensing  breaks parameter degeneracies; in the current study we are prior dominated in these systematics parameters, but in the $3\times2$pt case this is significantly less true \citep{y3-3x2ptkp}. 

We choose uninformative priors on all of the five IA parameters, motivated by the relative lack of constraints reported in the literature for the TATT model (with the exception of \citealt{samuroff18}). Since IAs are highly sensitive to the composition of the galaxy population, as well as physical characteristics like redshift and luminosity, deriving appropriate priors is a complicated matter. We consider whether these IA parameters are detected by the data in Sec. \ref{sec: IA constraints}. 

It is finally worth remarking that marginalizing posterior distributions from a high-dimensional parameter space down to 1D is prone to projection effects that can significantly displace the full-dimensional best-fit from the mean (or peak) of the 1D parameter distributions. The magnitude of these effects can be non-trivially affected by the choice of priors, particularly on the parameters that are not well constrained by cosmic shear data alone. We again refer the reader to \citet{y3-generalmethods}, where tests of these effects using the priors given in Table \ref{table: priors} can be found.

\begin{table}
	\centering
	\vspace{1cm}
    \caption{A summary of the priors used in the fiducial Y3 analysis. The top seven rows are cosmological parameters, while those in the lower sections are nuisance parameters cooresponding to astrophysics and data calibration. We fix $w=-1$ when analyzing \lcdm. Priors are either uniform (U) or normally-distributed, $\mathcal{N}(\mu,\sigma)$.} \label{table: priors}
\begin{tabular}{cc}
\hline 
Parameter & Prior\tabularnewline
\hline 
\hline
\bf{Cosmological Parameters} & \\
\omegam & $\mathrm{U}[0.1, 0.9]$ \tabularnewline
\as & $\mathrm{U}[0.5,  5.0]\times 10^{-9}$ \tabularnewline
\omegab &$\mathrm{U} [0.03, 0.07]$ \tabularnewline
\ns & $\mathrm{U}[0.87, 1.07]$   \tabularnewline
$h$ &   $\mathrm{U}[0.55, 0.91]$ \tabularnewline
$\Omega_{\nu}h^2$ & $\mathrm{U}[0.6, 6.44]\times 10^{-3}$   \tabularnewline
$w$ & $\mathrm{U}[-2, -0.333]$ \tabularnewline
\hline
\bf{Calibration Parameters} & \\
$m_{1}$ & $\mathcal{N}(-0.0063,0.0091)$ \tabularnewline
$m_{2}$ & $\mathcal{N}( -0.0198,0.0078)$ \tabularnewline
$m_{3}$ & $\mathcal{N}( -0.0241,0.0076)$ \tabularnewline
$m_{4}$ & $\mathcal{N}(-0.0369, 0.0076)$ \tabularnewline
$\Delta z_{1}$ & $\mathcal{N}(0.0,0.018)$ \tabularnewline
$\Delta z_{2}$ & $\mathcal{N}(0.0,0.015)$ \tabularnewline
$\Delta z_{3}$ & $\mathcal{N}(0.0,0.011)$ \tabularnewline
$\Delta z_{4}$ & $\mathcal{N}(0.0, 0.017)$ \tabularnewline
\hline
\bf{Intrinsic Alignment Parameters} & \\
$a_1$ & $\mathrm{U}[-5, 5]$ \tabularnewline
$a_2$ & $\mathrm{U}[-5, 5]$ \tabularnewline
$\eta_1$ & $\mathrm{U}[-5, 5]$ \tabularnewline
$\eta_2$ & $\mathrm{U}[-5, 5]$ \tabularnewline
$b_{\rm TA}$ & $\mathrm{U}[0, 2]$ \tabularnewline
\hline
\bf{Shear Ratio Parameters} & \\
$\Delta z^{\rm lens}_{1}$ & $\mathcal{N}(-0.009,0.007)$
\tabularnewline
$\Delta z^{\rm lens}_{2}$ & $\mathcal{N}(-0.035,0.011)$
\tabularnewline
$\Delta z^{\rm lens}_{3}$ & $\mathcal{N}(-0.005,0.006)$
\tabularnewline
$\delta ^{\rm lens}_{z,1}$ & $\mathcal{N}(0.975,0.062)$
\tabularnewline
$\delta ^{\rm lens}_{z,2}$ & $\mathcal{N}(1.306,0.093)$
\tabularnewline
$\delta ^{\rm lens}_{z,3}$ & $\mathcal{N}(0.870,0.054)$
\tabularnewline
$b^{1-3}_{g}$ & $\mathrm{U}[0.8, 3]$ 
\tabularnewline
\hline 
\end{tabular}
\end{table}

\section{Model robustness tests on synthetic data}\label{section: unblinding}

We now test the assumptions of the model set out in the previous section using contaminated mock data. The idea is that we generate data with astrophysical effects that are \textit{not contained} in the model, and seek to quantify the level of bias they can cause to parameters of interest. These tests fall into two different categories, both involving synthetically generated $\xi_\pm$: noiseless analytic data (which is produced using the theory pipeline), and simulated data (measured using mock catalogs from N-body simulations). Chronologically, all of the tests described in this Section were performed \textit{before unblinding}.

\subsection{Analytically generated data}

We analyze these data with the fiducial model, priors and analysis choices (including scale cuts), to verify that the inferred cosmological constraints are biased by less than 0.3$\sigma$ in the $S_8 \times \omegam$ plane. Fig.~\ref{fig: systematics} shows the systematic contributions to the contaminated data vectors for a number of the tests described below.

\subsubsection{Nonlinear matter power spectrum}

By construction, the scale cuts defined in Sec. \ref{section: scale cuts} are conservative and mitigate most of the impact of small-scale modeling uncertainties in the 3-dimensional gravity-only matter power spectrum. To show that these uncertainties are negligible, we generate a synthetic cosmic shear data vector using the gravity-only Euclid Emulator \citep{euclidemu}. Though it can be used at a relatively restricted range of cosmologies, the emulator is the most accurate method for estimating the nonlinear $P_\delta(k)$ currently available to the community. The deviation between this data vector and the fiducial prediction is shown in Fig. \ref{fig: systematics} (dot-dashed red), and is at most $\sim1$ percent on the scales included in our analysis. We verify that the bias in cosmological parameters when analysing these mock data with the fiducial \lcdm model is below 0.3$\sigma$ \citep{y3-generalmethods}. 

In addition to the test of the nonlinear growth model, we also generate synthetic data vectors using matter power spectra derived from the \blockfont{Horizon-AGN}, \blockfont{Eagle} and \blockfont{MassiveBlack-II} hydrodynamic simulations (Fig.~\ref{fig: systematics}, \citealt{horizonagn, eagle, khandai15}). The $P_\delta$ obtained from these simulations contain the imprint of baryonic feedback processes. Although these effects are notoriously difficult to model, the simulations are useful as a measure of the range of uncertainty in the baryonic contributions. That is, if we can demonstrate insensitivity to a range of (at least semi-) realistic scenarios, then that offers some reassurance that our cosmological results are unaffected. We confirm that the resulting biases are significantly smaller than 0.3$\sigma$ \citep{y3-generalmethods}. At some level this result is expected, since our fiducial scale cuts are constructed based on the \blockfont{OWLS-AGN} model, which is a relatively extreme feedback scenario. It does, however, demonstrate that our scale cut prescription is generally conservative, and is not fine-tuned to the specific redshift or scale dependence of \blockfont{OWLS-AGN}.

\subsubsection{Higher-order shear contributions}\label{sec: higher order shear}

We also verify that the effects of reduced shear, source sample magnification and source clustering do not affect the cosmic shear constraints significantly. These generally enter as higher-order corrections. For the case of reduced shear and source magnification, the correction to the convergence angular power spectrum has the form \citep{Dodelson2006,Schmidt2009,Shapiro2009}:
\begin{multline}
\Delta C^{ij}_{\kappa}(\ell) = \\ 2(1+C^i_{s}) 
\int\frac{d^{2}\ell_{1}}{(2\pi)^{2}}\cos2\phi_{\ell_{1}}B^{ij}_{\kappa}(\vec{\ell}_{1},\vec{\ell}-\vec{\ell}_{1},-\vec{\ell})
\end{multline}
where $B^{ij}_\kappa$ is the convergence bispectrum, $C^i_s$ is a magnification coefficient related to the slopes of the flux and size distributions \citep{Schmidt2009} and $\phi_{\ell_1}$ is the angle between modes $\vec{\ell}$ and $\vec{\ell}_1$. We show in Fig. \ref{fig: systematics} this contribution to the shear power spectrum for the source magnification coefficients $C_s^{1...4}=(-1.17,-0.64,-0.55,0.80)$ 
as measured by \citet{y3-2x2ptmagnification} in each redshift bin. 
We note that the theoretical modeling of non-linear scales in the matter bispectrum is a dominant source of uncertainty in the calculated correction (see for instance \cite{Lazanu2016}). In particular, we utilize the non-linear (\blockfont{Halofit}) power spectrum in the tree-level bispectrum description, but future analyses will possibly benefit from improved approaches, such as \blockfont{BiHalofit} \citep{Takahashi2020_bihalofit}.  
Still, the largest potential biases as seen in Fig. \ref{fig: systematics} are $<5\%$, smaller than our error bars ($\gtrsim 15\%$). We also find, like \citet{y3-gglensing} demonstrate for galaxy-galaxy lensing, that the reduced shear contribution to cosmic shear as computed perturbativelly overestimates the impact of that correction when compared to measurements of reduced shear on \textsc{buzzard} simulations. The leading sources of higher-order corrections and their contribution for DES Y3 are computed in more detail in \citet{y3-generalmethods}. Their results confirm that these effects are negligible in terms of parameter constraints, and provide a guide for future DES and next generation analyses.

\subsubsection{Intrinsic Alignments}\label{section:simulated_ia_tests}

\begin{figure}
	\includegraphics[width=\columnwidth]{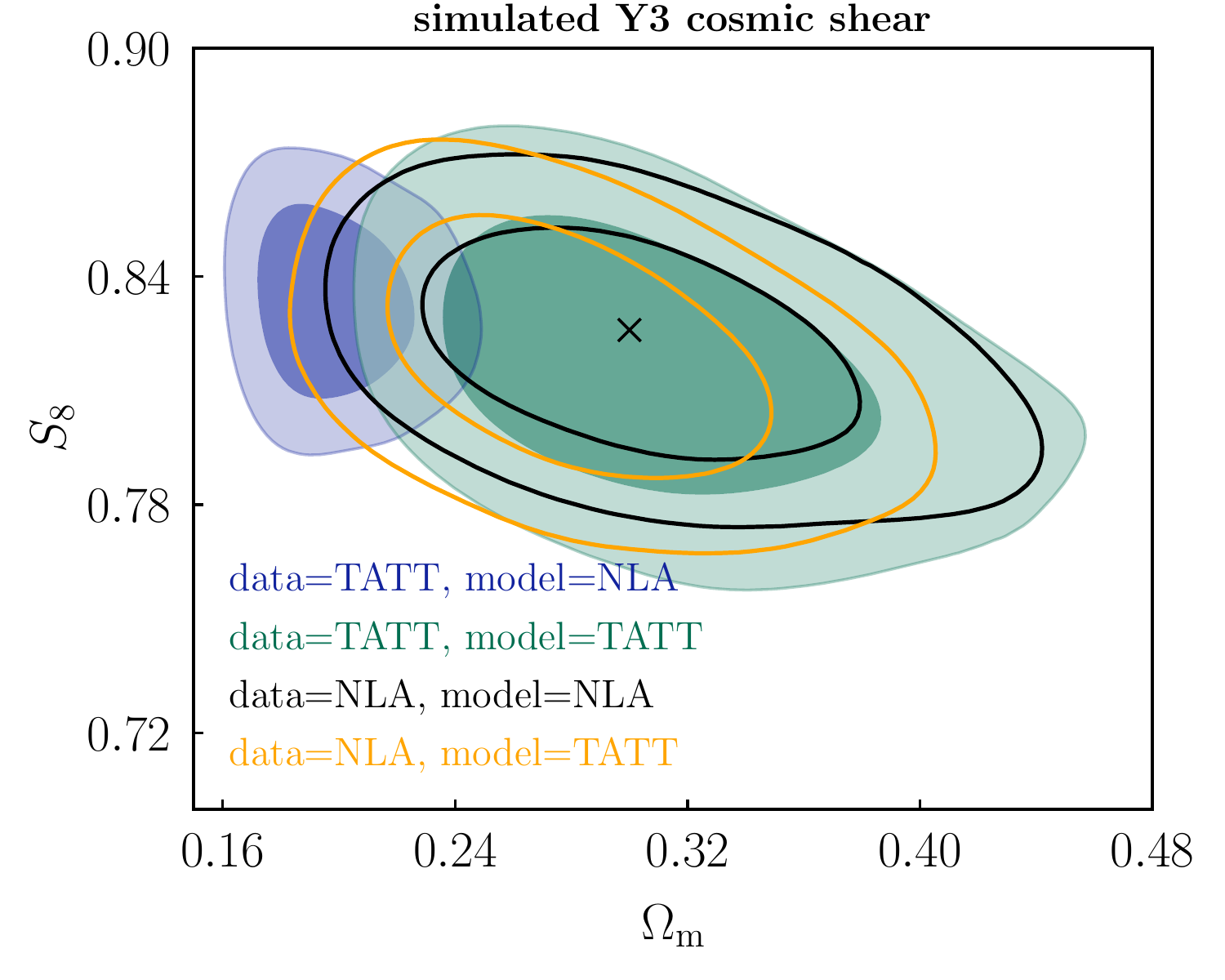}
    \caption{Posteriors from the analysis of \textit{synthetic data vectors with differing input IA signals}, with a known input cosmology (black cross). Strikingly, if IAs in DES Y3 are present at the level of those in the synthetic TATT data vector (chosen to match the DES Y1 result of \cite{samuroff18}), the NLA model is not sufficient to recover the true cosmology (blue contours). We explore whether this scenario is actually reproduced in real data in Sec. \ref{sec: IA model selection}. All other model/data combinations recover the input cosmology to within significantly less than $1\sigma$.}
    \label{fig: NLA and TATT permutations}
\end{figure}

Prior to unblinding of the real data, we carried out a series of tests designed to verify the robustness and sufficiency of our fiducial choice of intrinsic alignment model. Our expectation is to find, using synthetic data, a set of model choices that prevent biases coming from an incorrect characterization of IA. To this end, we generate two simulated data vectors: one including full TATT contributions with input parameters given by the mean posterior IA constraints from the Y1 $3\times2$pt analysis of \citealt{samuroff18} $(a_1=0.7,a_2=-1.36,\eta_1=-1.7,\eta_2=-2.5,b_{\rm TA}=1)$, and an another with the NLA subset of those parameters $(a_1=0.7,\eta_1=-1.7,a_2=0, b_{\rm TA}=0)$.
Our results can be summarized as:
\begin{itemize}
\item{Fitting the NLA data vector with either the NLA or TATT model results in no significant biases. Projection effects in the 1D $S_8$ posterior are negligible in both cases. The more complex model weakens the constraint by increasing the $S_8$ error bar by $14\%$. See Appendix \ref{sec: appendix NLA vs TATT} for a discussion of this on real data, where we find a somewhat similar loss in constraining power.}

\item{Likewise, and as expected, fitting the TATT data vector using TATT recovers the input cosmology and IA parameters.}

\item{Fitting the TATT data vector with the simpler NLA model results in significant residual bias in $\sigma_8$, $S_8$ and \omegam. We also see artificially tightened constraints in this case, as the posteriors on some parameters begin to hit the prior edges, and a significantly degraded best-fit $\chi^2$.}

\end{itemize}
\noindent
The findings above are summarized in Fig.\ref{fig: NLA and TATT permutations}, and a more detailed discussion of these tests, and what they reveal about the systematic error budget, can be found in Appendices \ref{sec:redshift_appendix} and \ref{sec: appendix NLA vs TATT}. The configuration which presented significant biases (NLA modeling of TATT synthetic data) should be considered as a sufficiency test of the simpler IA model, which NLA \textit{fails to meet at the synthetic data level}. This means that, given our expectation from the DES Y1 results of \citet{samuroff18}, and given our deliberate choice of relying solely on synthetically generated data to make a model selection at the blinded analysis stage, our findings provide a strong argument for adopting TATT as our fiducial choice. In summary, previous results on DES Y1 data do not allow us to rule out alignments at the level present in the simulated TATT data vector used in these tests, and we thus choose TATT as our fiducial model. We explore in Sec. \ref{sec: IA model selection} whether this pre-unblinding expectation of the inadequacy of NLA to fit our data is actually realized and what level of IA complexity our data seems to point to. Foreshadowing that discussion, post-unblinding we find NLA to be compatible with DES Y3 data, and that our results for $S_8$ are robust to the choice of the IA model. 

Our fiducial setup makes the explicit assumption that the redshift evolution of IAs in DES Y3 can be adequately described by a power law (Eq. \ref{eq:theory:tatt_c1} and \ref{eq:theory:tatt_c2}). To test this particular analysis choice, we produce synthetic data using the recipe set out in (\citealt{krause2016}; see their Sec. 4.1). In brief, we assume a particular luminosity scaling for the linear alignment strength $\beta_1$, which is combined with a measured luminosity function from the Deep Extragalactic Evolutionary Probe2 (DEEP2; \citealt{DEEP2}) spectroscopic survey, and the measured red fraction $f_{\rm red}(z)$, to predict an IA signal with a non-trivial redshift dependence. Both the redshift and luminosity evolution of red and red+blue galaxies are thus extrapolated (not measured) down to the limiting magnitude over the redshift range of the DES sample.  Our baseline is to use the best fitting $\beta_1$ from \citet{joachimi11} and assume zero alignments in blue galaxies. Note that while constraints on the luminosity dependence of the $a_2$ component of the TATT model do exist in the literature, they are still relatively weak, and consistent with $\beta_2=0$; for this reason, we restrict ourselves to the NLA case. 
The results are described in more detail in \citet{y3-generalmethods}, which also includes similar tests in the context of $2\times2$pt and $3\times2$pt analyses. 
With our fiducial input parameters, we find biases of up to $0.5\sigma$. Although larger than our $0.3\sigma$ passing requirement, it is worth bearing in mind that this test relies on a number of  assumptions and extrapolations and is quite possibly unrealistically stringent. Given this, and in the absence of a stronger test, we maintain our fiducial model and decide to test for the redshift scaling of the $a_1$ parameter \textit{a posteriori}, after unblinding, in Sec. \ref{section: robustness to modeling}.

Finally, we assess whether the interaction between erroneous calibration of photo-$z$s and their uncertainties and our IA model can lead to biased estimates of cosmology. A rationale for this test is that photo-$z$ mischaracterization can produce amplitude shifts in cosmic shear correlations. These shifts can potentially be degenerate with IA model parameters, and bias their posterior distributions, which in turn can lead to biases in cosmology as discussed, e.g., by  \citet{wright20}. Another example is when $n(z)$ distribution tails are underestimated in the data, leading to an erroneous assumption of small overlap between redshift bins and consequently a suppressed II contribution, which is more sensitive to this overlap than the lensing signal is (see also Appendix C of \citealt{fortuna20}).

We seek to test our robustness to such redshift-IA interactions by reanalysing the synthetic TATT data vector described above repeatedly using alternative (incorrect) redshift distributions in the modeling. For this test, we use the fiducial analysis setup including redshift error as parameterised by free $\Delta z$ shift parameters. Rather than introducing analytic distortions to the fiducial $n(z)$, we draw alternative realizations from the ensemble generated by the SOMPZ pipeline \citep*{y3-sompz}. This should naturally capture plausible cases of redshift error, beyond what can be captured by a simple shift in the mean $z$. We select three redshift realizations representative of a range of cases in terms of the change in the predicted data vector ($\Delta\chi^2$). The results are shown in more detail in Appendix \ref{sec:redshift_appendix}, but in brief we find only small ($<0.3 \sigma$) shifts in $S_8$ in the different scenarios. Even in the most extreme case, where $\Delta \chi^2\sim14$, the input IA parameters and cosmology are accurately constrained. In summary, at the level of synthetically generated data with plausible fluctuations in $n(z)$, we do not see evidence that the relatively flexible IA model is absorbing redshift error to a degree that could bias cosmology or IA parameters.

\subsection{Mock data from N-body simulations}\label{sec: sims and mocks}

While most model tests are performed using noiseless analytic data vectors, a subset of tests make use of mock catalogs from N-body simulations. We employ two sets of simulations: \buzzard~ \citep{DeRose2019} and the MICE-Grand Challenge Galaxy and Halo Light-cone catalog\footnote{\url{https://cosmohub.pic.es/home}} \citep{fosalba15}. With \buzzard, a thorough testing of the DES Y3 measurement, photo$-z$ and likelihood pipelines is performed not only for cosmic shear but also for the other DES $3\times2$pt probes \citep{y3-simvalidation}. With MICE, we focus specifically on testing the IA model.

\subsubsection{Buzzard v2.0}

The \textsc{Buzzard v2.0} simulations (\buzzard\ hereafter) are a suite of 18 simulated galaxy catalogs built on $N$-body lightcone simulations (\citealt{DeRose2019}; \citealt{DeRose2021}; \citealt{Wechsler2021}). 
The $N$-body simulations are produced using the \textsc{L-Gadget2} $N$-body code, a memory optimized version of \textsc{Gadget2} \citep{springel05}. The initial conditions for the simulations are generated at $z=50$ using \textsc{2LPTIC} and linear power spectra computed by \textsc{CAMB} at the \buzzard~ flat \lcdm~cosmology: $(\sigma_8, \ns, h, \omegam, \omegab) =(0.82,0.96, 0.7, 0.286, 0.046)$. Each simulation is run on three different and independent boxes  with sizes (1.0, 2.6 and 4.0)$^3\,h^{-3}\mathrm{Gpc}^3$ containing (1400, 2048 and 2048)$^3$ particles respectively, and a lightcone with footprint area over 10,000 deg$^2$ is produced from each of the sets. Galaxies are included in these lightcones with properties such as position, ellipticity and spectral energy distribution using the \textsc{Addgals} algorithm, with details specified in \cite{Wechsler2021}.

Weak lensing quantities are introduced via ray-tracing \citep{Becker2013}, and the complete $3\times2$pt cosmology analysis of DES Y3 is reproduced in these simulations \citep{y3-simvalidation}. For the validation of cosmic shear, \buzzard\ provides a way to verify that several astrophysical effects ignored in our modeling are, as expected, sub-dominant. In addition to higher order corrections from reduced shear and source sample magnification which are also examined in this paper, \buzzard~ mocks contain source galaxy clustering, anisotropic redshift distributions across the survey area and multiple-plane lensing deflections. We point the reader to \citet{y3-simvalidation} for more details.

\subsubsection{Intrinsic Alignments Testing with MICE}

The MICE Grand Challenge run is an $N-$body gravity-only simulation with $4096^3$ collisionless
particles with masses $2.927 h^{-1} M_\odot$ in a $3072 h^{-1}$ Mpc box using a flat $\Lambda$CDM cosmology
with $(\sigma_8, \ns, h, \omegam, \omegab) =(0.8,0.95, 0.7,0.25, 0.044)$.
Dark matter halos are identified as friends-of-friends groups and populated with synthetic galaxies
using a hybrid method of Halo Occupation Distribution modeling and Abundance Matching
\citep{fosalba15b, Crocce15, fosalba15, Hoffmann15, Carretero15}.

Although MICE does not account for baryonic feedback and other hydrodynamic processes related to galaxy formation and evolution, an intrinsic galaxy alignment signal is implemented in the mock using a semi-analytic technique (Appendix \ref{sec: MICE appendix}; see \citealt{kai} for a detailed overview). In brief, shapes and orientations are assigned to galaxies based on a combination of their color and luminosity, and the spin and orientation of the host halos. The model is conceptually similar to those of \citet{Okumura09} and \citet{Joachimi13a, Joachimi13b}, and has been tuned to match the distribution of galaxy axis ratios from COSMOS observations \citep{Scarlata07,  Laigle16} in bins of redshift, stellar mass and color.
Various diagnostic tests have been carried out, and it has been shown to reproduce the projected intrinsic galaxy-shape correlation $w_{g+}$ of luminous red galaxies in the spectroscopic BOSS LOWZ sample \citep{Singh16} as a function of luminosity \citep{kai}.

The MICE IA mock is particularly valuable for our purposes, since it provides an alternative IA prediction with its own redshift dependence, which does not assume either NLA or TATT. At the time of writing, however, only a single realization of the IA mock is available, and we thus we cannot apply the more stringent criteria of \citet{y3-simvalidation} for validating the DES Y3 pipeline on simulations. For this reason, passing the tests presented in this subsection was not an unblinding prerequisite. 

We describe in Appendix \ref{sec: MICE appendix} how a source sample is created with the MICE mock which reproduces basic DES Y3 specifications. The measurement of cosmic shear correlation functions $\xi_{\pm}$ is made using \textsc{TreeCorr} in a similar way to on the real Y3 data. Since the MICE catalog contains the gravitational shears $\gamma_{1,2}$ (G) and the intrinsic shapes $\epsilon_{1,2}$ (I), we can estimate the GG, GI and II components separately if desired. We thus measure two data vectors: a version which contains the GG signal only (the ``baseline'') and another version in which G+I shears are summed together (\citealt{Seitz1997} Eq. 3.2). In this second case, the two-point correlations correspond to GG+GI+II (a ``contaminated'' data vector including the noisy IA signal). With the shape noise and effective number density of MICE galaxies per redshift bin determined in Appendix \ref{sec: MICE appendix}, we construct an analytic covariance using \blockfont{CosmoLike} \citep{cosmolike}, but rescale the shape noise terms so that the figure-of-merit of the fiducial model constraints in the $S_8 \times \omegam$ plane is similar to that of the other simulated cosmic shear tests. We employ the fiducial scale cuts and nuisance parameters. 

With the same likelihood pipeline used in other tests, we find that the bias (distance between posterior peaks) in the $S_8 \times \omegam$ plane between the baseline and contaminated data vectors is 0.6$\sigma$ for the fiducial (TATT) IA model. A similar run, but using NLA as the model, results in a 0.3$\sigma$ bias between baseline and contaminated data. Since the more flexible TATT model should, in principle, be able to capture any IA scenario that NLA can, we ascribe this difference to competing shifts due to the statistical fluctuations and the simpler IA model, which partially cancel. Furthermore, the posteriors from $a_1$ and $\eta_1$ coming from the two models are fully consistent.

One important caveat here is that we have only a single noise realization; the size of the biases quoted for NLA and TATT therefore cannot be decoupled from fluctuations due to shape noise and cosmic variance. This point complicates the interpretation of our results. That said, there is a qualitative difference between the MICE+IA simulated data vectors and the other synthetic $\xi_\pm$ vectors: while all other data are analytically generated using either the NLA or TATT model, the MICE data are agnostic with respect to the analytic model and are tuned to fit observed IA signals (albeit at significantly lower redshifts). It is, then, reassuring that we obtain reasonable IA constraints, and the cosmological parameters $S_8$ and $\omegam$ are not catastrophically biased. Analysis of a greater number of MICE+IA mocks, in order to disentangle IAs from noise and cosmic variance, is an interesting extension left for future work, but beyond the scope of this paper.

\section{Parameter constraints}\label{section: baseline cosmology}

In this section we present the baseline cosmic shear constraints from DES Y3. Sec. \ref{sec: LCDM results} presents our main results in \lcdm, Sec. \ref{sec: wcdm results} shows we find no detection of the dark energy equation of state parameter $w$, and Sec. \ref{sec: IA model selection} presents a detailed comparison of the evidence for different IA models. A series of extended models, including parameterized deviations from General Relativity, and additional neutrino species, will be discussed in \citealt{y3-extensions}, which is in preparation. A number of external data sets are introduced in this section, details of which can be found in Sec. \ref{sec:external_data:datasets}. For a summary of our fiducial results, and of the robustness tests discussed in the next section, see Table \ref{tab:cosmology} and Fig.~\ref{fig: analysis variations}. To facilitate comparison of constraining power, we also show the 2D $S_8-\omegam$ Figure of Merit (FoM), which is defined as $\textrm{FoM}\equiv\det{\mathbf{C}_{S_8,\Omega_\textrm{m}}}^{-1/2}$, where $\mathbf{C}_{S_8,\Omega_\textrm{m}}$ is a sub-block of the parameter covariance matrix.

\begin{table*}\caption{A summary of cosmological constraints from DES Y3 cosmic shear. In each case in the top 9 rows, we show the posterior mean and $68\%$ confidence bounds on each $S_8$ and \omegam, as well as the Maximum Posterior $S_8$ value (denoted $\hat{S}_8$), the 2D $S_8-\omegam$ Figure of Merit and the goodness-of-fit. These top rows also include, as a default, the shear ratio likelihood. A visual summary of the $S_8$ constraints can be seen in Fig.~\ref{fig: analysis variations}. Constraints from KiDS-1000 and HSC are as nominally reported and were not re-processed under the DES Y3 priors and analysis choices, while Planck 2018 numbers have been obtained under DES Y3 cosmology priors.}
    \centering
\begin{tabular}{ccccccc}
\hline 
   & $S_8$ & $\hat{S}_8$ & $\sigma_8$ & $\omegam$ & FoM$_{S_8 \omegam}$ & $\chi^2/\mathrm{dof}$ \tabularnewline
\hline 
\hline 
\textbf{Fiducial \lcdm} & $0.759^{+0.023}_{-0.025}$ & $0.755$ & $0.783^{+0.073}_{-0.092}$ & $0.290^{+0.039}_{-0.063}$ & 927 & $237.7/222=1.07$\\
\lcdm-Optimized & $0.772^{+0.018}_{-0.017}$ & $0.774$ & $0.795^{+0.072}_{-0.076}$ & $0.289^{+0.036}_{-0.056}$ & 1362 & $285.0/268=1.06$ \\

No IAs & $0.774^{+0.017}_{-0.018}$ & $0.760$ & $0.775^{+0.071}_{-0.077}$ & $0.306^{+0.040}_{-0.061}$ & 1253 & $243.3/225=1.08$ \\

NLA & $0.773^{+0.020}_{-0.021}$ & $0.773$ & $0.791^{+0.070}_{-0.086}$ & $0.293^{+0.039}_{-0.056}$ & 1163 & $242.1/224=1.08$ \\

NLA, free $a_1$ per $z-$bin & $0.790^{+0.022}_{-0.020}$ & $0.783$ & $0.834^{+0.075}_{-0.082}$ & $0.275^{+0.035}_{-0.051}$ & 1144 & $246.3/221=1.11$ \\

$a_1>0$ prior & $0.755^{+0.022}_{-0.022}$ & $0.727$ & $0.731^{+0.064}_{-0.085}$ & $0.327^{+0.046}_{-0.066}$ & 881 & $238.7/222=1.07$\\

Fixed neutrino mass & $0.772^{+0.023}_{-0.023}$ & $0.748$ & $0.816^{+0.071}_{-0.094}$ & $0.275^{+0.040}_{-0.052}$ & 1063 & $238.3/222=1.07$\\

\wcdm~cosmology  & $0.735^{+0.023}_{-0.041}$ & $0.699$ & $0.723^{+0.071}_{-0.100}$ & $0.319^{+0.050}_{-0.071}$ & 497 & $237.5/222=1.07$\\

HMCode power spectrum & $0.772^{+0.026}_{-0.027}$ & $0.791$ & $0.793^{+0.088}_{-0.102}$ & $0.294^{+0.038}_{-0.070}$ & 827 & $236.8/222=1.06$ \\

NLA, \lcdm-Optimized, fixed neutrino mass & $0.788^{+0.017}_{-0.016}$ & $0.775$ & $0.825^{+0.078}_{-0.79}$ & $0.279^{+0.036}_{-0.053}$  & 1501 & $288.1/270=1.07$ \\

\hline 
DES Y1 & $0.780^{+0.027}_{-0.021}$ & - & $0.764_{-0.072}^{+0.069}$ & $0.319^{+0.044}_{-0.062}$ & 625 & $227/211 = 1.08$ \\
KiDS-1000 COSEBIs & $0.759^{+0.024}_{-0.021}$ & - & $0.838^{+0.140}_{-0.141}$ & $0.246^{+0.101}_{-0.060}$ & 650 & $85.5/70.5=1.21$\\
HSC Y1 $C_\ell$ & $0.780^{+0.030}_{-0.033}$ & - & - & $0.162^{+0.086}_{-0.044}$ & 461 & $45.4/53 = 0.86$ \\
HSC Y1 $\xi_\pm$ & $0.804^{+0.032}_{-0.029}$ & - & $0.766^{+0.110}_{-0.093}$ & $0.346^{+0.052}_{-0.100}$ & 402 & $162.3/167=0.97$ \\
Planck 2018 TT + TE + EE + lowE & $0.827^{+0.019}_{-0.017}$ & - & $0.793^{+0.024}_{-0.009}$ & $0.327^{+0.008}_{-0.017}$ & 3938 & - \\
\hline
\end{tabular}\label{tab:cosmology}
\end{table*}

\subsection{$\Lambda$CDM}\label{sec: LCDM results}

\begin{figure*}
	\includegraphics[width=\columnwidth]{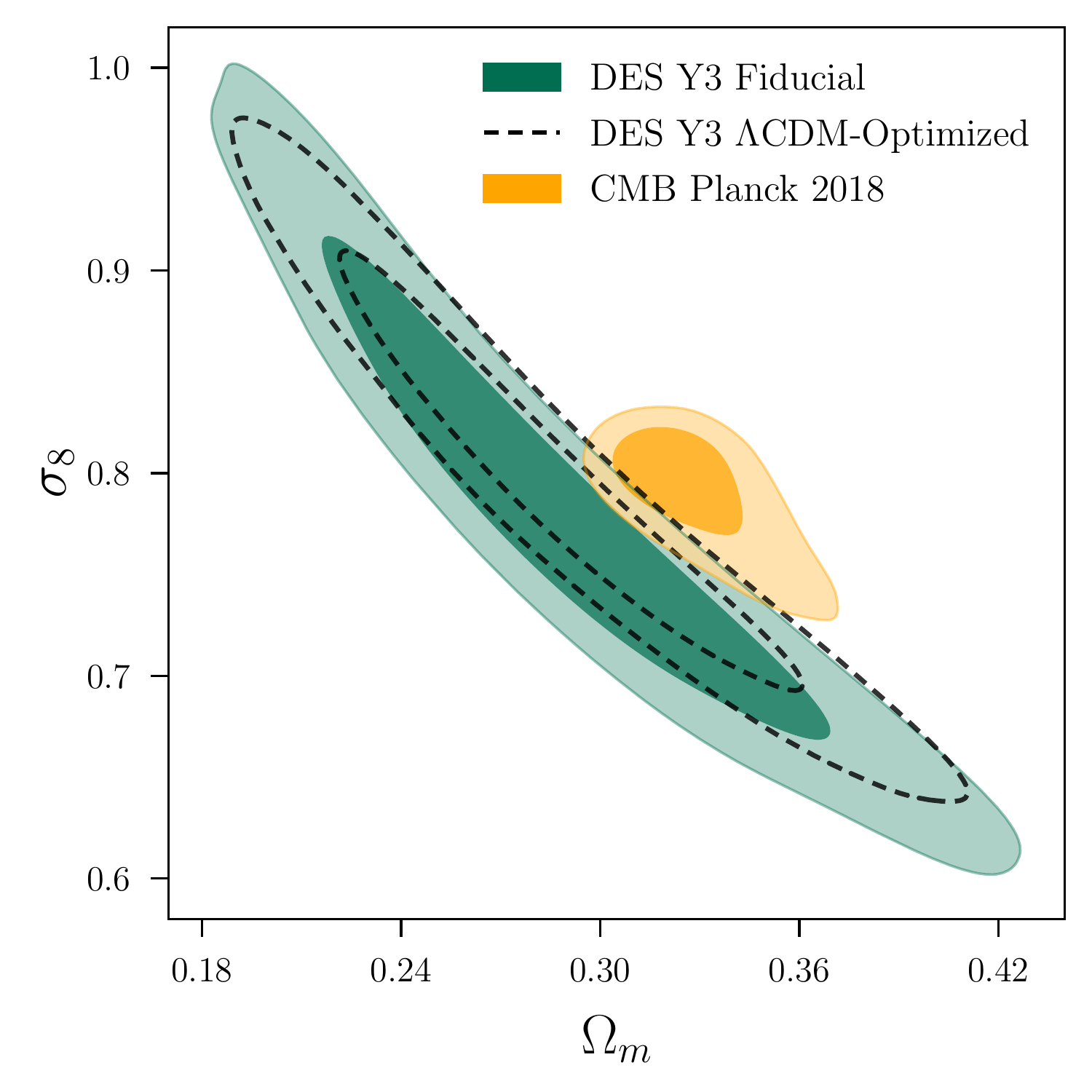}
	\includegraphics[width=\columnwidth]{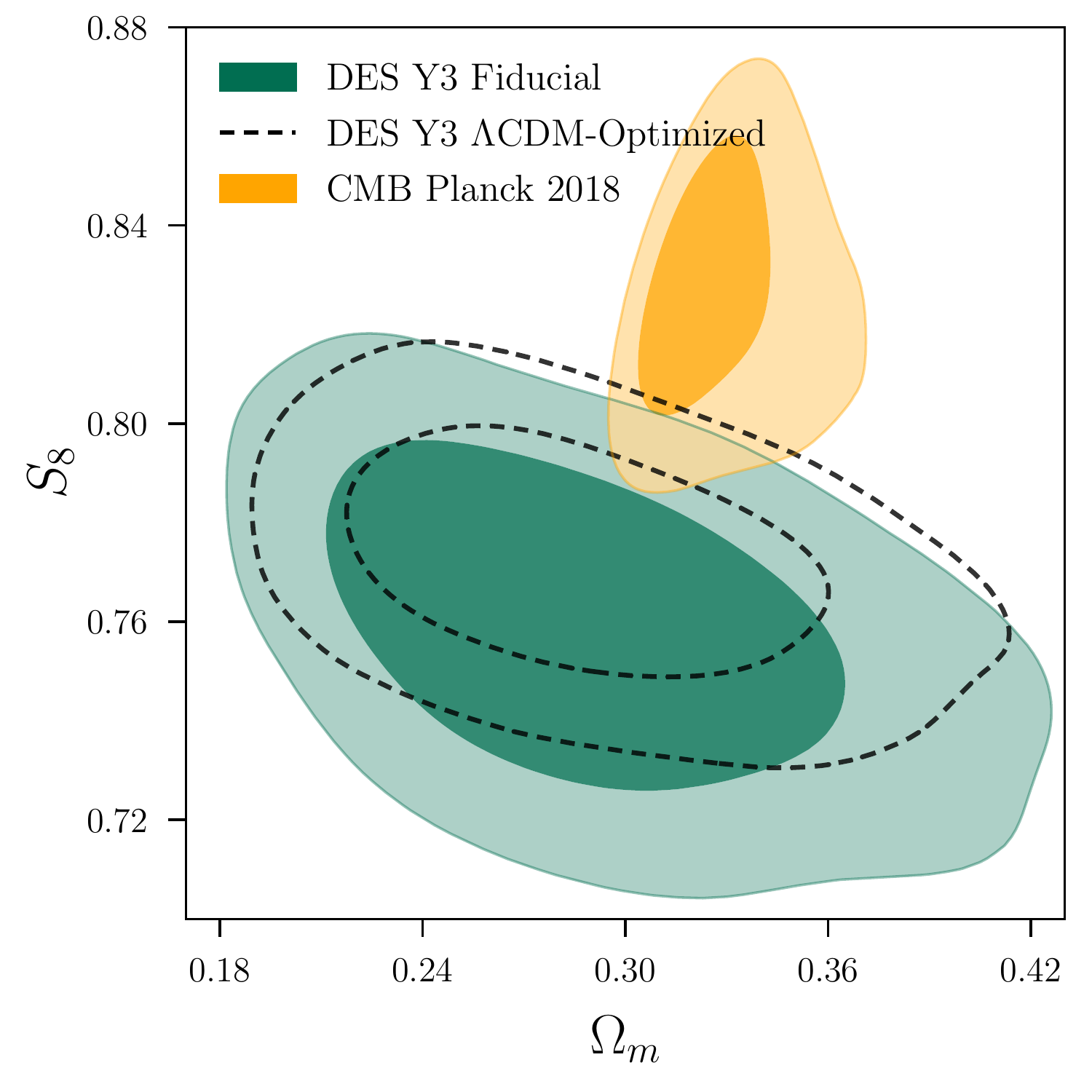}
    \caption{The posteriors of DES Y3 cosmic shear (green) and Planck 2018 (TT+TE+EE+lowE no lensing; yellow). The shaded green contours are the results from our fiducial analysis, as described in Sec. \ref{sec:model}, and constraints using scale cuts that are optimized for a \lcdm-only analysis are shown in dashed lines. In each case we show both $68\%$ and $95\%$ confidence levels.}
    \label{fig: LCDM}
\end{figure*}

The posteriors from our fiducial analysis are shown in Fig.~\ref{fig: LCDM}, along with our \lcdm optimized result (with scale cuts that maximize the constraining power in \lcdm, see Sec. \ref{section: scale cuts}; black dashed), alongside Planck 2018 TT+TE+EE+lowE (no lensing, \citealt{aghanim19}; yellow). As with almost all previous lensing analyses to date, DES Y3 favors a somewhat lower $S_8$, a pattern we discuss further in Sec. \ref{sec: WL agreement}.
The marginalized mean $S_8$ and $\omegam$ values in \lcdm are:
\begin{eqnarray*}
S_{8} & = & 0.759_{-0.025}^{+0.023}\,\quad (0.755)\quad\quad (\Lambda\textrm{CDM})\\
\Omega_{\rm m} & = & 0.290_{-0.063}^{+0.039}\,\quad (0.293)\quad\quad (\Lambda\textrm{CDM})
\end{eqnarray*}
where uncertainties are 68\% confidence intervals. These 1D constraints represent our best estimate for the uncertainties within our analysis framework. It is worth bearing in mind, however, that comparison of constraining power between different lensing surveys is complicated by differences in the analysis choices and priors \citep{chang19}. 
Our fiducial \lcdm~analysis has 222 effective degrees of freedom\footnote{We calculate the effective number of d.o.f. following \citet{Raveri_Hu_2019}, as the number of data points for $\xi_\pm$ after scale cuts (227) minus the number of effective parameters ($N_\textrm{p,eff}\approx5$), given by $N_\textrm{p,eff}\equiv N_\textrm{p,nom}-\Tr{\mathbf{C}^{-1}_\Pi\mathbf{C}_\textrm{p}}$, where $N_\textrm{p,nom}$ is the nominal number of free parameters and $\mathbf{C}_\Pi$, $\mathbf{C}_\textrm{p}$ are respectively the covariances of the prior and posterior. As $N_\textrm{p,eff}$ can change with different modeling choices, we re-compute it whenever necessary.}. At the maximum posterior (MAP) point we obtain $\chi^2/\mathrm{dof} = 237.7/222 = 1.07$, and a corresponding $p-$value of 0.22.   

One important question arising from Fig. \ref{fig: LCDM} is the extent to which the DES and Planck results are consistent with each other. Given that we are considering a complex $28$ dimensional parameter space, assessing agreement purely using projected contours can be misleading (see \citealt{y3-tensions} for discussion). We present a more rigorous quantitative discussion of possible tensions with external data sets in Sec. \ref{sec: tension metric}.

Different angular scales of the cosmic shear correlation function have slightly different sensitivities to cosmological parameters \citep{jain1997}. The degeneracy between $\sigma_8$ and \omegam~is such that the best constrained combination is $S_8=\sigma_8(\omegam/0.3)^\alpha$ with $\alpha\approx0.5$, but we can also determine the exponent $\alpha$ directly from the data. We carry out a Principal Component Analysis of the projected \lcdm posteriors to obtain the exponent value that most effectively decorrelates $\sigma_8$ and \omegam. We find in our fiducial \lcdm TATT analysis that $\alpha=0.586$. To avoid confusion, we call the corresponding lensing amplitude $\Sigma_8$ and find:    
\begin{eqnarray*}
\Sigma_{8}\equiv\sigma_{8}(\Omega_m/0.3)^{0.586}  =  0.756^{+0.021}_{-0.021}\quad (0.729)\quad (\Lambda\mathrm{CDM}).
\end{eqnarray*}
Also shown in Fig.~\ref{fig: LCDM} (black dashed) are the results of our optimized cosmic shear analysis. The details of, and justification for, this additional analysis can be found in Sec. \ref{section:optimized_cuts}, but the key idea is to use a set of scale cuts that are tuned to maximise the constraining power of cosmic shear alone in \lcdm. The result is a tighter constraint in the $S_8- \omegam$ plane in \lcdm:
\begin{eqnarray*}
S_{8}  =  0.772^{+0.018}_{-0.017}\; \quad (0.774)\quad (\Lambda\mathrm{CDM}\textrm{ Optimized}).
\end{eqnarray*}
As can be seen in Fig.~\ref{fig: LCDM}, the gain in constraining power is asymmetric about the posterior peak, which has the effect of shifting the mean $S_8$ up slightly. We consider the impact of this in terms of statistical consistency in Sec.  \ref{sec: tension metric}. The extra data points increase the number of effective degrees of freedom to 268, giving a goodness-of-fit at the maximum posterior of $\chi^2/\mathrm{dof}=285.0/268=1.06$, with a $p$-value of 0.22, similar to the fiducial analysis.

\subsection{$w$CDM}\label{sec: wcdm results}

A simple extension to \lcdm is to free the dark energy equation-of-state parameter (previously fixed to $w=-1$). The prior bounds on $w$ in Table \ref{table: priors} are chosen such that cosmic acceleration is ensured by $w<-1/3$, and ``phantom'' models with $-2<w<-1$ are allowed. To assess the statistical preference of our data for this extended model, we evaluate the evidence ratio (see Sec. \ref{section: 2pt_like}) between \lcdm and \wcdm and find 
\begin{eqnarray*}
R_{w/\Lambda} = 0.94 \pm 0.22,
\end{eqnarray*}
which, based on the Jeffreys scale, is inconclusive in terms of preference for the model with free $w$. We thus find no evidence for (or against) a departure from \lcdm using DES Y3 cosmic shear data alone. 

While $w$ is unconstrained by cosmic shear data  within our priors, we can still constrain $S_8$ in that parameter space, and we show its marginalized value in Fig. \ref{fig: analysis variations}. We find a shift towards lower lensing amplitude, qualitatively implying that this model extension would not help to provide a solution to the differences with respect to Planck.

\subsection{Intrinsic Alignments}\label{sec: IA constraints}

In addition to the main cosmological results, it is also interesting to consider what our data can tell us about IAs. Since we see no evidence that redshift error is biasing our IA results (see Appendix \ref{sec:redshift_appendix}), the IA parameters have a physical interpretation, and are potentially useful for future lensing analyses. For details about how changes in the IA model can affect the inferred cosmology see Sec. \ref{sec: IA robustness on data}.

\subsubsection{IA Model Constraints}

\begin{figure*}
	\includegraphics[width=1.5\columnwidth]{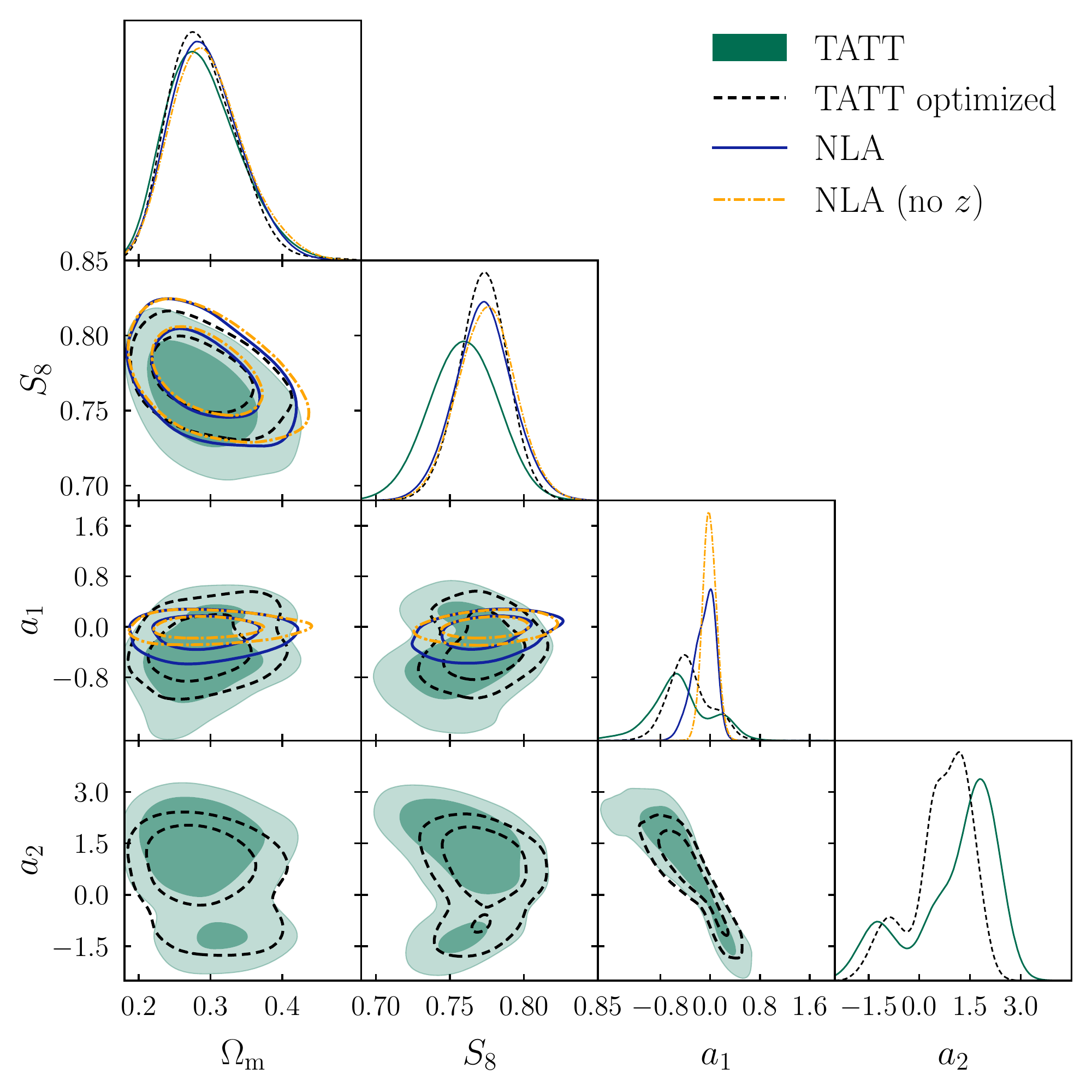}
    \caption{Posterior parameter constraints from the fiducial \lcdm~cosmic shear analysis (TATT, green), which includes shear ratios as a fiducial choice. The solid blue (NLA) and dot-dashed yellow (NLA without redshift evolution) contours show the results of the same cosmic shear analysis, but using alternative (simpler) IA models. We also include our \lcdm~optimized TATT result (dashed black). As before, both $68\%$ and $95\%$ confidence levels are shown. We find that different IA models lead to consistent results and similar goodness-of-fit (see Table \ref{table: IA model selection}), and the main difference between them is the change in constraining power.}
    \label{fig: IAs_tatt_nla}
\end{figure*}

In Fig. \ref{fig: IAs_tatt_nla} we show the IA posteriors in both TATT and NLA, as well as in the optimized \lcdm analysis and NLA with a fixed redshift evolution parameter (for the numerical values, see Table \ref{table: IA model selection}). Note that there are additional parameters in the full model (redshift indices and $b_{\rm TA}$), which are only weakly constrained and so are not shown here (see Appendix \ref{sec: appendix contour plot} for the full posterior). 
In all model scenarios, the data favor a smaller IA signal than DES Y1, although the values are consistent within the uncertainties (see \citealt{samuroff18}'s Fig.~12), resulting in a total GI+II contribution that is only at the level of a few percent of the cosmological signal (as shown in Fig.~\ref{fig: measured xipm}). In the fiducial TATT case, the data favor mildly negative $a_1$, combined with $a_2>0$:

\begin{eqnarray*}
a_1=-0.47^{+0.30}_{-0.52}\,\quad (-0.73)\quad\quad (\textrm{TATT } \Lambda\textrm{CDM}) \\ a_2=1.02^{+1.61}_{-0.55}\,\quad \quad (1.88)\quad\quad  (\textrm{TATT } \Lambda\textrm{CDM}).
\end{eqnarray*}
We observe asymmetric error bars in the posteriors of $a_1$ and $a_2$ and also find slight evidence for a secondary peak with reversed signs of the parameters. This hint of bimodality in the IA subspace can be readily understood within our model: contributions to power spectra in equations (\ref{eq:dEtot})-(\ref{eq:BBtot}) that scale as $a^2_1$, $a^2_2$ or $a_1a_2$ are insensitive to positive/negative sign flips. This degeneracy is further enabled in our present regime, where IA signals are small compared to the measurement errors, and terms proportional to $a_1$ and $a_2$ partially cancel each other when they have opposite signs. We also note that the secondary peak is less pronounced in the optimized analysis shown in Fig. \ref{fig: IAs_tatt_nla}, and \citet{y3-cosmicshear1} shows that the addition of the shear ratios data in our fiducial constraints has already contributed to suppressing the bimodal feature. We thus interpret this doubly-peaked posterior as an internal degeneracy of the model that is broken as statistical power increases. Note that forcing a prior such that $a_1>0$, eliminating one of the peaks, also leads to essentially unchanged cosmology results (see below and in Sec. \ref{sec: IA robustness on data}).

It is worth bearing in mind that the $a_1=a_2=0$ point lies within the bounds of the 2D $1\sigma$ contour, and it is plausible that the negative best-fit $a_1$ is simply the result of a small true IA amplitude and noise (see also Appendix \ref{sec: MICE appendix}). We also confirm that rerunning our fiducial analysis with a prior that forces $a_1>0$ (in line with expectation from observations of red galaxies and the theory of tidal alignment) shifts the IA constraints, and results in a negative $a_2$, as consistent with our Y1 results, but does not appreciably alter the confidence contours in the $S_8-\omegam$ plane.
If we restrict ourselves to the simpler (two-parameter) NLA model, the data still favor negative $a_1$, although at lower significance ($a_1=-0.09^{+0.20}_{-0.13}$). Again, this is lower than the Y1 results, both from cosmic shear alone and in combination with galaxy--galaxy lensing and galaxy clustering ($a^{3\times2\mathrm{pt}}_1=0.49^{+0.15}_{-0.15}$, $a^{1\times2\mathrm{pt}}_1=1.03^{+0.45}_{-0.57}$; see \citealt{y1keypaper} and \citealt{samuroff18}'s Table 5). The impact of the optimized \lcdm~analysis, which includes extra small-scale information, is to tighten the contours in the $a_1-a_2$ plane (compare the shaded green and black dashed lines in Fig. \ref{fig: IAs_tatt_nla}), and it does not qualitatively change our conclusions here.

One detail worth noting is that all of the posteriors presented in this paper include a contribution from shear ratios. The small-scale ($\sim2-6 \; \mathrm{Mpc}/h$) galaxy--galaxy lensing information significantly improves our ability to constrain IAs (see Fig. 10 of  \citealt*{y3-shearratio}). That said, it has been demonstrated that the use of $\gamma_t$ on large (instead of small) scales for the SR likelihood does not substantially change the favored IA scenario \citep*{y3-shearratio}. It has also been shown that removing the SR likelihood altogether results in consistent, though broader, constraints in the $a_1-a_2$ plane \citep{y3-cosmicshear1}.   

\begin{figure}
	\includegraphics[width=\columnwidth]{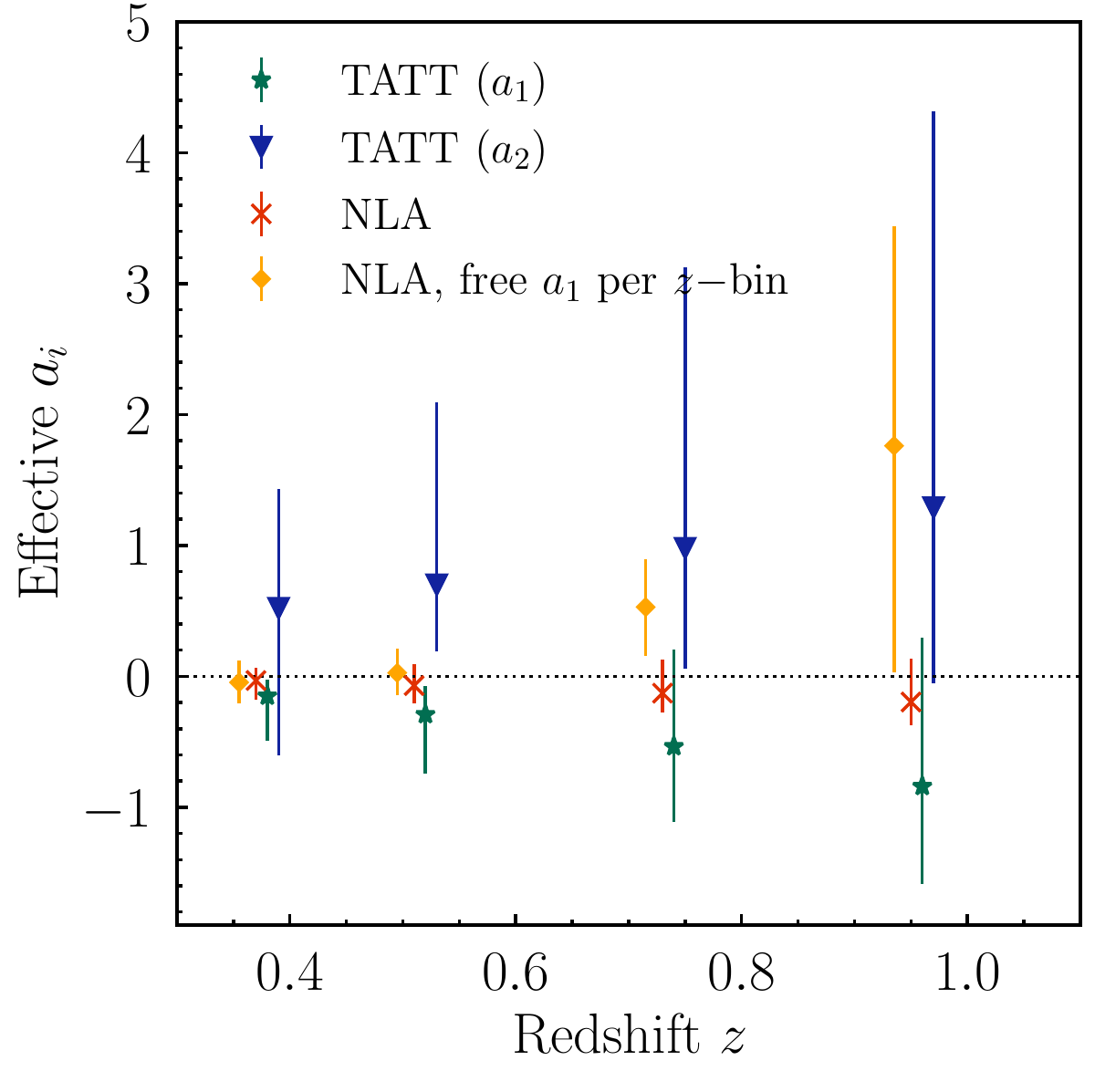}
    \caption{Effective IA amplitudes $a_{1,2}$ as a function of redshift. Note that the green and blue points (both labelled TATT) are the product of the same analysis, in which $a_1$ and $a_2$ were varied simultaneously. We report no clear evidence for redshift evolution in $a_1$, and relatively good agreement between models.}
    \label{fig: zevo}
\end{figure}

We illustrate the redshift dependence of the inferred IA signal in Fig.~\ref{fig: zevo}, which can be compared to the analogous version from DES Y1 in Fig.~16 of \citet{TroxelY1}, and also to Fig. 7 of \citet{y3-3x2ptkp}. In IA models where we use a parametric redshift dependence (TATT and NLA), one can derive effective amplitudes for each redshift bin $j$ as $a^j_i=a_i(1+\bar{z}_j/1+z_0)^{\eta_i}$, where $i\in(1,2)$. The points and error bars in Fig.~\ref{fig: zevo} are the mean and marginalized $68\%$ confidence contours of these derived parameters. In all cases the tidal alignment amplitude $a_1$ is consistent between models to $\sim 1 \sigma$.

\subsubsection{IA Model Selection}\label{sec: IA model selection}

In addition to the basic results using variations of NLA and TATT in the IA implementation, we also perform a more rigorous model comparison in order to determine which IA model preferred by the data in a statistical sense. We explain this concept in more detail below.

First, we rank order models that are subspaces of TATT by their complexity and step up that list one at a time, usually by including a new parameter. At each step $i$, we re-run a chain to obtain the best-fit $\chi^2$ and  evidence of the model. We then compute the evidence ratios (see Sec. \ref{section: 2pt_like}) between the models at step $i$ and $i-1$, and between the model at step $i$ and the fiducial TATT model. In doing this, we are in principle able to determine at which point in the complexity ``ladder'' our data stops justifying the addition of extra IA parameters. 

The order of the model complexity we use is not rigorously defined, but it follows the logic of the perturbative modeling approach. The simplest possible case is a model with no IA. We then consider tidal alignment (TA) less complex than models with tidal torquing (TT), and within these categories, models without redshift dependence are less complex.
Thus, the simplest IA model under consideration has $a_1$ free and all else held fixed, followed by $a_1$ with a free power-law parameter for redshift  evolution, $\eta_1$, and so on up to the (fiducial) 5-parameter TATT (see Table \ref{table: IA model selection}).

In order to minimize sampling noise in the estimate of the best-fit $\chi^2$, we set \blockfont{polychord} to output 10$\times$ more samples than its default. We verify that sampling the parameter space with a maximum posterior finder leads to $\chi^2$ values that are insignificantly different from our best-fit ($\Delta \chi^2 <1$ ). We obtain the evidences as a standard output of \blockfont{polychord}. At each step, we also report the constraints on the marginalized IA parameters and their 68\% confidence intervals.

With this method, we aim to find how many of the TATT parameters meaningfully improve the fit to the data. The evidence ratios are especially suited for this purpose because even though a model with an extra IA parameter might be  \textit{constrained}, the Occam factor in the evidence ratios can still penalize that model if the required prior volume is excessive when compared to a simpler model, Eq.~\eqref{eq: evidence ratio}. We note that according to our analysis choices, this procedure can only be carried out \textit{after} the data and constraints are unblinded, and so we could not have applied the same reasoning before unblinding (e.g. as a strategy to optimize the selection of the fiducial IA model). 

We have performed this test with SR data vectors built using both DES Y3 lens samples (\textsc{redMaGiC} and \textsc{MagLim}), as well as without including the SR likelihood at all, in which case all information comes from $\xi_\pm(\theta)$ on relatively large angular scales. For these three cases we find that the ``no IA model'' has the highest evidence, although it is only ``weakly preferred'' over the other models, based on the Jeffreys scale, reflecting the fact that our inferred IA amplitudes are consistent with zero. We find in all cases that, once the tidal torquing term $a_2$ is free, the evidence ratios seem to suggest a preference for a redshift evolution. Apart from that, we find that the different SR lens samples yield slightly different levels of preference for the intermediate models such as NLA and NLA without $z$-evolution. Our results are shown in Table \ref{table: IA model selection}, where stepping down line-by-line corresponds to increasing the IA model complexity, and for ease of interpretation we show only the case \textit{without} the inclusion of the shear ratio likelihood (so note that the IA constraints presented in the table \textit{do not exactly match the constraints reported elsewhere} in this work).

\begin{table*}
\caption{Marginalized constraints on IA parameters, best-fit $\chi^2$ and evidence ratio $R$ metrics for IA models of increasing complexity on $\xi_\pm(\theta)$ \textit{without including the shear ratio likelihood}. In each row, the evidence ratios assume TATT \& the model in the row above are in the denominator (so $R>1$ implies preference for the model defined in the corresponding row). Marginalized constraints are defined here as the mean of posteriors and their 68\% uncertainties. Empty values in the table correspond to parameters that are fixed to zero.}\label{table: IA model selection}
\centering
\resizebox{\textwidth}{!}{
\begin{tabular}{c|ccccccccc}
\hline 
IA Model (free parameters) & $\chi^{2}$/d.o.f & log Evidence & $R$ (w.r.t. TATT) & $R$ (w.r.t. above) & $a_{1}$ & $\eta_{1}$ & $a_{2}$ & $\eta_{2}$ & $b_{\textrm{TA}}$\tabularnewline
\hline 
\hline 
No IAs & 240.6 / 225 & $3215.79\pm0.11$ & $9.48\pm1.66$ & N/A & - & - & - & - & - \tabularnewline

NLA no $z$-evo. ($a_{1}$) & 238.6 / 224 & $3213.89\pm0.12$ & $1.42\pm0.30$ & $0.15\pm0.03$ & $0.34_{-0.23}^{+0.25}$ & - & - & - & - \tabularnewline

NLA ($a_{1},$ $\eta_{1}$) & 238.3 / 224 & $3214.07\pm0.13$ & $1.70\pm0.36$ & $1.19\pm0.24$ & $0.36_{-0.36}^{+0.43}$ & $1.66_{-1.05}^{+3.26}$ & - & - & - \tabularnewline

TA ($a_{1},$ $\eta_{1}$, $b_{\textrm{TA}}$) & 238.8 / 224 & $3213.87\pm0.13$ & $1.38\pm0.25$ & $0.81\pm0.14$ & $0.27^{+0.35}_{-0.31}$ & $2.10^{+2.89}_{-0.71}$ & - & - & $0.83^{+0.31}_{-0.82}$ \tabularnewline

No $z$-evo. ($a_{1},$ $a_{2}$, $b_{\textrm{TA}}$) & 238.6 / 223 & $3211.81\pm0.14$ & $0.17\pm0.03$ & $0.12\pm0.02$ & $0.18^{+0.21}_{-0.30}$ & - & $0.10^{+0.55}_{-0.57}$ & - & $0.80^{+0.29}_{-0.78}$ \tabularnewline

No $a_{2}$ $z$-evo. ($a_{1},$ $\eta_{1}$, $a_{2}$, $b_{\textrm{TA}}$) & 238.2 / 223 & $3212.09\pm0.14$ & $0.23\pm0.04$ & $1.32\pm0.26$ & $-0.02^{+0.71}_{-0.31}$ & $2.17^{+2.82}_{-0.70}$ & $-0.27^{+0.59}_{-0.50}$ & - & $0.87^{+0.38}_{-0.83}$ \tabularnewline

\textbf{TATT }($a_{1},$ $\eta_{1}$, $a_{2}$, $\eta_{2}$, $b_{\textrm{TA}}$) & 233.1 / 222 & $3213.54\pm0.13$ & 1 & $4.28\pm0.83$ & $-0.24_{-0.41}^{+0.98}$ & $2.38_{-0.61}^{+2.62}$ & $0.63_{-1.89}^{+1.93}$ & $3.11_{-0.31}^{+1.77}$ & $0.87_{-0.84}^{+0.38}$\tabularnewline
\hline 
\end{tabular}
}
\end{table*}

In summary, this test, as well those shown in Fig.~\ref{fig: IAs_tatt_nla}, suggest that simpler IA models are a sufficient assumption for modeling the DES Y3 data. While this result is different from those in the pre-unblinding robustness tests which led to our choice of fiducial model (see Fig. \ref{fig: NLA and TATT permutations}), it is not inconsistent with expectations. Our earlier simulated tests used the best-fit TATT parameter values from the DES Y1 measurements. At these values, the impact of higher order IA contributions was large enough to require the full TATT model. However, the uncertainty in these Y1 measurements was fairly large, and they remain consistent with the overall smaller IA amplitudes found in this Y3 analysis. Smaller amplitudes allow for a less complex IA model, and indeed our current data appear to marginally prefer it.

Our results are consistent with other recent results, including \cite{wright20,fortuna20}, which also suggest low IA amplitudes. While the tests presented here provide additional information to the community for planning future analyses, we emphasize that further study is needed on the underlying IA behavior of typical source galaxies and the interaction with other elements on the model, including baryonic feedback and photometric redshifts, as well as the impact of noise.
We have physical reasons to believe that higher-order IA contributions exist at some level, and the lack of a significant detection here does not imply that these terms can be safely ignored at increased precision or for different source samples. We encourage careful testing of IA model sufficiency in future analyses to avoid cosmological biases.

\section{Model robustness tests on real data}\label{section: robustness to modeling}

\begin{figure*}
	\includegraphics[width=1.6\columnwidth]{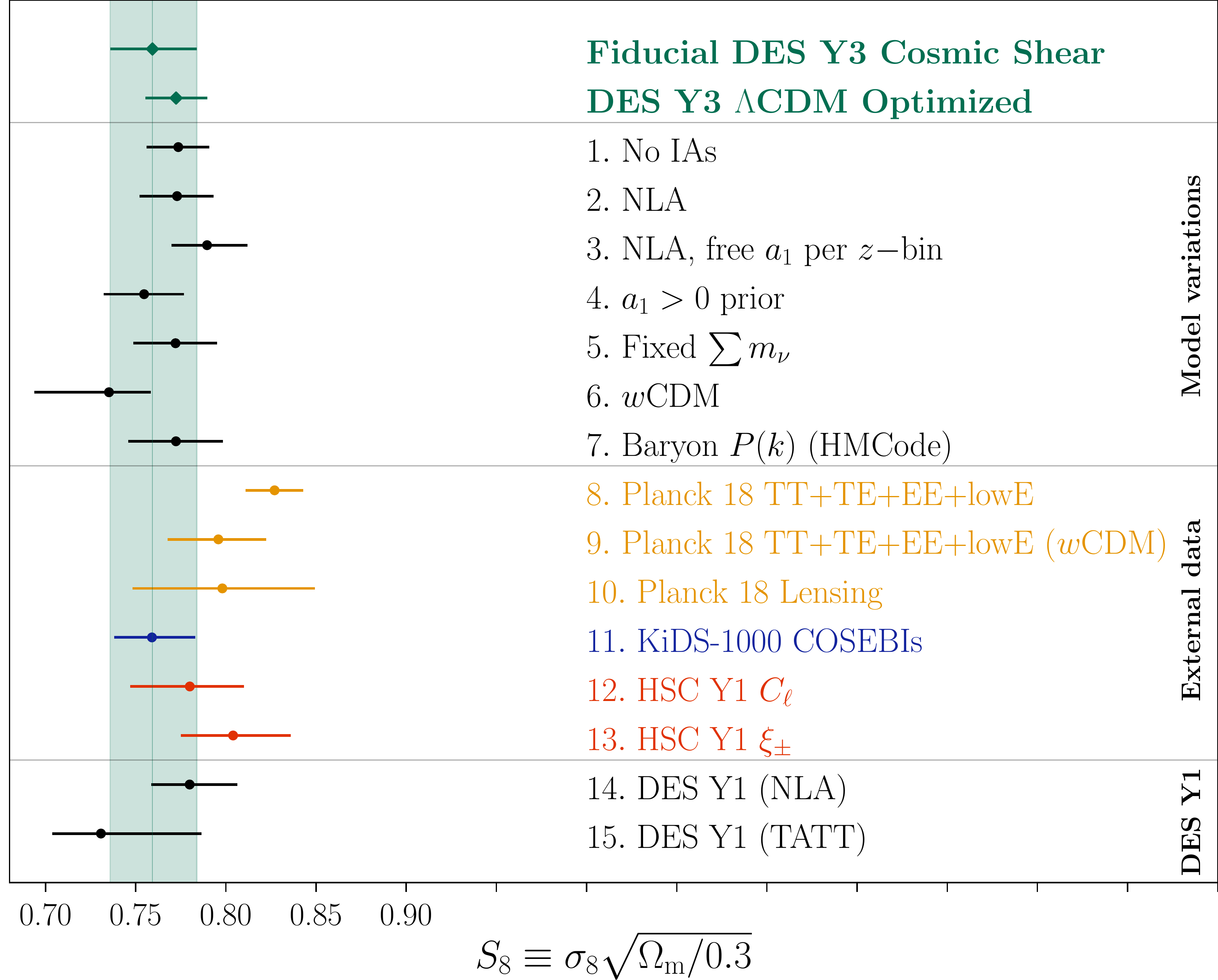}
    \caption{A summary of marginalized 1D constraints on $S_8$ shown in this paper. In the upper-most panel we show the main cosmological results of this paper: the DES Y3 cosmic shear constraints using fiducial and \lcdm Optimized scale cuts. The panel below (marked ``model variations", rows 1-7) shows a range of modified analyses, designed to test the robustness of the fiducial result, which are detailed in Sec.s \ref{section: baseline cosmology} \& \ref{section: robustness to modeling}). In the lower two panels we show equivalent constraints on $S_8$, both from external data (rows 8-13), and DES Y1 (rows 14-15). Points and error bars correspond to the marginal posterior mean and $68\%$ confidence interval on $S_8$, with the exception of KiDS \& HSC (rows 11-13) for which we report nominal headline results. Rows 1-10 are obtained using the DES Y3 fiducial analysis choices (including cosmology priors), while external lensing and DES Y1 rows 11-15 are \textit{not re-processed} to match exactly the Y3 model, prior and scale cuts.}
    \label{fig: analysis variations}
\end{figure*}

In this section we present a number of analysis permutations, with the aim of stress testing the results described in the previous section. The tests fall naturally into a number of groups, which are discussed in more detail in the following paragraphs. For an overall summary, see Fig. \ref{fig: analysis variations}. 

This exercise is distinct from the tests in Sec. \ref{section: unblinding}; whereas there we were seeking to validate our model implementation \textit{prior to unblinding}, we are now seeking to test the robustness of our unblinded results to reasonable variations to the analysis choices.
Our focus here is on various aspects of the theory modeling. For a complementary raft of internal-consistency and data oriented tests, we refer the reader to our companion paper \citep{y3-cosmicshear1}. One can also find extensive catalog-level tests in \citet*{y3-shapecatalog}, which includes a null detection of shear B-modes in the Y3 data using both COSEBIs and pseudo-$C_\ell$s. Further tests of the PSF model and photo-$z$ catalogs are described in other Y3 papers (\citealt{y3-piff}; \citealt*{y3-sompzbuzzard}, \citealt*{y3-sompz}). Also note that the details of the Y3 methodology, including the tests described in this section, were chosen prior to unblinding, and are not informed by the results described in Sec. \ref{sec: LCDM results}. 

\subsection{Baryonic $P(k)$ \& Neutrinos }

One significant source of systematic uncertainty is in the modeling of the small- to intermediate-scale matter power spectrum. Uncertainties at high $k$ arise both as a result of nonlinear clustering, a process that is typically calibrated using $N-$body simulations, and baryonic physics. Although we have verified our insensitivity to reasonable changes in baryonic and nonlinear growth scenarios in Sec. \ref{section: scale cuts}, the tests there use noiseless synthetic data. 

In order to assess the dependence of our results on our particular choice of $P_\delta(k)$ model, we repeat our fiducial \lcdm~analysis, with an alternative power spectrum estimator. That is, instead of \blockfont{HaloFit}, which is our baseline choice, we use the halo model of \citet{mead2015}, which has a nominal accuracy of $5\%$ at $k < 10 h \mathrm{Mpc}^{-1}$,
as assessed by comparison with the Coyote Universe simulations. The halo model includes baryonic effects via two additional parameters: an amplitude $B$ which governs the halo concentration-mass relation, and $\eta_0$, which is referred to as the halo bloating factor. 
We marginalize over both with wide flat priors $\eta_0=[0.4,1]$, $B=[1,7.5]$. These priors are wider than those utilized by, e.g., \citet{asgari20} and are intentionally chosen that way so we are agnostic with respect to the baryonic effects tested in \citet{mead2015}. Another difference with respect to the analysis of \citet{asgari20} is that we free both parameters independently, as opposed to assuming a linear relation between them. 

\begin{figure}
	\includegraphics[width=\columnwidth]{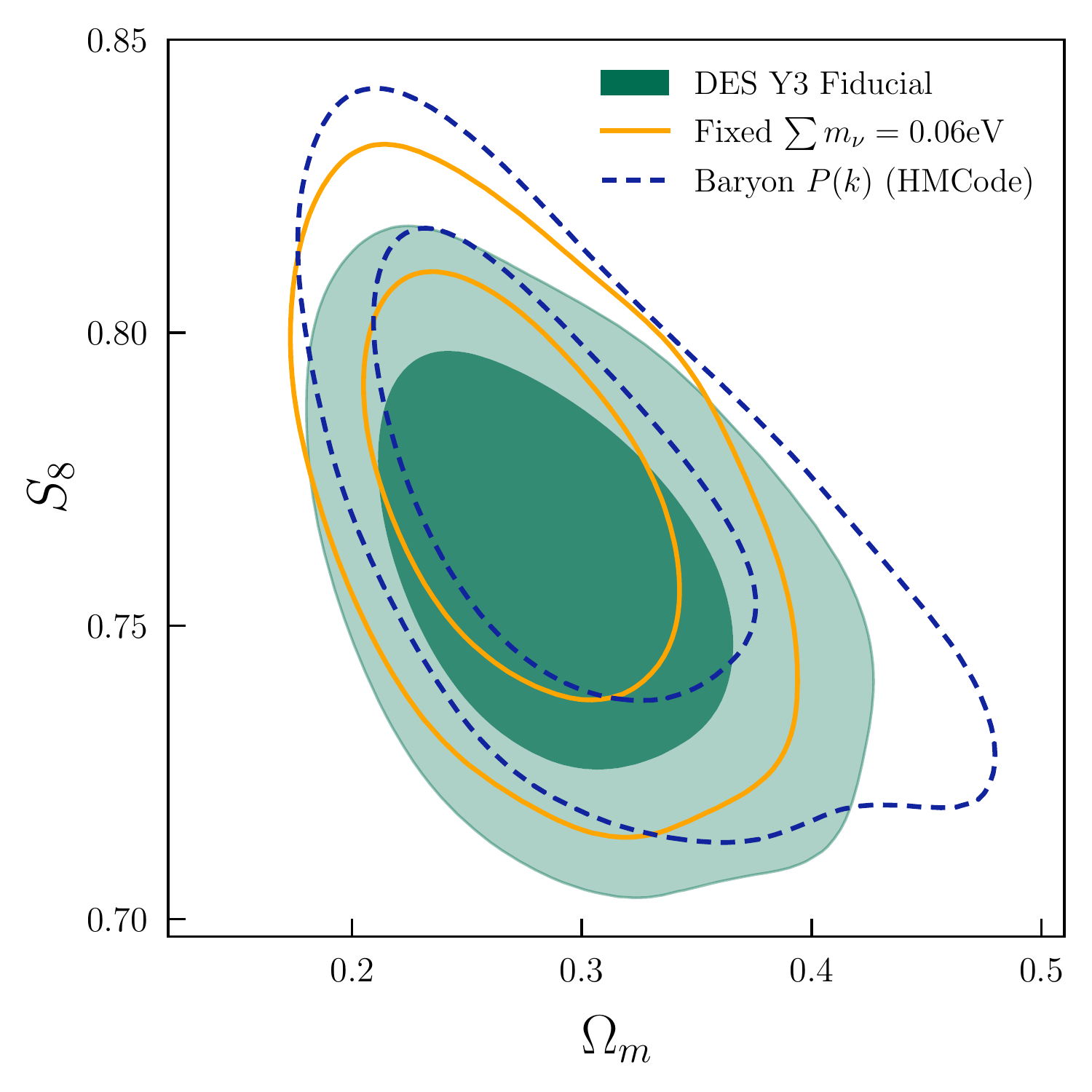}
   \caption{The marginal posterior distribution in the $S_8\times \omegam$ plane for the model variations on the baryonic power spectrum (HMCode with 2 free parameters) and neutrino mass density (fixed at minimum mass). The results of our fiducial analysis is also shown (green). We observe consistency between these cases and our fiducial analysis, and do not interpret shifts as pointing to insufficiency in the analysis (see text).}
    \label{fig: hmcode and neutrinos}
\end{figure}

The results of this exercise are shown in Fig.~\ref{fig: analysis variations} (labelled ``HMCode $P(k)$") and in Fig.~\ref{fig: hmcode and neutrinos}. As shown there, there is a small 0.5$\sigma$ shift in $S_8$ due to this substitution. We do not, however, interpret this as the correction of residual baryonic physics present in the data, but rather a result of projection effects. Indeed, we verify that the halo-model parameters $B, \eta_0$ are completely unconstrained by the data within the prior bounds; this is a consequence of our conservative scale cuts, which are designed to remove angular scales on which baryonic feedback processes (and so the halo-model parameters) enter. The small shift is, therefore, simply an artifact of projecting the high-dimensional posteriors onto the 1D plane. To support this point, we verify that shifts in the same direction appear when analysing a gravity-only synthetic data vector with the same wide baryon priors. We see no significant improvement in the goodness-of-fit of our data when employing \textsc{HMcode} (a $\chi^2$ of 236.8 \textit{vs.} our fiducial 237.7, with 2 added parameters).

In addition to the tests described above, we also repeat our fiducial analysis with the neutrino mass density parameter $\Omega_\nu h^2$ fixed. For this exercise, we set the sum of the neutrino masses, $\sum m_\nu$, to the lower limit obtained by oscillation experiments, assuming the normal (non-inverted) hierarchy, of 0.06eV \citep{capozzi14, esteban19}.
We maintain our fiducial choice of 3 degenerate neutrino species \citep{y3-generalmethods}. The neutrino mass parameter is not constrained by cosmic shear alone, so the small upward shift of around $\sim 0.2\sigma$ we observe in the 1D marginal $S_8$ constraint shown in Fig. \ref{fig: analysis variations} is again thought to be an effect of the projection of the high-dimensional parameter space with a different prior volume. 

We show the 2D parameter constraints in the $S_8$--$ \omegam$ plane under the variations of assumptions in the baryonic power spectrum and the neutrino mass density in Fig.~\ref{fig: hmcode and neutrinos}. We note that despite the small differences in marginalized $S_8$ constraints, these analysis variations are still fully consistent with our fiducial analysis.   

\subsection{Intrinsic Alignments}\label{sec: IA robustness on data}

In this section we explore how plausible variations in our IA modeling can propagate to cosmology. In addition to the fiducial TATT model, we consider a number of alternatives, which are detailed below (see also Fig.~\ref{fig: analysis variations}). The variations cover a range of complexity scenarios, from the simplest case of null IAs (i.e all parameters fixed to zero), through NLA and TATT to a version with added flexibility in redshift.

In Fig.~\ref{fig: IAs_tatt_nla} we consider our baseline result (green), alongside the equivalent using the two-parameter NLA model (blue; \citealt{TroxelY1} and \citealt{hikage2019}'s fiducial choice), and an even simpler one-parameter version with no redshift dependence (dot-dashed yellow; \citealt{asgari20}'s fiducial choice). Although only at the level of $\sim 0.5 \sigma$, we see a shift in $S_8$ when switching between NLA and TATT, which is roughly consistent with the findings of \citet{samuroff18} in DES Y1. This shift is not accompanied by a large improvement in the $\chi^2$ per degree of freedom, but there is some preference for the simpler mode in terms of Bayesian evidence (see Sec. \ref{sec: IA constraints} and Table \ref{table: IA model selection}). It is also notable that the variants of NLA with and without redshift variation give virtually identical cosmology results, primarily because the extra $\eta_1$ parameter is poorly constrained, and relatively uncorrelated with $S_8$. We also to note briefly that the impact of switching IA models also slightly rotates the best constrained direction in parameter space, with $\alpha=0.576$ for NLA and $\alpha=0.586$ in the fiducial TATT, when one defines the lensing amplitude as $\Sigma_8=\sigma_8(\omegam/0.3)^\alpha$. 

Given the negative $a_1$ and hints of bimodality seen in Fig.~\ref{fig: IAs_tatt_nla}, we run a version of the Y3 cosmic shear analysis with a restrictive prior $a_1=[0,5]$.\footnote{Our baseline prior bounds $a_1=[-5,5]$ were designed to be uninformative and avoid prior edge effects. It is, however, reasonable to expect that $a_1>0$, in the absence of systematics.} This results in almost no change to the best-fit $S_8$ (compare lines 1 and 6 in Fig.~\ref{fig: analysis variations}), although it does tighten the error bar slightly by restricting the posterior (row 6 of Fig.~\ref{fig: analysis variations}), which implies that the fact that our fiducial results encompass a region of negative IA space is not driving our cosmological constraints in a particular direction.

We also present a case with a flexible version of the NLA model, shown in Fig.~\ref{fig: analysis variations} (line 5, labelled ``NLA, free $a_1$ per $z$-bin"). The basic idea here is to test how limiting our assumption of power law redshift dependence is in this context, since we have no first-principles reason to expect IAs should obey this particular scaling. Interestingly, adding flexibility via extra TATT parameters, and via the ability to deviate from a power law redshift evolution, is seen to push $S_8$ in opposite directions relative to the basic NLA (c.f. lines 1, 4 and 5 in Fig.~\ref{fig: analysis variations}). While we find a shift of around $1\sigma$ w.r.t TATT (0.5$\sigma$ w.r.t to NLA), the per-bin NLA model is significantly disfavored. The Bayesian evidence ratios strongly favor TATT and NLA over the per-bin NLA model (with $R=237\pm60$ and $202\pm49$ respectively), and in the latter model we additionally see a significantly worse $\chi^2$ (increased by about 10 with respect to TATT while the effective number of degrees of freedom is decreased by 1). 

Taken as a whole, the tests described above paint a consistent picture. Apart from NLA per-bin, the IA modifications we explored cause shifts no greater than $\sim 0.5 \sigma$ in $S_8$, even in the most extreme scenario where we neglect IAs entirely. In NLA per-bin, we find a shift of $1\sigma$ with respect to our fiducial constraints, but also find that this model is disfavored by evidence ratios and goodness-of-fit tests. We therefore do not believe we that our IA modeling is \textit{insufficient} to describe the data and, through the IA model selection described in Sec. \ref{sec: IA model selection}, actually have evidence that it might be simplified in future analyses. We do find a change in constraining power between NLA and TATT, with NLA (at fiducial scale cuts) being comparable to TATT at the \lcdm-Optimized scale cuts. This suggests that our uncertainty budget is dominated by the lack of knowledge in modeling astrophysics on small cosmological scales. This is verified by our companion \citet{y3-cosmicshear1}, Sec. XII.

\section{Comparison with external data}\label{sec:external_data}

After having verified that DES-only results are robust with respect to changes in astrophysics modeling assumptions, we now seek to place our results in the context of the wider field and to quantify tension with respect to subsets of external data.

\subsection{External Data Sets}\label{sec:external_data:datasets}

We describe a number of external data sets (see Table \ref{tab:external_data}), to which we will compare our results. Where appropriate, we recompute the cosmological posteriors in order to facilitate a meaningful comparison. The marginalized $S_8$ constraints from the external data sets that meaningfully constrain this parameter alone are shown in the lower half of Fig. \ref{fig: analysis variations}. The data sets we consider in this paper are largely common to those described in \citet{y1keypaper}, \citet{y1extensions} and \citet{TroxelY1}, with some more recent updates. Among these, as described in Sec. 1, the KiDS and HSC surveys are Stage-III WL surveys like DES. The full set of surveys we consider is:
\begin{itemize}
    \item \textbf{KiDS-1000}: The KiDS weak lensing data comprise roughly 1000 square degrees and 21M galaxies ($n_{\rm eff}\sim6.2$ / arcmin$^2$), and are described in \citet{giblin20}. In their latest results papers, \citep{asgari20} present cosmic shear analyses using three different statistics; since they clearly designate their COSEBIs-based results as their fiducial analysis, we compare with these here. Note that we do not recompute the posteriors, but rather compare with the published results. In particular, one should be aware that our choice of IA model, nonlinear power spectrum and cosmological priors all differ from the fiducial analysis of \citet{asgari20}.
    \item \textbf{HSC Y1}: The first year HSC lensing data are drawn from 137 square degrees, but go significantly deeper than either KiDS or DES, reaching $n_{\rm eff}\sim22$ /arcmin$^2$. The data are described in \citet{mandelbaum17}, and are calibrated using image simulations \citep{mandelbaum17b}. As in the case of KiDS, we do not reanalyze the two-point data, instead comparing with the published results. Since, at the time of writing, there is no reason to prefer one over the other, we show both the real space analysis of \citet{hamana20} and the power spectrum analysis of \citet{hikage2019} in Fig. \ref{fig: analysis variations}. The differences between the two are thought to be statistical, due to different Fourier mode sensitivities. For the sake of clarity, we choose to show the latter in the visual comparison of Fig.~\ref{fig:external_lensing}.
    \item \textbf{eBOSS}: We include spectroscopic baryon acoustic oscillation (BAO) measurements from eBOSS \citep{allam20}. The BAO likelihood is assumed to be independent of DES, but we do recompute the posterior in our choice of cosmological parameter space (i.e. sampling \as~and $\omegam$, and with the sum of the neutrino masses free).

    \item \textbf{Pantheon Supernovae}: We also include the luminosity distances from Type Ia supernovae from the Pantheon sample \citep{scolnic18}; this data set includes 279 type Ia supernovae from the PanSTARRS Medium Deep Survey $(0.03 < z < 0.68)$ and samples from SDSS, SNLS and HST. The final Pantheon catalogue includes 1048 SNe, out to $z = 2.26$.
    \item \textbf{Planck 2018 Main}: These data are the final release from the Planck Cosmic Microwave Background (CMB) experiment \citep{aghanim19}. We incorporate the primary $TT$ data on scales $30<\ell<2508$, and also the joint temperature and polarization measurements ($TT+TE+EE+BB$) at $2<\ell<30$. As in previous analyses, we recompute the CMB likelihood in our fiducial parameter space, including neutrinos.
    \item \textbf{Planck 2018 Lensing}: We also consider CMB lensing from the Planck survey \citep{aghanim18} as a separate data set. These data probe an intermediate redshift $z\sim2$, which is slightly higher than DES and somewhat lower than CMB temperature and polarization.
\end{itemize}

\begin{table*}\label{tab:external_data}
    \centering
    \begin{tabular}{c|cccc}
    \hline 
        Data set  & Type & Median Redshift & Area [sq. deg] & Source \\
        \hline 
        \hline 
        DES Y3         & WL                 &  0.62               &   4143         & This work  \\
        KiDS-1000      & WL                 & 0.54            &   777          & \citet{asgari20} \\
        HSC Y1         & WL                 & 0.81            &   137          & \citet{hamana20}, \citet{hikage2019}  \\
        eBOSS          & BAO                & 0.7             &   6813         & \citet{allam20} \\
        Pantheon       & SNe                 & 1.0             &  -             & \citet{scolnic18} \\
        Planck Main    & CMB $TT+TE+EE+$lowE  & 1090            & full-sky         & \citet{aghanim19}    \\
        Planck Lensing & CMB Lensing        & 2.0             & full-sky         & \citet{aghanim18} \\
       \hline
    \end{tabular}
    
    \caption{A summary of the external data sets used in this paper. More details can be found in Sec. \ref{sec:external_data} and the references listed.}

\end{table*}

\subsection{Quantifying Tension}\label{sec: tension metric}

 Assuring that the data collected by two different experiments have a quantified degree of agreement is crucial when performing a joint probes analysis. Therefore, for DES Y3, we have presented an thorough study on the ability of different tension metrics to identify inconsistencies amongst cosmological parameters measured by different experiments. We used simulated data to compute the predictions of different tension metrics belonging to two classes: Evidence-based methods and parameter-space methods. A robust way to quantify possible tensions is by combining  those two types of metrics, as they answer somewhat different, but yet complementary, questions \citep{y3-tensions}.

In DES, we utilize priors that are deliberately wide and uninformative such that "DES-only" constraints can be obtained. In this case, assessing tension utilizing the Bayes Ratio, a broadly used evidence-based tension metric, can produce misleading results since it is largely dependent on the prior volume as discussed in \citet{Handley:2019wlz}. To avoid this problem, for our choice of an evidence-based metric we compute the Bayesian Suspiciousness \citep{Handley:2019b} instead, a method that corrects for the prior dependence.

Consider two independent data sets A and B. The motivation behind Suspiciousness is that the Bayes Ratio $R$ can be divided into two parts: the first one captures the prior dependence, i.e.  the probability of the data sets matching given the prior width, which is quantified by the information ratio $I$ :
\begin{equation}
    \log I \equiv \mathcal{D}_A + \mathcal{D}_B - \mathcal{D}_{AB},
\end{equation}
where
\begin{equation}
    \mathcal{D}_D \equiv \int \mathcal{P}_D \log \left( \frac{\mathcal{P}_D}{\Pi} \right) \ \dd \theta,
\end{equation} 
is the Kullback--Leibler Divergence \citep{Kullback:1951}, that can be understood as the amount of information that has been gained going from the prior $\Pi$ to the posterior $\mathcal{P}$. The lower index $D$ denotes the data set from which the posterior is derived ($A$, $B$, or the joint data vector $AB$). The second part is the so-called Bayesian Suspiciousness $S$, which is the part of the Bayes Ratio left after subtracting the dependence on the prior, leaving only on the actual differences between the posteriors: 
\begin{equation}
    \log S = \log R - \log I.
\end{equation}

All the quantities required to compute the Suspiciousness metric are provided by a single nested sampling chain, just as the ones required to compute the Bayes Ratio are, meaning the computational cost is the same in both cases. The necessary tools are implemented in the python package {\tt anesthetic}\footnote{
\url{https://github.com/williamjameshandley/anesthetic}
}
\citep{anesthetic}.

In addition, we calculate a Monte Carlo estimate of the probability of a parameter difference, set out in~\citet{Raveri:2019gdp}, as our parameter space-based method. It relies on the calculation of the parameter difference probability density $\mathcal{P}(\Delta \theta)$ which, in the case of two uncorrelated data sets, is simply the convolution integral: 
\begin{equation} \label{Eq:ParameterDifferencePDF}
\mathcal{P}(\Delta \theta) = \int_{V_p} \mathcal{P}_A(\theta) \mathcal{P}_B(\theta-\Delta \theta) d\theta,
\end{equation}
where $V_p$ is the prior support and $\mathcal{P}_A$ and $\mathcal{P}_B$ are the two posterior distributions of parameters. 
The probability of an actual shift in parameter space is obtained from the density of parameter shifts:
\begin{equation} \label{Eq:ParamShiftProbability}
\Delta = \int_{\mathcal{P}(\Delta\theta)>\mathcal{P}(0)} \mathcal{P}(\Delta\theta) \, d\Delta\theta,
\end{equation}
which is the posterior mass above the contour of constant probability for no shift, $\Delta\theta=0$. 

Usually we only have discrete representations of the posterior samples and, since we are working with high dimensional parameter spaces, the posterior samples cannot be easily interpolated to a obtain a continuous function. Therefore, the integral in Eq. (\ref{Eq:ParamShiftProbability}) needs to be performed using a Monte Carlo approach. 

The idea is to compute the Kernel Density Estimate (KDE) probability that $\Delta\theta=0$ and the same at each of the samples from the parameter difference posterior. Then, the estimate of the integral in Eq.~(\ref{Eq:ParamShiftProbability}) is given by the number of samples with non-zero KDE probability, divided by the total number of samples. These computations are done using the {\tt tensiometer}\footnote{\url{https://github.com/mraveri/tensiometer}} code.

\newpage
\subsubsection{Compatibility between Low-$z$ and High-$z$ Data}\label{sec: highz vs lowz}

We now assess the consistency between high- and low-redshift cosmological probes. As in \citet{y3-3x2ptkp}, we conclude that two data sets are statistically consistent if the $p$-value implied by our  tension metrics is greater than 0.01. This same standard value was utilized in our internal consistency tests in \citet{y3-cosmicshear1}. In order to quote tension metrics, we first recompute the low-$z$ (here defined as the combination BAO + SNe) and Planck 2018 posteriors with priors on cosmological parameters matched to those of DES Y3, as defined in Table \ref{table: priors}. In addition to assessing tension relative to our fiducial Y3 analysis, we also explore whether switching to the \lcdm-Optimized scale cuts significantly changes our findings, since the one-dimensional result in Fig. \ref{fig: analysis variations} can offer hints, but not quantify the agreement between data sets.

\begin{figure}
	\includegraphics[width=\columnwidth]{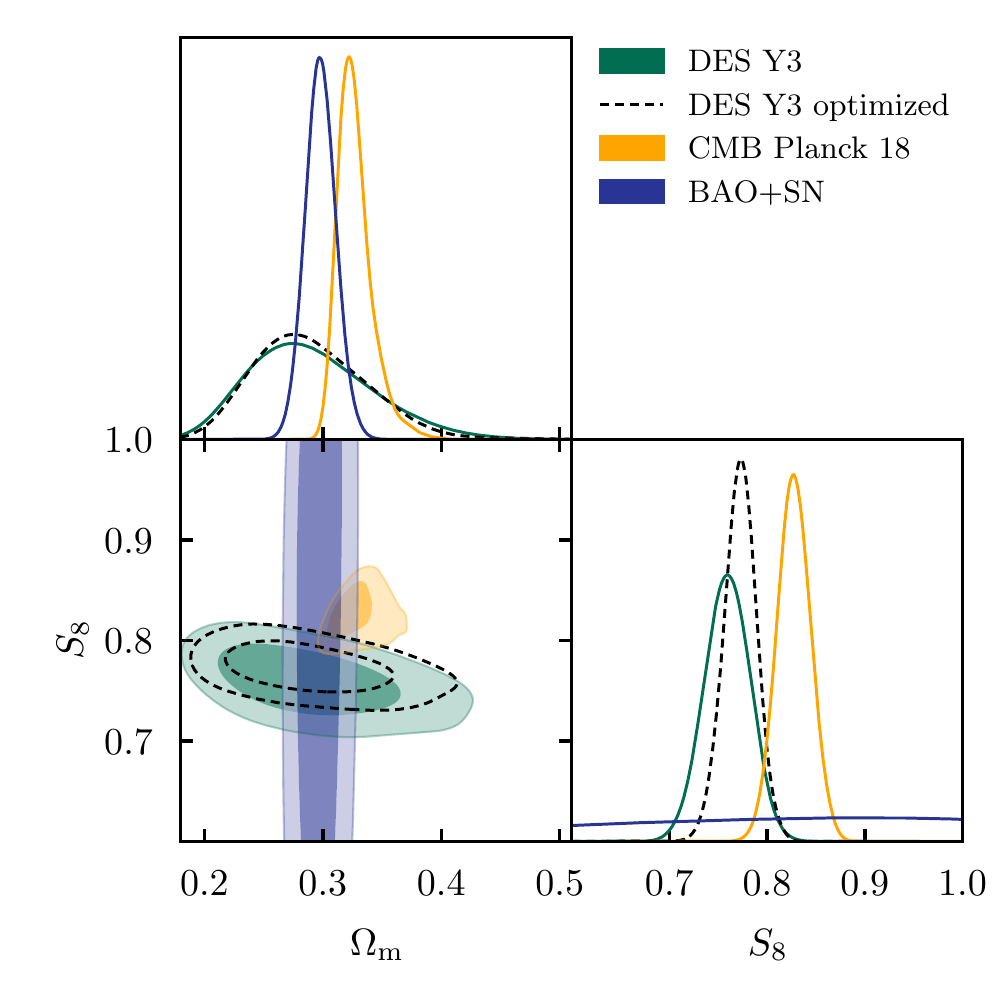}
    \caption{DES Y3 and external data constraints from low and high redshift probes in \lcdm. We present our fiducial and \lcdm optimized constraints (green and black-dashed) in comparison with Planck 2018 (TT+EE+TE+lowE, no lensing; yellow) and the combination of BAO and type Ia SNe (SDSS BOSS and Pantheon respectively; blue). In all cases we show both the $68\%$ and $95\%$ confidence limits. We find no evidence for statistical inconsistency between DES Y3 cosmic shear and either external data set. }
    \label{fig: TATT_NLA_vs_external_LCDM}
\end{figure}

Fig.~\ref{fig: TATT_NLA_vs_external_LCDM} shows the posteriors from DES as well as low-$z$ and Planck 2018 in \lcdm. In the full parameter space, for DES Y3 cosmic shear \textit{versus} BAO+SNe in \lcdm we find:
\begin{eqnarray*}
\textrm{Suspiciousness:} & \quad{\normalcolor 0.5\sigma\pm 0.3\sigma}\quad\small{(\textrm{DES \textit{vs} Low-}z\,\Lambda\textrm{CDM})}\\
\textrm{Parameter shift:} & \quad\qquad 0.4\sigma\quad\quad\;\;\small{(\textrm{DES \textit{vs} Low-}z\,\Lambda\textrm{CDM})},
\end{eqnarray*}
which implies $p(0.5\sigma)=0.62>0.01$, and we conclude these data sets are consistent. This confirms more rigorously what is seen qualitatively in Fig.~\ref{fig: TATT_NLA_vs_external_LCDM}.

We similarly quantify the compatibility between DES Y3 cosmic shear and Planck 2018 in~\lcdm:
\begin{eqnarray*}
\textrm{Suspiciousness:} & \quad{\normalcolor 2.3\sigma\pm 0.3\sigma}\quad\small{(\textrm{DES \textit{vs} Planck }\Lambda\textrm{CDM})}\\
\textrm{Parameter shift:} & \qquad\quad 2.3\sigma\quad \quad\;\small{(\textrm{DES \textit{vs} Planck }\Lambda\textrm{CDM})},
\end{eqnarray*} 
with a $p$-value $p(2.3\sigma)=0.02>0.01$. We thus find that, according to our tension metrics, these data sets are consistent. We additionally explore the consistency of DES Y3 cosmic shear with Planck 2018 in our \lcdm-Optimized setup and find:
\begin{eqnarray*}
\textrm{Suspiciousness:} & \quad{\normalcolor 2.0\sigma\pm 0.4\sigma}\quad\small{(\textrm{DES optim. \textit{vs} Planck }\Lambda\textrm{CDM})}\\
\textrm{Parameter shift:} & \qquad\quad 2.1\sigma\quad \quad\;\small{(\textrm{DES optim. \textit{vs} Planck }\Lambda\textrm{CDM})},
\end{eqnarray*} 
which yields approximately $p(2.0\sigma)=0.05>0.01$, again implying that these data sets are consistent. That result can be expected, at least qualitatively, by inspecting Figs.~\ref{fig: analysis variations} and \ref{fig: TATT_NLA_vs_external_LCDM}: while the constraint on $S_8$ is improved in the optimized analysis, there is also a small shift towards Planck.

\begin{figure*}
	\includegraphics[width=\columnwidth]{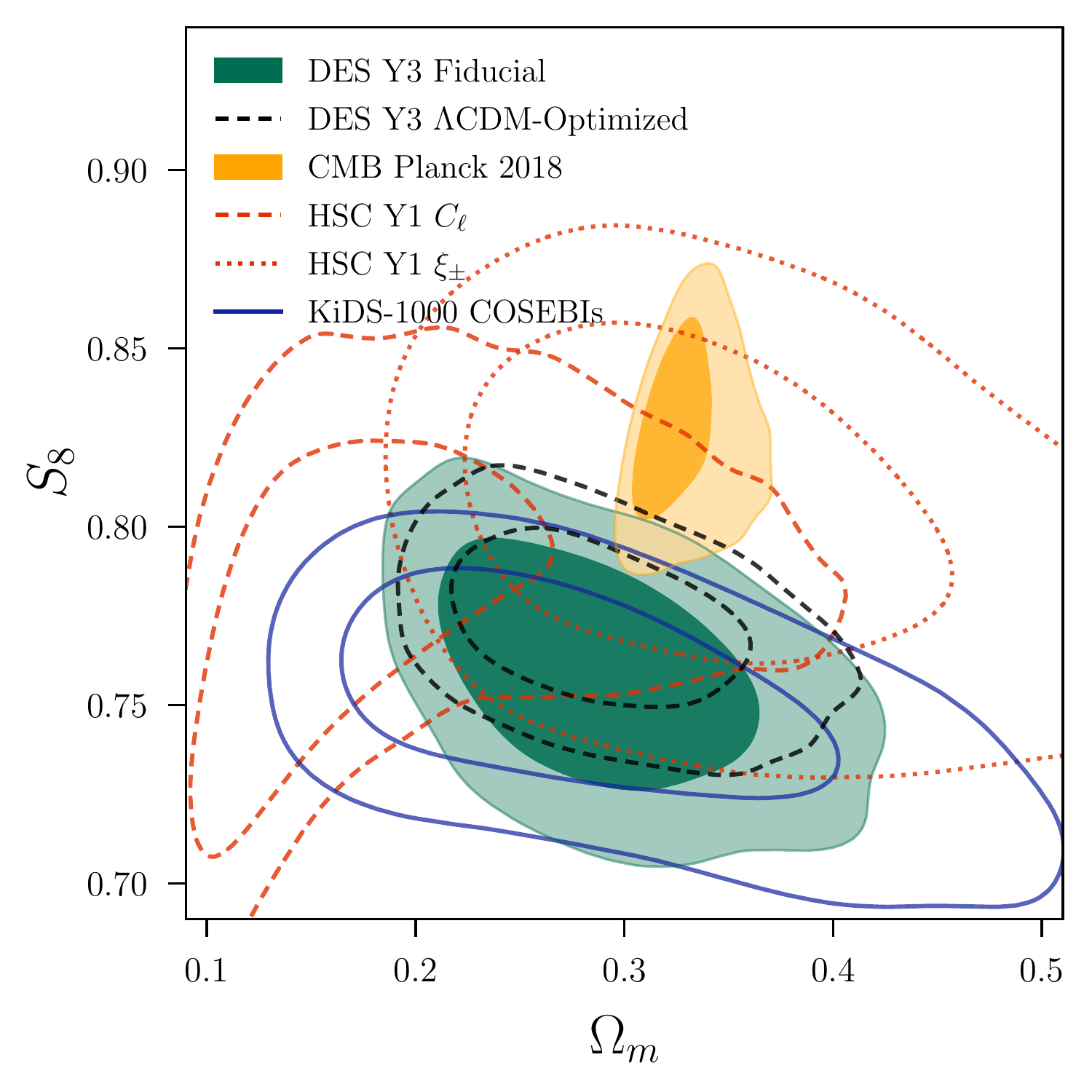}
	\includegraphics[width=\columnwidth]{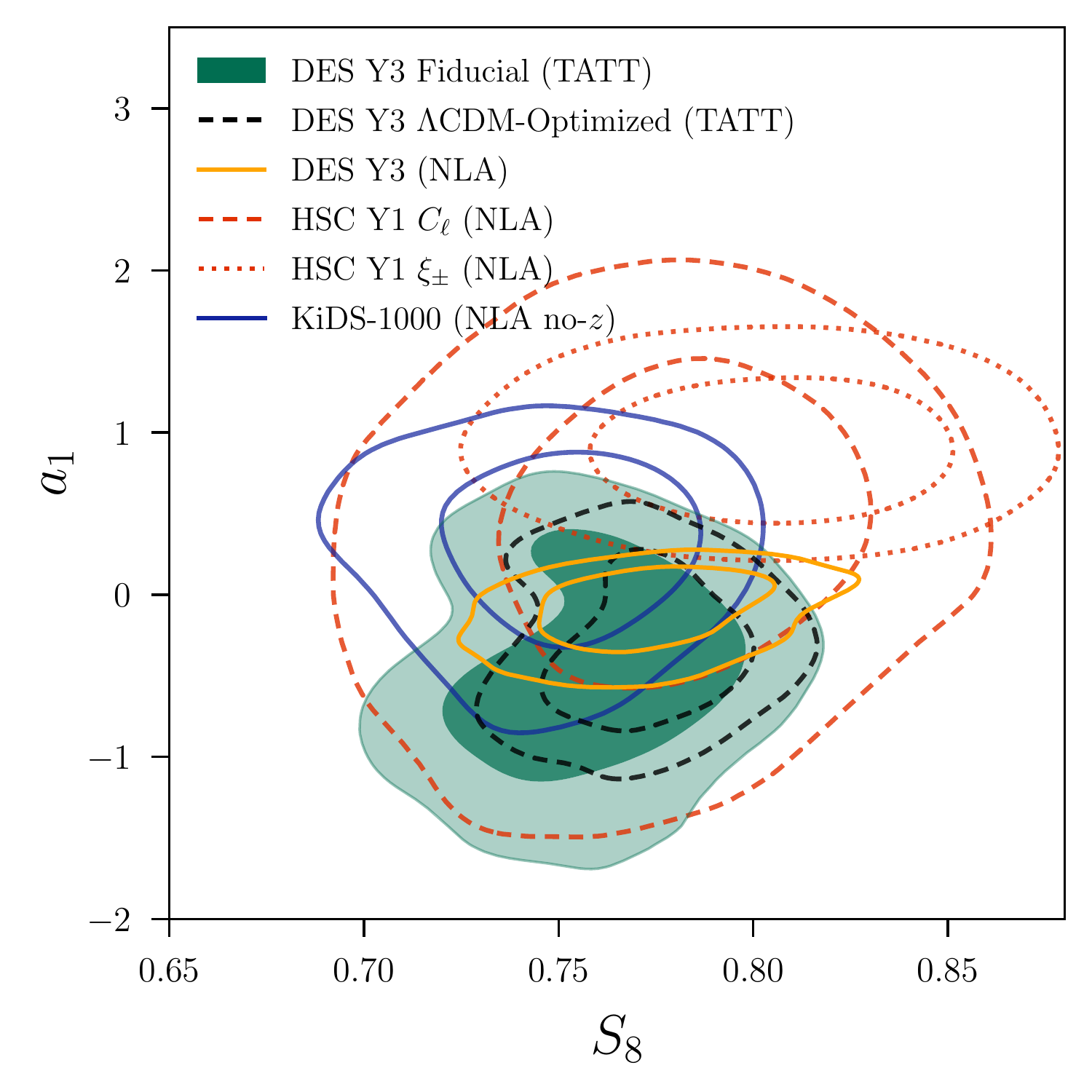}
    \caption{Posterior constraints on cosmology (left panel) and intrinsic alignment  parameters (right panel) from contemporary weak lensing surveys. With the exception of Planck 2018, which is re-analyzed with DES Y3 priors, the weak lensing posterior data above are plotted as published by each collaboration, therefore a direct, quantitative comparison is complicated due to differing parameter priors, modeling and calibration choices. An overall qualitative agreement is seen in both panels, within approximately 1$\sigma$, with lensing surveys predicting similarly low $S_8$ and consistent $a_1$ amplitude.}
    \label{fig:external_lensing}
\end{figure*}

\subsubsection{Comparing Weak Lensing Surveys}\label{sec: WL agreement}

We show our fiducial DES Y3 \lcdm~results (green shaded), alongside those of a number of contemporary weak lensing surveys in Fig. \ref{fig:external_lensing}. Also shown (yellow) are the equivalent constraints from most recent Planck CMB data release (without lensing). 
Most notably, all the published cosmic shear analyses (including earlier, less constraining, results that are not shown here) favor lower $S_8$ than Planck, to varying degrees of significance. Although the cosmic shear surveys are independent, both in the sense that there is limited overlap between the catalogues (see Fig. \ref{fig: footprint}), and that the measurement and analysis pipelines are largely separate, it is remarkable that their $S_8\times\omegam$ posteriors overlap significantly, especially given that these weak lensing analyses are carried out blind. 

That said, the naive comparison of the surveys is complicated by the fact that there are significant differences in the analysis choices underlying the published results, and unifying them becomes a crucial task \citep{chang19}. Though it is quite possible to compute the metrics discussed in the previous section, interpreting them would be complicated, as it would be difficult to determine if any apparent tension were real, or due to differences in priors (see the discussion in \citealt{y3-tensions}).  
A proper quantitative comparison would require matching priors and other analysis choices, and would ideally involve input from the different collaborations; such an exercise would be in line with the recommendations of \citealt{chang19} and is left for future work.

We can, however, compare our findings with the tension analyses that the various collaborations choose to present. For example, in DES Y1 cosmic shear the discrepancy relative to Planck 2018 was assessed to be at the level of $1.0\sigma$ and $0.5\pm0.3 \sigma$ using the parameter shift and Suspiciousness metrics respectively \citep{y3-tensions}. 
These numbers can be compared directly with our findings in Sec. \ref{sec: highz vs lowz}. We see that the discrepancy between cosmic shear and Planck 2018 has increased substantially in DES Y3 in comparison with DES Y1, from  $1.0\sigma$ to 2.3$\sigma$. The latter number is, coincidentally, the same difference found in the comparison between Planck and the $3\times2$pt analysis of DES Y1 (including galaxy clustering and galaxy-galaxy lensing).

The KiDS-1000 cosmic shear analysis of \citet{asgari20} opted to quantify their agreement with Planck using the Bayes ratio. They find ``substantial" evidence for disagreement, with the caveat that the performance of this metric is known to be prior dependent. Though they also compute a Suspiciousness, they do not quote an interpretable value in terms of $\sigma$, due to difficulties in computing effective dimensionalities (see their App. B3).
The approach of HSC differs slightly between their two published analyses. \citealt{hamana20} (their  Sec 6.6) rely on a comparison of projected contours, reporting no tension due to the apparent overlap in the $\sigma_8\times \omegam$ plane. On the other hand, \citet{hikage2019} employ both the Bayes ratio and also differences in the log-likelihood at the MAP point in parameter space (see their Sec 6.3 and \citealt{raveri19}). They report no evidence for inconsistency based on these metrics.

In the right-hand panel of Fig.~\ref{fig:external_lensing} we also show the IA model constraints from the various surveys. Although they are roughly consistent (to within $\sim 1 \sigma$), it is worth being careful here. In all cases but KiDS-1000, the IA model has more flexibility than the one amplitude (an additional redshift scaling in HSC and DES Y1, and the extra TATT parameters in Y3), which means, for a given $a_1$, the predicted IA signal is not necessarily identical between the surveys. It is also true that, unlike cosmological parameters, IAs are \emph{expected} to be dependent on both the galaxy selection and the shape measurement method. Differences, while interesting from the modeling perspective, are not necessarily a cause for concern.

\section{Conclusions}\label{sec:conclusions}
 
This paper and its companion, \citet{y3-cosmicshear1}, present together the cosmological constraints from cosmic shear with over $\sim100$ million galaxies from Dark Energy Survey Year 3 (DES Y3) shape catalogs, which cover 4143 square degrees. We model the cosmic shear signal in \lcdm and constrain the lensing amplitude with 3\% precision, finding $S_8=0.759^{+0.023}_{-0.025}$. Our best constraint on \lcdm is of $S_8=0.772^{+0.018}_{-0.017}$, at 2\% precision, when optimizing the angular scales while still maintaining biases under control.  Our results qualitatively agree well with previous lensing analyses based on the KiDS-1000, DES Y1 and HSC data sets, all of which favor lower $S_8$ than the most recent CMB measurements. Using quantitative tension metrics based on the full parameter space, we report a $2.3 \sigma$ difference between our DES Y3 results and Planck 2018 (a $p$-value of 0.02), and we consider these data sets to be statistically consistent. At our present constraining power, we do not report a meaningful constraint on the equation-of-state parameter of dark energy in \wcdm. That is, the cosmic shear likelihood does not significantly add information about the parameter $w$ beyond its prior. 

Throughout this paper, we have focused on the aspects of the DES Y3 cosmic shear analysis relating to astrophysical modeling. The overarching philosophy that guided this work and its conclusions can be summarized as follows. Firstly, in a blinded fashion, we test our modeling assumptions on synthetic data obtained from theory pipelines and N-body mocks, in order to validate our fiducial analysis choices. We then obtain cosmology constraints from shear data that are verified to be internally consistent and calibrated to high confidence by our companion paper \citep{y3-cosmicshear1}. Finally, we proceed to relax our fiducial theory assumptions and vary model parameterizations, finding that the results to be robust to these variations, and then compare our constraints with those of external probes. The main conclusions of this paper are:

\begin{itemize}

  \item Informed by previous intrinsic alignment (IA) studies in DES Y1 data, we account for IAs with a 5-parameter model that includes tidal alignments and tidal torquing (TATT). We present tests of this modeling assumption on analytically generated data as well as realistic MICE (N-body) simulations, and find that TATT is able to capture complex IA signals, as well as that of simpler models such as NLA (Fig. \ref{fig: NLA and TATT permutations});

  \item We select angular scales conservatively to mitigate the effect of baryonic physics, the dominant small-scale systematic for this analysis. We show that our scale cuts suppress baryonic contamination as inferred from hydrodynamic simulations with varying levels of feedback strength, and utilize a gravity-only model for the matter power spectrum. We verify in synthetic data that residual biases due to baryons are well below 0.3$\sigma$ in the $\Omega_\textrm{m}\times S_8$ plane, and that our selected scales are insensitive to uncertainties in the nonlinear power spectrum and in higher order shear contributions (Fig. \ref{fig: systematics});
  
    \item The tension metrics we use in the full parameter space yield a $2.3\sigma$ discrepancy with respect to Planck (2018), with the leading contribution to the tension being our lower $S_8$. External probes at low-z (BAO+SNe), which are sensitive to $\Omega_m$, are within $0.5\sigma$ under the same metrics. We regard both these probes as statistically consistent with DES cosmic shear. We also find qualitative agreement between our data and external lensing results from HSC Y1 and KiDS-1000, all of which yield lower nominal $S_8$ values in comparison with Planck 2018 (Figs. \ref{fig: LCDM}, \ref{fig: TATT_NLA_vs_external_LCDM} and left panel of \ref{fig:external_lensing});

  \item We demonstrate that our posteriors on IA and cosmological parameters are consistent within $1\sigma$ as we vary the parameterization of intrinsic alignments by simplifying our fiducial model from TATT to NLA with free/fixed redshift dependence. This is also true when we allow for free $a^i_1$ amplitudes on each redshift bin $i$. We demonstrate that our posteriors on cosmological parameters are stable when inference is carried out at fixed neutrino mass and with free baryonic feedback parameters (Figs.~\ref{fig: zevo}, \ref{fig: analysis variations} and \ref{fig: hmcode and neutrinos});

\item We perform a detailed Bayesian evidence-based model selection for intrinsic alignments, finding that our data shows a weak preference for simpler (and better constrained) parameterizations. We additionally find that the best-fit IA amplitudes in DES Y3 are smaller than those in Y1. However, the two are consistent with each other, and also with other lensing surveys, although quantitative statements are challenging without further studies of e.g. sample selections and model differences. In combination with the fact that posteriors on $S_8\times\Omega_\textrm{m}$ from simpler IA models are consistent with our fiducial choice (TATT), these findings point to less complex parameterizations, such as NLA, being a sufficient and unbiased description of our data (Fig.~\ref{fig: IAs_tatt_nla}, Table \ref{table: IA model selection} and right panel of Fig.~\ref{fig:external_lensing}).   
\end{itemize}

Additionally, we share a summary of the main conclusions of our companion paper \citep{y3-cosmicshear1}, demonstrated with the same Y3 cosmic shear analysis, and point the reader to that paper for details:
\begin{itemize}
        \item The analysis is shown to be robust to the choice of redshift calibration sample, either photometric of spectroscopic,  methodology and the modeling of redshift uncertainty, within $\sim0.5\sigma$. 
    \item We model the impact of blending using state-of-the-art image simulations and show that our posteriors are stable to this correction, within $0.5\sigma$.
 \item The analysis passes all internal consistency tests, finding that the cosmology is stable across redshifts, angular scales and measurement statistics. 
 \item The impact of additive shear systematics, such as PSF contamination and B-modes is assessed and found to be negligible for the analysis. 
 \item We investigate the limiting factors of the cosmic shear constraints and find observational systematics are subdominant compared to systematics due to modeling astrophysical effects. 
\end{itemize}

The results presented in this study show that cosmic shear has reached the requirements of percent level precision: every source of systematic uncertainty must be controlled to better than a percent of the signal, since our statistical uncertainty on the amplitude parameter $S_8$ is three percent. When the first detections of cosmic shear were published in 2000, it was far from a given that this level of precision was achievable. With the full DES survey awaiting analysis, and the much larger surveys from Euclid and the Rubin Observatory's LSST starting in a few years, it is interesting to consider prospects for further qualitative advances. 

On the theoretical side, we are limited by our ability to describe the astrophysics of small scales, in particular baryons and IA. Baryonic physics is the driver of our scale cuts, and we will need to rely on advances in hydrodynamic simulations, constrained by complementary observations such as Sunyaev-Zel'dovich (SZ) maps, otherwise the necessary cuts and/or marginalization will cause more information to be discarded. Conversely, we have found that intrinsic alignments are likely to contribute less than the most conservative forecasts, but more studies are required to extended our findings to arbitrary galaxy samples. Parallel advances in shear estimation and calibration and the photo$-z$s of source galaxies are required to keep pace with statistical errors. Beyond cosmic shear, a suite of statistical measures that capture the non-Gaussian information in the shear field have been developed, such as peak statistics and three-point correlations. These will add complementary information and also serve as cross-checks on the robustness of the signal. We can look forward to new cosmological tests with the application of these approaches to DES and future surveys.

Finally, we reiterate that, while the assumption of \lcdm as the ultimate end-to-end model connecting the early- and late- Universe has withstood another test, our result should be understood within a broader context. It is still an open puzzle that modern weak lensing surveys, independently and in blind analyses, find a lower lensing amplitude than predicted by the CMB, and the difference between DES cosmic shear itself with respect to Planck has increased from $1.0\sigma$ in DES Y1 to $2.3\sigma$ in Y3. More effort is required to quantify the agreement between lensing studies, especially towards unifying and homogenizing their analyses and exploiting their complementarities.

\section*{Acknowledgements}
LFS and BJ were supported in part by the
US Department of Energy grant DE-SC0007901.
We thank Rachel Mandelbaum for many useful discussions during the writing of this paper. We would like to kindly ask the readers to cite this manuscript with dual-authorship as Secco \& Samuroff \textit{et al.} whenever possible.

Funding for the DES Projects has been provided by the U.S. Department of Energy, the U.S. National Science Foundation, the Ministry of Science and Education of Spain, 
the Science and Technology Facilities Council of the United Kingdom, the Higher Education Funding Council for England, the National Center for Supercomputing 
Applications at the University of Illinois at Urbana-Champaign, the Kavli Institute of Cosmological Physics at the University of Chicago, 
the Center for Cosmology and Astro-Particle Physics at the Ohio State University,
the Mitchell Institute for Fundamental Physics and Astronomy at Texas A\&M University, Financiadora de Estudos e Projetos, 
Funda{\c c}{\~a}o Carlos Chagas Filho de Amparo {\`a} Pesquisa do Estado do Rio de Janeiro, Conselho Nacional de Desenvolvimento Cient{\'i}fico e Tecnol{\'o}gico and 
the Minist{\'e}rio da Ci{\^e}ncia, Tecnologia e Inova{\c c}{\~a}o, the Deutsche Forschungsgemeinschaft and the Collaborating Institutions in the Dark Energy Survey. 

The Collaborating Institutions are Argonne National Laboratory, the University of California at Santa Cruz, the University of Cambridge, Centro de Investigaciones Energ{\'e}ticas, 
Medioambientales y Tecnol{\'o}gicas-Madrid, the University of Chicago, University College London, the DES-Brazil Consortium, the University of Edinburgh, 
the Eidgen{\"o}ssische Technische Hochschule (ETH) Z{\"u}rich, 
Fermi National Accelerator Laboratory, the University of Illinois at Urbana-Champaign, the Institut de Ci{\`e}ncies de l'Espai (IEEC/CSIC), 
the Institut de F{\'i}sica d'Altes Energies, Lawrence Berkeley National Laboratory, the Ludwig-Maximilians Universit{\"a}t M{\"u}nchen and the associated Excellence Cluster Universe, 
the University of Michigan, the National Optical Astronomy Observatory, the University of Nottingham, The Ohio State University, the University of Pennsylvania, the University of Portsmouth, 
SLAC National Accelerator Laboratory, Stanford University, the University of Sussex, Texas A\&M University, and the OzDES Membership Consortium.

The DES data management system is supported by the National Science Foundation under Grant Numbers AST-1138766 and AST-1536171.
The DES participants from Spanish institutions are partially supported by MINECO under grants AYA2015-71825, ESP2015-88861, FPA2015-68048, SEV-2012-0234, SEV-2016-0597, and MDM-2015-0509, 
some of which include ERDF funds from the European Union. IFAE is partially funded by the CERCA program of the Generalitat de Catalunya.
Research leading to these results has received funding from the European Research
Council under the European Union's Seventh Framework Program (FP7/2007-2013) including ERC grant agreements 240672, 291329, and 306478.
We  acknowledge support from the Australian Research Council Centre of Excellence for All-sky Astrophysics (CAASTRO), through project number CE110001020.

This manuscript has been authored by Fermi Research Alliance, LLC under Contract No. DE-AC02-07CH11359 with the U.S. Department of Energy, Office of Science, Office of High Energy Physics. The United States Government retains and the publisher, by accepting the article for publication, acknowledges that the United States Government retains a non-exclusive, paid-up, irrevocable, world-wide license to publish or reproduce the published form of this manuscript, or allow others to do so, for United States Government purposes.

Based in part on observations at Cerro Tololo Inter-American Observatory, 
National Optical Astronomy Observatory, which is operated by the Association of 
Universities for Research in Astronomy (AURA) under a cooperative agreement with the National 
Science Foundation.
This research manuscript made use of Astropy \cite{astropy:2013,astropy:2018}, GetDist \cite{getdist} and Matplotlib \cite{matplotlib}, and has been prepared using NASA's Astrophysics Data System Bibliographic Services.



\bibliographystyle{mnras_2author}
\bibliography{cosmic_shear_bibliography.bib,des_y3kp.bib}



\appendix
\section{Construction of the MICE source catalog and results}\label{sec: MICE appendix}

In this section, we discuss the construction of a mock weak lensing source catalog with the MICE N-body simulation. Significantly, in addition to cosmological shear, these mock catalogs contain a realistic IA component, making them a good testing ground for our modeling; this is the primary motivation for using them here, as discussed in Sec. \ref{sec: sims and mocks}. The implementation of the IA signal into MICE is described in far more detail in \citet{kai}, and is summarized below. 

Each MICE galaxy has a (projected) shape assigned to it:
\begin{equation}\label{eq:mice:gamma}
    \gamma = \gamma_{\rm G} + \gamma_{I} + \mathcal{\epsilon}.
\end{equation}
\noindent
The cosmological part of this is obtained via ray tracing (see \citealt{fosalba15}); $\gamma_I$ is the coherent intrinsic component, which is generated via semi-analytic modeling (see below and \citealt{kai}), and $\mathcal{\epsilon}$ is random Gaussian shape noise.

For the purposes of our test, it is desirable that the overall constraining power of the mock catalog (in terms of the figure-of-merit in the $S_8\times\Omega_\mathrm{m}$ plane) resembles that of the fiducial source catalog; to that end, we match the constraining power of Y3 cosmic shear by adding a shape noise component to the covariance matrix of the hypothetical MICE survey, generated with \blockfont{CosmoLike}. This scatter is chosen such that the uncertainty in the shape noise dominated regime is equivalent between MICE and \blockfont{metacal}: $\Delta C^{\kappa \kappa}_{\mathrm{\blockfont{metacal}}}(\ell) = \Delta C^{\kappa \kappa}_{\mathrm{MICE}}(\ell)$, with 
\begin{equation}
\Delta C^{\kappa\kappa}(\ell)\sim\frac{1}{\sqrt{f_{\textrm{sky}}}}\left(\frac{\sigma_{e}^{2}}{\bar{n}}\right)
\end{equation}
where $f_{\mathrm{sky}}$ is the observed fraction of the total sky area, $\bar{n}$ is the (effective) galaxy number density and $\sigma^2_e$ is the variance of the ellipticities. With $\bar{n}$ and $f_{\mathrm{sky}}$ fixed in both catalogs (our MICE catalog has approximately 130M galaxies over an area of $5191$ deg$^2$ while \blockfont{metacal} has 100M galaxies spanning an area of $4143$ deg$^2$), and $\sigma_e=0.261$ fixed in \blockfont{metacal}, we adjust $\sigma_{e,\mathrm{MICE}}$ so that the equality is reached, and verify that we achieve a comparable constraining power as DES Y3 in the 2D $S_8\times\Omega_\mathrm{m}$. The IA contribution to the MICE shears ($\gamma_I$) has, additionally, an irreducible noise coming from a randomization of galaxy orientations. This randomization is a component of the IA modeling
and has been tuned to match observed IA correlations from the BOSS LOWZ sample \citep{Singh16}.

It is also desirable that the source galaxy photometric properties in the mock also resemble those of galaxies in the fiducial DES Y3 catalog, since IAs can depend on galaxy color and luminosity. In summary, we verify that the distribution of $riz$ magnitudes and color-magnitude distributions are similar between MICE and \blockfont{Metacalibration}. It would also be beneficial if the redshift distributions $n(z)$ on the mock catalog were closely matched to what is found in real data (Fig. \ref{fig: nz}). Since the implementation in MICE of the SOMPZ framework utilized in DES Y3 (\citealt*{y3-sompz}; \citealt*{y3-sompzbuzzard}) is beyond the scope of this work, and since the goal of this model test should be agnostic to the exact shape of the redshift distributions, we utilize Directional Neighborhood Fitting (DNF) photo-$z$s \citep{DeVicente2016} for our mock $n(z)$s, as these are made available by default with the MICE data releases. In general, we find that DNF redshift distributions are narrower and less overlapping across bins than the fiducial DES Y3 $n(z)$s.

With the mock catalog constructed, we then apply the same pipelines that were used for measuring the tomographic cosmic shear data vector $\xi^{ij}_\pm(\theta)$ and inferring cosmological parameters. Where possible we mirror the model choices and nuisance parameter marginalization used in the analysis of the real data. 

We measure 2 sets of shear correlation functions from the mock: a ``MICE Baseline'' data vector estimated from the IA-free GG signal (i.e. using only the $\gamma^G$ component of Eq. \ref{eq:mice:gamma}), and a ``MICE IA'' data vector which includes the full GG+GI+II signal, as well shape noise added at the catalog level. Since the former captures the cosmological signal (plus cosmic variance), it provides a useful fiducial case, relative to which we can gauge biases. The input cosmology is known for MICE, however the IA signal is, by construction, not exactly mapped into a set of NLA or TATT parameters. Thus, for a single realization of the mock, our expectations are: 1) that the constraints on the MICE Baseline GG-only vector are consistent with the input cosmology while at the same time the IA parameter constraints are consistent with zero; 2) that cosmology constraints on the full MICE IA data vector are not excessively biased with respect to the baseline constraints. 

Both our expectations are fulfilled, with MICE IA offset by  $\sim 0.6\sigma$ from the peak of Baseline. Our results are shown in Fig. \ref{fig:  mice_post}. Although this shift is larger than the $0.3\sigma$ tolerance we applied to previous tests on analytic data, it is worth bearing in mind that the nature of the test discussed here is slightly different. That is, we only have one realization of the noise in the IA vector (both cosmic variance and shape noise). While this test is valuable, in the sense that large shifts would be a sign of significant problems in the fidelity of our model, offsets within $\sim 1-2\sigma$ are entirely consistent with noise. 

\begin{figure}
	\includegraphics[width=\columnwidth]{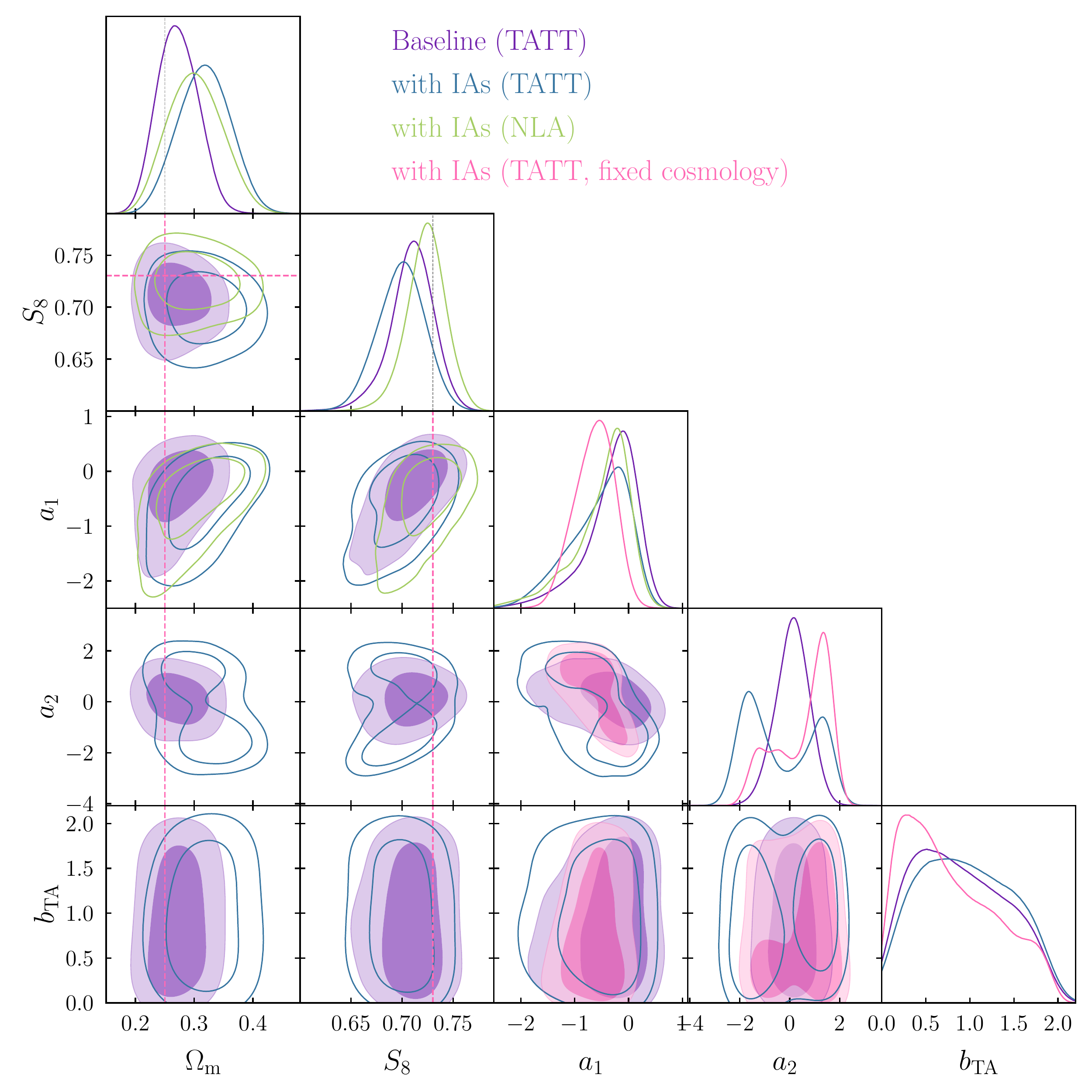}
    \caption{MICE posteriors: The baseline case (purple) includes only the GG part of the signal (no IAs) and IA parameters are consistent with zero. The contours labelled ``with IAs" (blue, green, pink) includes GG+GI+II, with the IA model indicated in parentheses. Note that two IA redshift power law indices are included in the fits, as per the fiducial Y3 model; they are not shown here as they are only very weakly constrained, and the contours are uninformative. Dashed pink lines are the input MICE cosmology.}
    \label{fig: mice_post}
\end{figure}

We also run the inference pipeline on the full IA data vector while fixing the cosmological parameters at the true MICE input to isolate the IA signal. Those are the ``reduced'' parameter space curves in Fig. \ref{fig: mice_post}. While constraints obtained this way are artificially tight, the exercise shows that the IA parameters measured in the reduced case are consistent within 1$\sigma$ with parameters obtained while varying cosmology. Additionally, we do find a marginal preference for a negative $a_1$ parameter in both cases that involve nonzero IA parameters. Although this is likely not a physical effect, and nor in this case can it be the result of unmodeled photo-$z$ error, it is quite conceivable that it is simply a result of a small but positive ``true" IA signal, combined with shape noise. 

This test on MICE is aimed at verifying that our fiducial model can recover unbiased cosmological constraints from a data vector that includes intrinsic alignments and that is not generated by an analytic pipeline. While we do obtain reasonable results and the present test has the rare benefit of including a realistic, simulated ($N-$body with semi-analytic model) IA signal, it is statistically weaker than a more complete validation of DES pipelines on simulations \citep{y3-simvalidation}, mainly due to the lack of independent realizations. A complete analysis on a relatively large number of MICE mocks is left for future work. 

\section{Robustness of IA to Unmodeled Redshift Error}\label{sec:redshift_appendix}

In this Appendix we set out in more detail the tests on analytic data introduced in Sec. \ref{section:simulated_ia_tests}. These tests, which are based on noiseless analytic data, are designed to test the specific question of whether our choice of IA model behaves well in the presence of realistic errors in the redshift distributions. We focus here on synthetic data vectors which do not contain real data complexity, in which case a thorough testing of redshift calibration is more complicated and performed by our companion \cite{y3-cosmicshear1}. 
We use the ensemble of 6000 SOMPZ realizations of $n(z)$'s, described in \citet*{y3-sompz}; for each sample of the Y3 redshift distributions, we compute a simulated cosmic shear data vector $\xi_{\pm}^{\mathrm{sample}}(\theta)$ at a fixed set of input cosmological parameters. Using the data covariance, we can then calculate the $\Delta \chi^2$ between $\xi_{\pm}^{\mathrm{sample}}(\theta)$ and a similarly generated cosmic shear data vector, but with the fiducial (mean) $n(z)$ as input, $\xi_{\pm}^{\mathrm{mean}}(\theta)$. We find that the distribution of $\Delta \chi^2$ obtained for the 6000 samples peaks close to zero ($\Delta \chi^2<1.0$, corresponding to small perturbations of the data vector) and has a long tail out to more extreme cases ($\Delta \chi^2\sim50.0$). 
Given our finite computing resources, we do not run nested sampling chains on all of the 6000 scenarios, but rather choose three of increasing severity, corresponding to the $60$th, $95$th and $99$th percentiles of the $p(\Delta \chi^2)$ distribution. 

We next translate these $\Delta \chi^2$ errors into cosmological biases by analysing the three $\xi_{\pm}^{\mathrm{sample}}(\theta)$ vectors with our fiducial setup. Although that setup includes redshift nuisance parameters $\Delta z^i$, it cannot explicitly account for perturbations in the $n(z)$ shape that don't map well into the first moment of the $n(z)$ distribution. More extensive testing of this is presented in \citet{y3-cosmicshear1}; \citet*{y3-hyperrank}. 

\begin{figure}
	\includegraphics[width=\columnwidth]{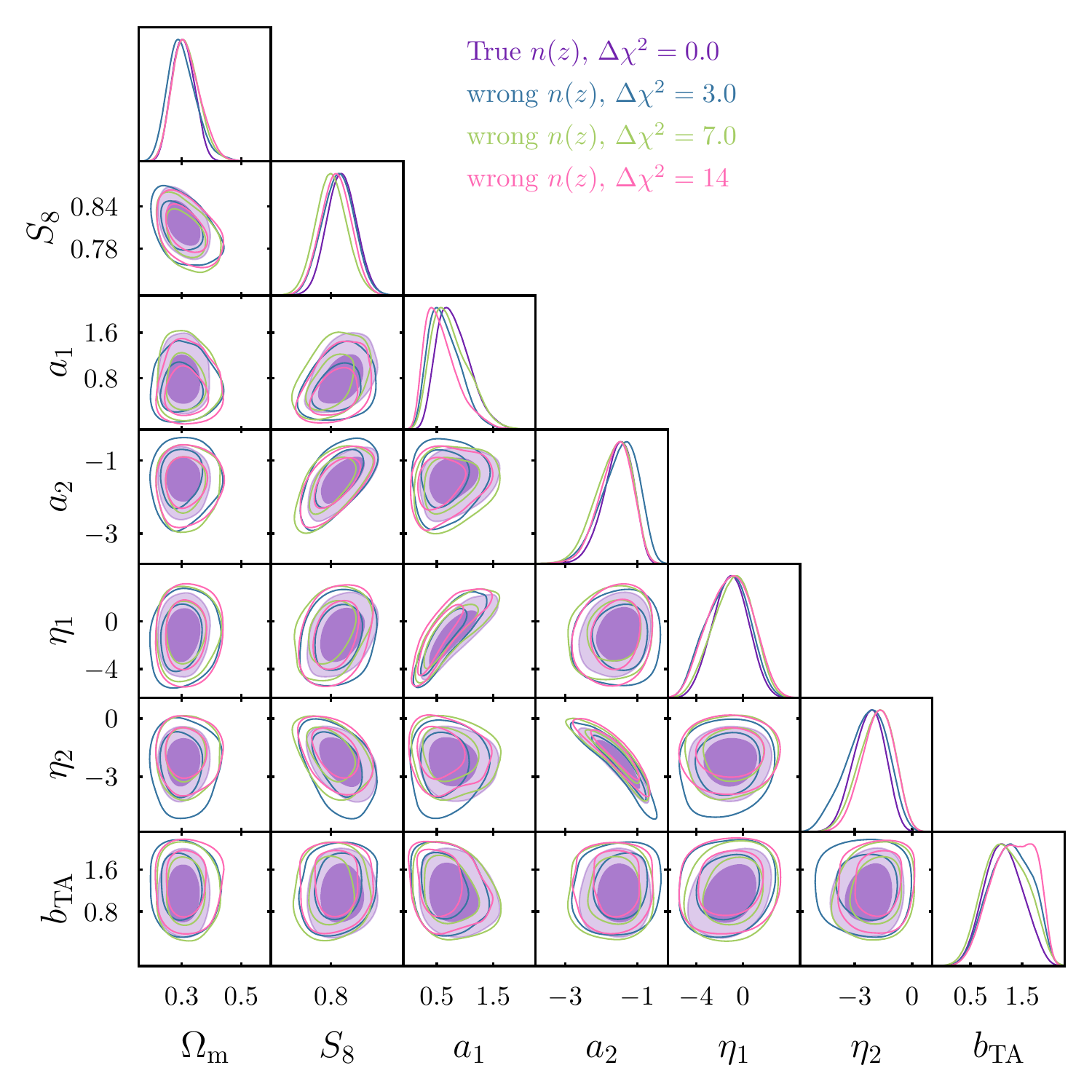}
    \caption{Analysis of synthetic data generated with the inclusion of unmodeled redshift errors. The purple contour shows the baseline posterior, obtained by analysing a simulated TATT data vector with the correct $n(z)$; the three other colours show redshift error scenarios of varying extremity. In each case the difference in $\chi^2$ induced by the redshift error, at the input point in parameter space, is indicated in the legend.}
    \label{fig:iaxnz}
\end{figure}

The results of this exercise are shown in Fig.~\ref{fig:iaxnz}. Each of the unfilled contours represents a different scenario of redshift error, and can be compared with the solid purple (which has no redshift error). In this limited testing scenario with synthetic data we see no large biases in the IA model parameters or $S_8$, which are recovered correctly, and no hints that the TATT model parameters are absorbing the unmodeled error. 

\section{Comparing NLA and TATT in DES Y3}
\label{sec: appendix NLA vs TATT}

In this appendix we explore in more detail the differences between our fiducial IA model for Y3 (TATT), and the more commonly used NLA model. As discussed in Sec. \ref{sec:ias:modeling}, TATT is a physically-motivated model containing NLA, such that in some limit ($a_2,b_{\rm TA} \rightarrow 0 $), the two are the same. 
Amongst various other permutations, we fit both the two-parameter NLA model and the five-parameter TATT model to the Y3 cosmic shear data. Using the more complex IA model results in a slight widening of the contours, as can be seen in Fig. \ref{fig: IAs_tatt_nla}, and also a rotation in the direction of the $\sigma_8-\omegam$ degeneracy. These two effects combine to produce a $\sim17\%$ difference in the projected error bar on $S_8$ between the two models. When considering the 2D Figure of Merit in the $S_8-\omegam$ plane, the difference is $\sim 25\%$.

\begin{figure*}
	\includegraphics[width=1.6\columnwidth]{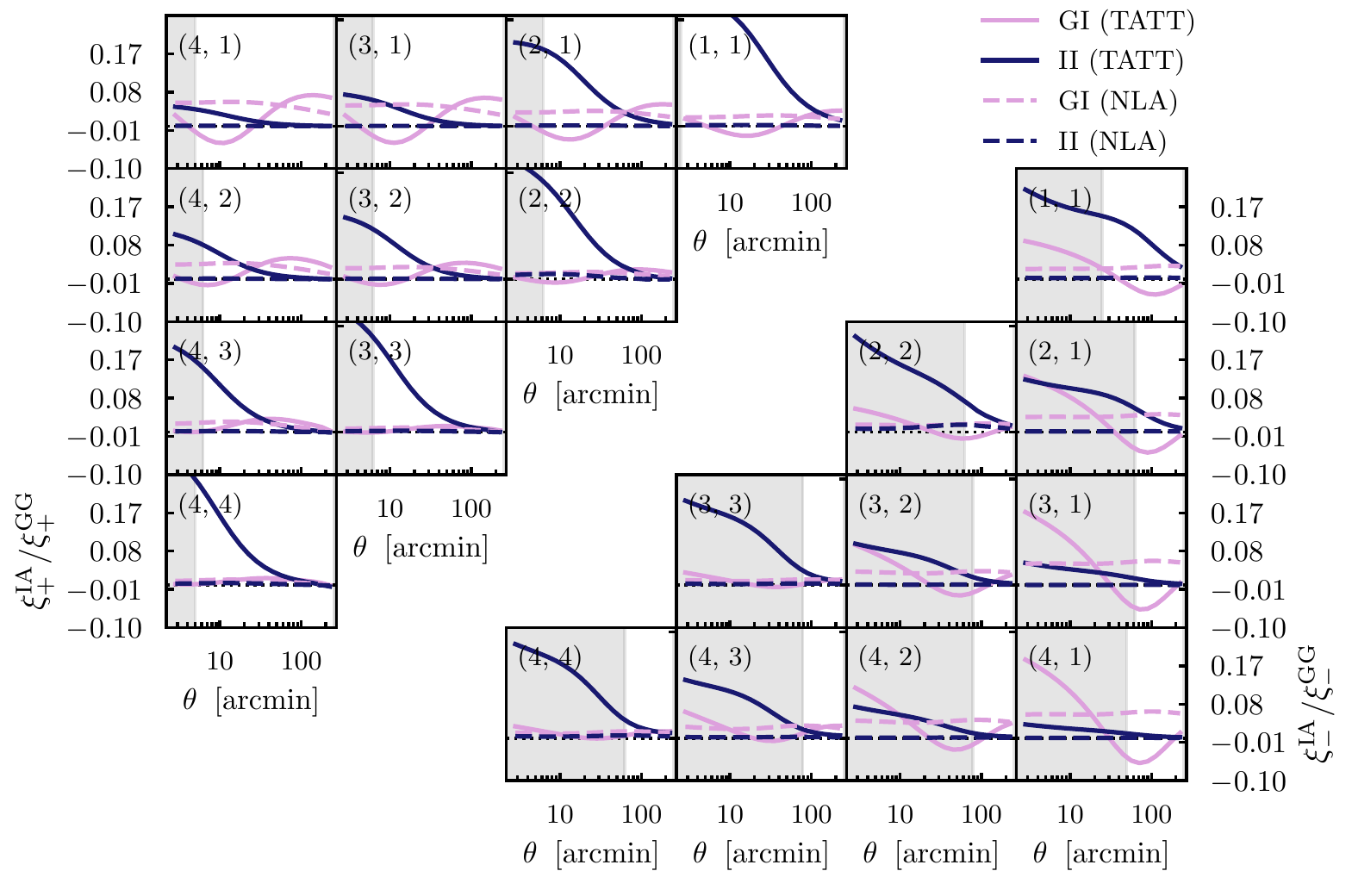}
	\includegraphics[width=1.6\columnwidth]{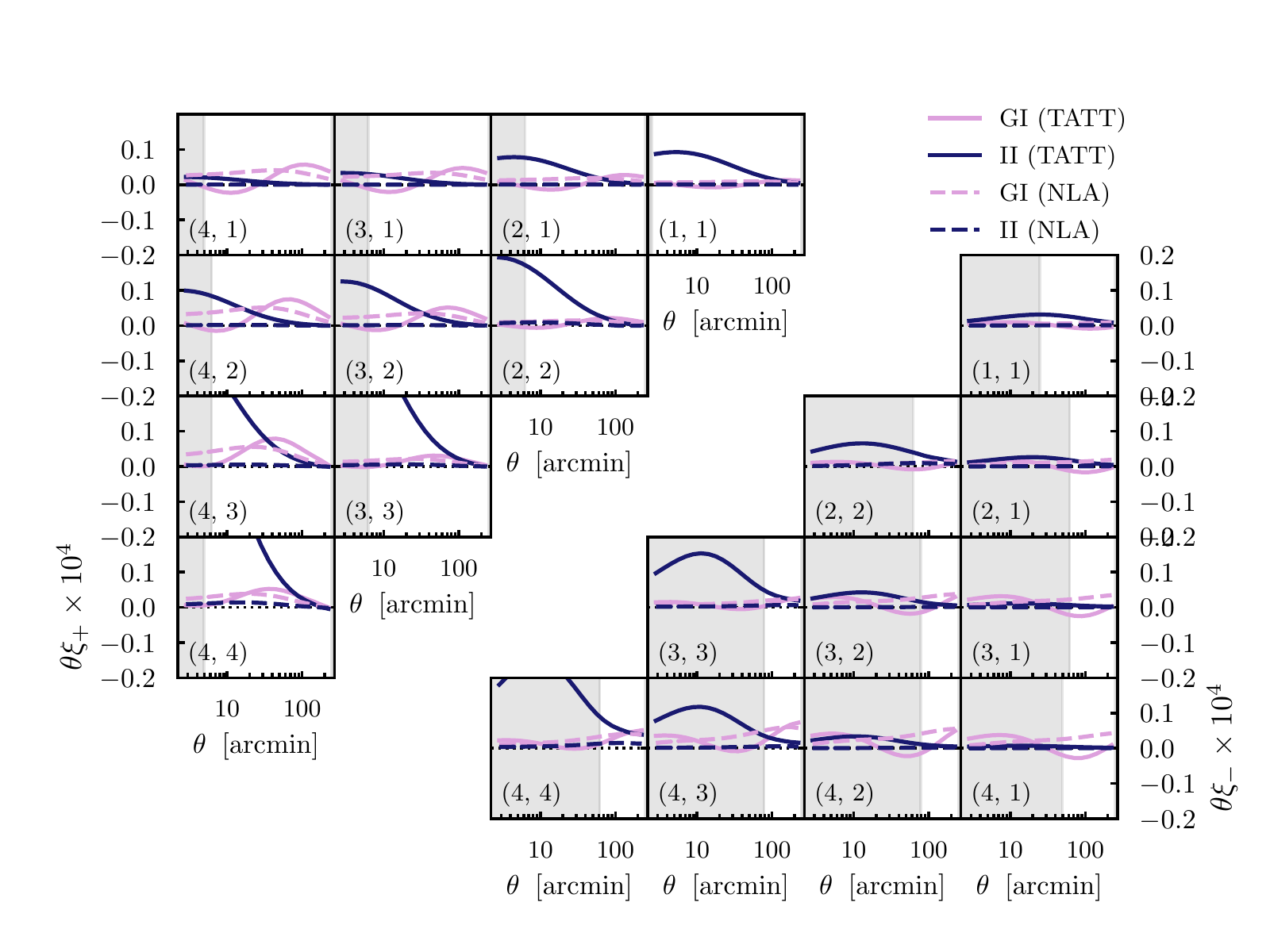}
    \caption{The best fitting IA predictions, based on NLA and TATT analyses. In the two panels we show the fractional IA contributions relative to the cosmological signal (top), and the absolute contributions (bottom). Both IA components, GI and II, are plotted separately (blue and light purple). As before, shaded regions indicate scales discarded from the analyses.}
    \label{fig:nla_tatt_dvs}
\end{figure*}

In addition to the difference in constraining power, NLA and TATT result in qualitatively different predictions for the IA signal. We illustrate this in Fig.~\ref{fig:nla_tatt_dvs}, which shows the theory IA contributions, generated using the respective best fitting IA parameters from the two chains. Both the absolute signal, and the contribution relative to the cosmological shear (GG) are included in the two panels. Although the choice of model does not change the conclusion that IAs are subdominant in all bins (at the level of a few percent), the shape and sign of the IA correlations do differ somewhat. Strikingly, TATT predicts a much stronger II signal than NLA, which is largely driven by the tidal torque ($a_2$) part of the model.   
It is also worth remarking here that although they look different, neither of these IA scenarios has been convincingly ruled out by direct measurements. That is, if one generates projected intrinsic alignment correlations (see e.g. \citealt{mandelbaum06} Sec. 3.2 for the full definition) $w_{g+}$ and $w_{++}$ predictions, using the IA power spectra from these fits, the results are within the range allowed by measurements on faint blue and red galaxies at low redshift from SDSS \citep{mandelbaum06, Singh15}.

\section{Full Posterior Constraints}\label{sec: appendix contour plot}

\begin{figure*}
	\includegraphics[width=2.0\columnwidth]{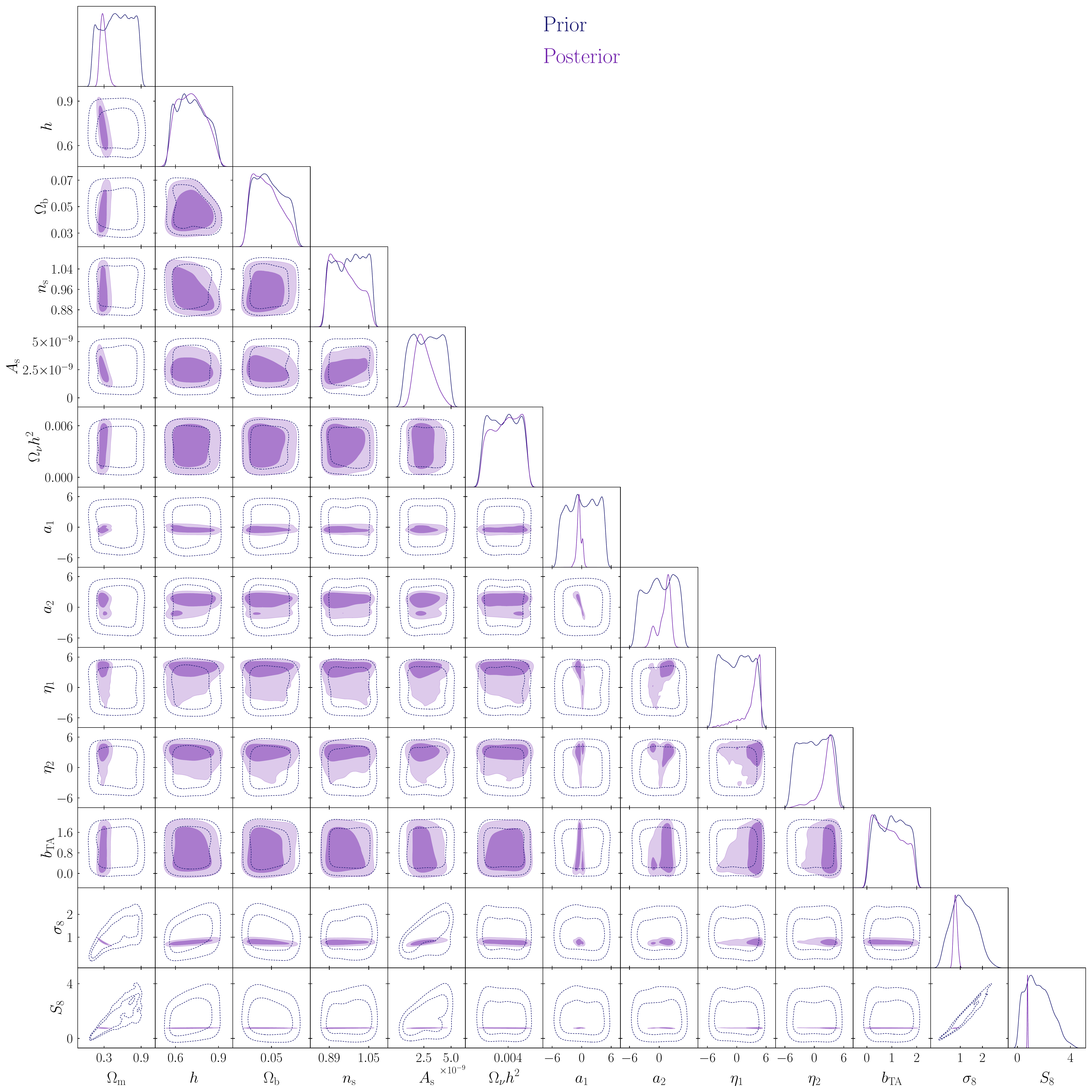}
    \caption{Posterior distributions (purple solid) and priors (blue dashed) of our analysis. We intentionally leave out hard-boundary corrections on the distributions above for ease of visualization. Our 1D marginalized priors are flat in the cosmological parameters described in Table~\ref{table: priors}, but sampling  and 1-dimensional marginalization can make them look noisy and non-uniform.}
    \label{fig:posteriors_all}
\end{figure*}

We show a set of 2D projections of our cosmological and IA parameter posteriors in Fig. \ref{fig:posteriors_all}, along with their priors. We note that $\sigma_8$ and $S_8$ are derived parameters which we do not sample over, so projections make their apparent priors non-uniform. Many of the parameters in our analysis are either prior dominated, or not of physical interest, and so are not shown in the main part of the paper. They are included here for completeness.  

\section{Changes After Unblinding}

Three changes to our analysis happened after unblinding, neither of them affecting significantly the DES Y3 cosmic shear cosmological constraints, particularly $S_8$. The first is a planned modification to the covariance matrix: after the tests described in Sec. \ref{section: unblinding} succeed and cosmology constraints are obtained, the analytic covariance matrix is re-computed at the best-fit (maximum posterior) parameters of the full $3\times2$pt data, and our cosmological constraints are updated with new nested sampling chains.   

The second change involves the DES Y3 lens sample, which only enters the cosmic shear data via lensing ratios (Sec. \ref{section:SR}). We substitute the original \textsc{redMaGiC} sample by the magnitude-limited \blockfont{Maglim} sample. We point the reader to \citet{y3-3x2ptkp} for further details and motivation for this change.

The third change is decision for presenting the \lcdm-Optimized scale cuts in cosmic shear as one of the main DES Y3 results. While fundamentally the optimized scale cuts also lead to robust $<0.3\sigma$ shifts in our simulated analysis and therefore could have been a reasonable pre-unblinding choice on a $\xi_\pm$-only analysis, this was only decided after unblinding. We report that here for transparency.

\label{lastpage}
\end{document}